\documentclass[]{aa} 

\usepackage[breaklinks, colorlinks, citecolor=blue, linkcolor=blue]{hyperref}
\usepackage{amsmath}
\usepackage{mathtools}
\usepackage{graphicx}
\usepackage{txfonts}
\usepackage{float}

\makeatletter
\DeclareRobustCommand{\ion}[2]{%
  \text{#1\,\check@mathfonts\fontsize\sf@size\z@\selectfont #2}%
}
\makeatother

\begin{document}

   \title{Resolved stellar population properties of PHANGS-MUSE galaxies}


   \author{
   I.~Pessa\inst{\ref{aip},\ref{mpia}}\and 
   E.~Schinnerer\inst{\ref{mpia}} \and 
    P.~Sanchez-Blazquez\inst{\ref{ucm}} \and 
    F.~Belfiore\inst{\ref{inaf}} \and 
    B.~Groves\inst{\ref{ICRA},\ref{Canb}} \and
    E.~Emsellem\inst{\ref{eso},\ref{lyon}}  \and
    J.~Neumann\inst{\ref{mpia},\ref{Portsmouth}} \and   
   A.~K.~Leroy\inst{\ref{ohio}} \and 
   F.~Bigiel\inst{\ref{bonn}} \and
   M.~Chevance\inst{\ref{zah},\ref{cool}} \and
    D.~A.~ Dale\inst{\ref{wyo}} \and 
    S.~C.~O.~Glover\inst{\ref{zah}} \and 
    K.~Grasha\inst{\ref{Canb}} \and 
    R.~S.~Klessen\inst{\ref{zah},\ref{zw}} \and 
    K.~Kreckel\inst{\ref{rechen}} \and
    J.~M.~D.~Kruijssen\inst{\ref{cool}} \and
    F.~Pinna\inst{\ref{mpia}} \and
   M.~Querejeta\inst{\ref{oan}} \and 
   E.~Rosolowsky\inst{\ref{alb}} \and
   T.~G.~Williams\inst{\ref{mpia}, \ref{ox}}
   }
   \institute{Leibniz-Institut f\"ur Astrophysik Potsdam (AIP), An der Sternwarte 16, 14482 Potsdam, Germany\label{aip}\\
        \email{ipessa@aip.de}
        \and Max-Planck-Institute for Astronomy, K\"onigstuhl 17, D-69117 Heidelberg, Germany\label{mpia}
        \and Departamento de F\'isica de la Tierra y Astrof\'isica, Universidad Complutense de Madrid, E-28040 Madrid, Spain \label{ucm}
        \and INAF — Osservatorio Astrofisico di Arcetri, Largo E. Fermi 5, I-50125, Florence, Italy\label{inaf}
        \and International Centre for Radio Astronomy Research University of Western Australia 7 Fairway, Crawley WA 6009 Australia\label{ICRA}
        \and Research School of Astronomy and Astrophysics, Australian National University, Canberra, ACT 2611, Australia\label{Canb}      
        \and European Southern Observatory, Karl-Schwarzschild-Stra{\ss}e 2, 85748 Garching, Germany\label{eso}
        \and Univ Lyon, Univ Lyon1, ENS de Lyon, CNRS, Centre de Recherche Astrophysique de Lyon UMR5574, F-69230 Saint-Genis-Laval France\label{lyon}
         \and Institute of Cosmology and Gravitation, University of Portsmouth, Burnaby Road, Portsmouth, PO1 3FX, UK\label{Portsmouth}
         \and Department of Astronomy, The Ohio State University, 140 West 18th Avenue, Columbus, OH 43210, USA\label{ohio}
         \and Argelander-Institut f\"ur Astronomie, Universit\"at Bonn, Auf dem H\"ugel 71, D-53121 Bonn, Germany\label{bonn}
         \and Universit\"at Heidelberg, Zentrum f\"ur Astronomie, Institut f\"ur theoretische Astrophysik, Albert-Ueberle-Stra{\ss}e 2, D-69120, Heidelberg, Germany\label{zah}
         \and Cosmic Origins Of Life (COOL) Research DAO, coolresearch.io\label{cool}
         \and Department of Physics and Astronomy, University of Wyoming, Laramie, WY 82071, USA\label{wyo}
         \and Universit\"at Heidelberg, Interdisziplin\"ares Zentrum f\"ur Wissenschaftliches Rechnen, Im Neuenheimer Feld 205, D-69120 Heidelberg, Germany\label{zw}
         \and Astronomisches Rechen-Institut, Zentrum f\"ur Astronomie der Universit\"at Heidelberg, M\"onchhofstra{\ss}e 12-14, D-69120 Heidelberg, Germany\label{rechen}
         \and Observatorio Astronómico Nacional (IGN), C/Alfonso XII, 3, E-28014 Madrid, Spain\label{oan}
         \and Department of Physics, University of Alberta, Edmonton, AB T6G 2E1, Canada\label{alb}
         \and Sub-department of Astrophysics, Department of Physics, University of Oxford, Keble Road, Oxford OX1 3RH, UK\label{ox}
        %
        }

   \date{Received XX, YY; Accepted XX XX, YY}

 
  \abstract
  {
Analyzing resolved stellar populations across the disk of a galaxy can provide unique insights into how that galaxy assembled its stellar mass over its lifetime.
Previous work at  $\sim$1\,kpc resolution has already revealed common features in the mass buildup (e.g., inside-out growth of galaxies). However, even at approximate kpc scales, the stellar populations are blurred between the different galactic morphological structures such as spiral arms, bars and bulges.  Here we present a detailed analysis of the spatially resolved star formation histories (SFHs) of 19 PHANGS-MUSE galaxies, at a spatial resolution of $\sim100$~pc. We show that our sample of local galaxies exhibits predominantly negative radial gradients of stellar age and metallicity, consistent with previous findings, and a radial structure that is primarily consistent with local star formation, and indicative of inside-out formation. In barred galaxies, we find flatter metallicity gradients along the semi-major axis of the bar than along the semi-minor axis, as is expected from the radial mixing of material along the bar during infall. In general, the derived assembly histories of the galaxies in our sample tell a consistent story of inside-out growth, where low-mass galaxies assembled the majority of their stellar mass later in cosmic history than high-mass galaxies (also known as ``downsizing''). We also show how stellar populations of different ages exhibit different kinematics. Specifically, we find that younger stellar populations have lower velocity dispersions than older stellar populations at similar galactocentric distances, which we interpret as an imprint of the progressive dynamical heating of stellar populations as they age. Finally, we explore how the time-averaged star formation rate evolves with time, and how it varies across galactic disks. This analysis reveals a wide variation of the SFHs of galaxy centers and additionally shows that structural features become less pronounced with age.}

\maketitle

\section{Introduction}
\label{sec:intro}

Understanding the process by which galaxies assemble their stellar mass through cosmic time is a crucial aspect of galaxy evolution. This process can be modulated either by external (e.g., galaxy merging) or internal (e.g., secular evolution,   star formation) mechanisms \citep{Kormendy2013}. These different mechanisms ultimately lead to the wide variety of galaxy properties that we see in the present-day Universe, such as their morphologies and the demographics of their stellar populations. The study of stellar populations through the fossil method \citep{Tinsley1968}, which consists of reproducing the observed spectra with a linear combination of single stellar populations (SSPs) of known ages and metallicities, has been demonstrated to be a powerful tool in unveiling the assembly histories of galaxies 
\citep[e.g., ][]{Conroy2013,McDermid2015,Wilkinson2015, LopezFernandez2018, Zhuang2019, Neumann2020}. 

Although the investigation of stellar populations has been used as an approach to address the evolution of galaxies for many years \citep{Sarzi2005, Rogers2007}, large integral field spectroscopy (IFS or IFU) surveys, such as CALIFA \citep{Sanchez2012}, MaNGA \citep{Bundy2015}, and SAMI \citep{Croom2012}, have further improved these analyses, enabling the simultaneous extraction of spectra from different regions in nearby galaxies, with typical spatial resolutions of $\sim$1 kpc, which has allowed for the characterization of stellar populations and ionized gas properties across galactic disks. This acknowledges the fact that galaxies are extended and complex objects of which the stellar population and gas properties might change drastically from one specific region to another.

One of the most studied features in the context of spatially resolved stellar populations is the galactocentric radial distribution of their properties such as age and metallicity \citep[see, e.g., ][]{SanchezBlazquez2014, GonzalezDelgado2014, GonzalezDelgado2016, IbarraMedel2016, Zheng2017, GarciaBenito2017, LopezFernandez2018, Zhuang2019, Dale2020, Parikh2021}. These studies have revealed that radial gradients of stellar age and metallicity are primarily connected with the Hubble type of the galaxy \citep{Goddard2017, Parikh2021, Smith2022}, with late-type galaxies (LTGs) showing negative age and metallicity gradients (i.e., centers are older and more metal-rich), and early-type galaxies (ETGs) showing a nearly flat age gradient, and a negative metallicity gradient (albeit flatter than that found for LTGs). These works also report a dependence of the gradients on total stellar mass, where more massive galaxies exhibit steeper gradients.

In terms of galaxy evolution, negative age gradients point to an inside-out growth of galaxies, where the inner regions assembled their stellar mass earlier than outer regions. In this regard, \citet{IbarraMedel2016} found that LTGs would have a more pronounced inside-out formation mode, compared to ETGs (i.e., a more obvious difference between the formation times of inner and outer regions), and that low-mass galaxies (M$_{*}$ < $5\times10^{9}$) show a large diversity in their radial assembly history.

With respect to the origin of the metallicity gradients, \citet{Zhuang2019} find that they are a reflection of the local stellar mass surface density-metallicity relation \citep{GonzalezDelgado2014B}, and conclude that the spatial distribution of stellar populations within a galaxy is primarily the result of the insitu local star formation history (SFH), rather than being shaped by radial migration. However, \citet{Neumann2021} report that the local correlation between stellar metallicity and stellar mass surface density ($\Sigma_{*}$) shows an increased scatter toward the outer parts of galaxies, suggesting that something else, besides $\Sigma_{*}$, drives chemical enrichment at larger galactocentric distances (e.g., gas accretion, outflows). \citet{Zhuang2019} also find that low-mass LTGs show more commonly positive metallicity gradients, consistent with stellar feedback being more efficient at regulating the baryon cycle in the central regions of these galaxies.  

Beyond radial gradients of stellar age and metallicity, recent studies have aimed to study the assembly history of the galactic disk in a 2D manner \citep{Guerou2016, Peterken2019, Pinna2019a, Pinna2019b, Peterken2020}, by ``time-slicing'' galaxies across their lifetime (i.e., studying the spatial distribution across the galaxy of stars that were born at a given epoch). However, the limited spatial resolutions of large IFU surveys place limitations on this kind of study. At a spatial resolution of $\sim1$ kpc, much of the galactic structure is not resolved \cite[e.g., spiral arms typically have widths of a few hundred parsecs; ][]{Querejeta2021}. On the other hand, higher resolution measurements \citep[e.g., ][]{Guerou2016} are often limited by small sample sizes.

In this paper, we present what we learn from spatially resolved SFHs of a sample of LTGs from the Physics at High Angular resolution in Nearby GalaxieS (PHANGS\footnote{\url{http://phangs.org/}}) survey \citep{Leroy2021b} survey, measured at spatial scales of $\sim100$~pc, at which different morphological components are clearly resolved and can be studied separately. Thus, we address how the properties of stellar populations vary as a function of the local galactic environment and assess the importance/role of galactic structure in the assembly history of a sample of nearby star forming galaxies. It is worth noticing that while there are other optical IFU surveys with similar spatial resolution that allow the study of the stellar population properties of nearby galaxies, such as TIMER \citep{Gadotti2019} and Fornax3D \citep{Sarzi2018, Iodice2019}, they were designed to address specific science questions and their samples are therefore not representative of the population of star-formation main-sequence galaxies in the local Universe. Here we focus primarily on a detailed characterization of the radial trends measured in the stellar population properties of star forming galaxies, and how these trends correlate with global galactic properties. We also show variations in the SFH across the galactic disk of galaxies and study the correlation between kinematic properties and age of their stellar populations.

This paper is structured as follows. In Sec.~\ref{sec:data} we present the data and data products used in our analysis. In Secs.~\ref{sec:radil_profiles_ag_z},~\ref{sec:inside_out}, and~\ref{sec:radial_mix_bar}  we present and discuss the radial distribution of the stellar population properties of the galaxies in our sample, for different local galactic environments. In Sec.~\ref{sec:dynamical_heating} we show how the kinematic properties of stellar populations change as a function of stellar age, and in Sec.~\ref{sec:SFH_Disk} we show how the SFHs of galaxies varies across different galactic regions (i.e., not radially integrated). We present a summary and the conclusions of our analysis in Sec.~\ref{sec:summary_stpops}, and we validate a revised set of stellar templates to improve fitting to the youngest regions in Appendix~\ref{sec:improving_young}.

\section{Data}
\label{sec:data}

We use a sample of $19$ star-forming galaxies, all of them close to the star-forming main sequence of galaxies \citep[SFMS; e.g., ][]{Brinchmann2004, Daddi2007, Noeske2007}. These galaxies represent a subsample of the PHANGS survey \citep{Leroy2021b}, and are observed as part of the PHANGS-MUSE survey~\cite[PI: E.~Schinnerer;][]{Emsellem2021}. The PHANGS galaxies have been chosen to constitute a representative set of star-forming disk galaxies where most of the star formation is occurring in the local Universe. They have been selected to be at distances smaller than $20$~Mpc to resolve the typical scale of star-forming regions ($50{-}100$~pc), moderately inclined ($i<60^{\circ}$) to limit the effect of extinction and allow the identification of individual star-forming regions. Their stellar mass limit is defined  by $\log_{10}(M_{*}/M_{\odot})\gtrsim9.75$, which translates to nearly twice the mass of the LMC, aiming to avoid focusing on low metallicity, low mass galaxies, where the detection of CO becomes increasingly harder \citep[e.g., ][]{Schruba2012, Bolatto2013, Cormier2014, Schruba2017}. We note that this mass cutoff does not strictly apply to the PHANGS-MUSE subsample because a revision of the fiducial distance estimates \citep{Anand2021}. Table~\ref{tab:sample_ch3} summarizes the properties our sample. We use the global parameters reported by \citet{Leroy2021b} and the distance compilation of \citet{Anand2021}. The inclination values adopted are those reported by \citet{Lang2020}.

\subsection{VLT/MUSE}
We use data from the PHANGS-MUSE survey. This survey employs the Multi-Unit Spectroscopic Explorer \citep[MUSE;][]{Bacon2014} optical integral field unit (IFU) mounted on the VLT UT4 to mosaic the star-forming disk of $19$ galaxies from the PHANGS sample. 

The mosaics combine three to $15$ individual MUSE pointings. Each MUSE pointing provides a $1\arcmin \times 1\arcmin$ field of view sampled at $0\farcs2$ per pixel, with a typical spectral resolution of ${\sim}2.5$~\AA\ FWHM (${\sim}100$~km~s$^{-1}$) covering the wavelength range of $4800{-}9300$~\AA, and with angular resolutions ranging from ${\sim}0\farcs5$ to ${\sim}1\farcs0$ for the targets observed with and without adaptive optics (AO), respectively. The nine galaxies observed with AO are marked with a black dot in the first column of Table~\ref{tab:sample_ch3}. The total on-source exposure time per pointing for galaxies in the PHANGS-MUSE Large Program is $43$~min. Observations were reduced using recipes from the MUSE data reduction pipeline provided by the MUSE consortium \citep{Weilbacher2020}, executed with ESOREX using the python wrapper developed by the PHANGS team\footnote{\url{https://github.com/emsellem/pymusepipe}} \citep{Emsellem2021}. The point spread function of the different individual MUSE exposures that are mosaicked together have been homogenized to the largest FWHM within each mosaic \citep[i.e., dataset labeled as ``copt'' in the public release, see][for a detailed description of how the homogenization is done]{Emsellem2021}. The total area surveyed by each mosaic ranges from $23$ to $441$~kpc$^{2}$. Once the data have been reduced, we have used the PHANGS data analysis pipeline (DAP) to derive various physical quantities. Some of these outputs are described in Sec.~\ref{sec:stellarpops_maps}. The DAP is described in detail in \citet{Emsellem2021}. It consists of a series of modules that perform single stellar population (SSP) fitting and emission line measurements to the full MUSE mosaic.

\subsection{Stellar population property maps}
\label{sec:stellarpops_maps}

The PHANGS-MUSE DAP \citep{Emsellem2021} includes a stellar population fitting module, which uses a linear combination of SSP templates of known ages, metallicities, and mass-to-light ratios is used to reproduce the observed spectrum. This allows us to infer stellar population properties from an integrated spectrum, such as mass- or luminosity-weighted ages, metallicities, and total stellar masses, together with the underlying SFH.

The full mosaics are initially corrected for Milky Way extinction, using an $E(B{-}V)$ value obtained from the NASA/IPAC Infrared Science Archive\footnote{\url{https://irsa.ipac.caltech.edu/applications/DUST/}} \citep{Schlafly2011}, and assuming a \citet[][]{Cardelli89} extinction law. In detail, our spectral fitting pipeline performs the following steps:

First, we apply a Voronoi tessellation \citep{Capellari2003} to bin our MUSE data to a minimum signal-to-noise ratio (S/N) of ${\sim}35$, computed at the wavelength range of $5300{-}5500$~\AA. This value is chosen in order to keep the relative uncertainty in our mass measurements below $15\%$, even for pixels dominated by a younger stellar population. To do this, we tried different S/N levels to bin a fixed region in our sample, and we bootstrapped our data to have an estimate of the uncertainties at each S/N level.

We use then the Penalized Pixel-Fitting (pPXF) code \citep{Capellari2004, Capellari2017} to fit the spectrum of each Voronoi bin. We fit the wavelength range $4850{-}7000$~\AA, in order to avoid spectral regions strongly affected by sky residuals. We further mask the regions around strong emission lines, as well as regions of the spectrum particularly affected by sky line residuals.

We fit our data with a grid of templates consisting of $17$ ages, ranging from $6.3$~Myr to  $15$~Gyr, logarithmically-spaced, and four metallicity bins $\mathrm{[Z/H]} = [-0.7, -0.4, 0, 0.22]$. The templates are from the E-MILES \citep{Vazdekis2010, Vazdekis2012} database, assuming a \citet{Chabrier2003} IMF and Padova isochrones \citep{Girardi2000}. The E-MILES templates were originally computed to start from a minimum age of $63$ Myr (for the Padova isochrones). We have complemented this original set of templates with the young extension to the E-MILES SSP models presented in \citet{Asad2017}, adding five additional age bins, from $6.3$~Myr to $\sim40$~Myr. These young templates have been computed for the same metallicity bins, except for the most metal-rich one, which for the young-extension templates has [Z/H] $= 0.41$ (instead of $0.22$). 

This choice of templates is different from that used in \citet{Emsellem2021} for the same data. In Appendix~\ref{sec:improving_young} we present an exhaustive exploration of the parameter space in order to improve the outcome of the spectral fitting of the regions dominated by young (<400 Myr) stellar populations, and justify this choice. We also refer the reader to \citet{Emsellem2021} for a detailed description of this issue. 

We implemented a two-step fitting process. First, we fit our data assuming a \citet{Calzetti2000} extinction law to correct for internal attenuation. We then corrected the observed spectrum using the measured attenuation value before fitting it a second time, including a $12$th degree multiplicative polynomial instead of attenuation in this iteration. This two-step fitting process accounts for offsets between individual MUSE pointings. These offsets arise because the different MUSE pointings were not necessarily observed under identical weather conditions, and variations in the sky continuum levels can lead to subtle differences in the flux calibration of individual neighboring MUSE pointing. An analysis of the regions of the mosaic where different pointings overlap revealed that these variations yield differences in the measured stellar extinction levels on the order of  $\Delta$E(B-V) $\sim0.04$~mag \citep[see][for a detailed description of this issue]{Emsellem2021}. Therefore, in the first iteration of the spectral fitting, we measure a reddening value and correct the observed spectra accordingly, and in the second iteration, we use a high-degree multiplicative polynomial to correct for those nonphysical features and homogenize the outcome of the different pointings. 

To recover the kinematic properties of the observed spectra, the templates are shifted in velocity space and convolved to match observed spectral features. We fit the first four velocity moments, using a single stellar kinematic component, that is, the same velocity and velocity dispersion, $h3$ and $h4$ is applied to all the templates used for the spectral fitting. The spectral resolution of  the E-MILES templates is higher than that of the VLT/MUSE data within the wavelength range considered (although at $\sim7000 \AA$ they are  virtually the same). Thus, templates are convolved to match the spectral resolution of the data, using an appropriate wavelength-dependent kernel, before the fit.

The output of \texttt{pPXF} consists of a vector with the coefficients of the linear combination of templates that best reproduce the observed spectrum. Physically, these weights represent the mass fraction of stars with a given age and metallicity, and they are used to derive stellar mass surface densities, and both light- and mass-weighted ages and metallicities in each spaxel. The stellar mass surface density maps include contributions from live stars and remnants. The total weight for a given age bin, integrating for the different metallicities, also represents the SFH of a given spectrum. For each pixel, the average age and metallicity are computed as follow:
\begin{equation}
\label{eq:age}
 \langle \log \mathrm{age} \rangle = \frac{\Sigma_{i}\, \log(\mathrm{age}_{i})\,w_{i}}{\Sigma_{i}\,w_{i}}
\end{equation}
and
\begin{equation}
\label{eq:z}
 \langle \mathrm{[Z/H]} \rangle = \frac{\Sigma_{i}\, \mathrm{[Z/H]}_{i}\, w_{i}}{\Sigma_{i}\, w_{i}}~,
\end{equation}
where age$_{i}$ and [Z/H]$_{i}$ correspond to the age and metallicity of each template, and $w_{i}$ is its corresponding weight in the linear combination. To convert mass-weighted quantities to luminosity-weighted quantities, we use the mass-to-light ratio of each template in the $V$-band. We compute luminosity-fraction weights ($w_{i}^\mathrm{LumW}$) of a given template as
\begin{equation}
w_{i}^\mathrm{LumW} = \frac{w_{i}}{(M/L_{V})_{i}}~,
\end{equation}
where $(M/L_{V})_{i}$ corresponds to its mass-to-light ratio in the V-band. We can use these luminosity-fraction weights to calculate luminosity-weighted properties, following Equations~\ref{eq:age} and~\ref{eq:z}. 

Regularization is a standard approach to solve ill-posed problems \citep[see][]{Capellari2017} and it is commonly used to reduce the noise in the recovery of the stellar population parameters. However, exhaustive testing showed that using a fixed level of regularization on our dataset leads to strongly biased star-formation histories, with systematic differences across different regions. This is due to the presence of young star forming regions being inhomogeneously distributed across the disk, causing strong differences in the stellar ages from one region to another. Thus, we do not use regularization in our fitting, and instead we use Monte Carlo simulations to estimate the uncertainty in the recovered stellar population parameters. For each spectrum, we perform 20 Monte Carlo iterations, where in each iteration, we perturb the input spectrum assuming a Gaussian noise with a mean of zero and a standard deviation corresponding to the spectral error at each wavelength bin. The uncertainties of stellar population parameters are calculated as the standard deviation of their distributions produced by the Monte Carlo realizations. This is meant as a first-order estimate of the true uncertainties. In Appendix~\ref{app:uncertainties}, we show the distribution of the relative uncertainties in stellar mass surface density, stellar age, and metallicity, for our full sample. We also demonstrate that 20 Monte Carlo realizations is a reasonable compromise between estimating the width of the posterior distribution and computational time to achieve our results.
We only run Monte Carlo realizations for the second step of the fitting procedure, once the stellar extinction has been fixed. Hence, no error is computed for the stellar $E(B-V)$. Finally, we have identified foreground stars as velocity outliers in the SSP fitting, and we have masked those pixels for the analysis carried out in this paper.

\begin{table*}
\footnotesize
\centering
\begin{center}
\renewcommand{\arraystretch}{1.2}
\setlength{\tabcolsep}{1.2pt}
\begin{tabular}{lclccccrccccr}
\hline
\hline

Target & RA & Dec & Log$_{10}(\mathrm{M}_{*})$ & Log$_{10}(\mathrm{M}_{\mathrm{H}_{2}})$ & Log$_{10}(\mathrm{M}_{\mathrm{H}_{I}})$ & Log$_{10}({\mathrm{SFR}})$ & $\Delta$MS & Distance & Inclination & Mapped area & $R_{25}$ & Morphol.\\
 & (degrees) & (degrees) & $(M_{\odot})$ & $(M_{\odot})$ & $(M_{\odot})$ & $(M_{\odot}\,yr^{-1})$ & (dex) & (Mpc) & (degrees) & (kpc$^{2})$ & (") & Compon. \\
 
 (1) & (2) & (3) & (4) & (5) & (6) & (7) & (8) & (9) & (10) & (11) & (12) & (13)\\
\hline 
NGC0628 & $24.1739$ & $15.7836$ & 10.3 & $9.4$ & $9.7$ & $0.24$ & 0.18 & 9.84$\pm$0.61 & 8.9 & 98 & 296$\pm$76 & D,SA,C \\
NGC1087 & $41.6049$ & $-0.4987$ & 9.9 & $9.2$ & $9.1$ & $0.12$ & 0.33 & 15.85$\pm$2.08 & 42.9 & 128 & 89$\pm$23 & D,B \\
NGC1300$^{\bullet}$ & $49.9208$ & $-19.4111$ & 10.6 & $9.4$ & $9.4$ & $0.07$ & $-$0.18 & 18.99$\pm$2.67 & 31.8 & 366 & 178$\pm$46 & D,R,SA,B,C \\
NGC1365 & $53.4015$ & $-36.1404$ & 11.0 & $10.3$ & $9.9$ & $1.23$ & 0.72 & 19.57$\pm$0.77 & 55.4 & 421 & 360$\pm$93 & D,R,SA,B,C \\
NGC1385$^{\bullet}$ & $54.3690$ & $-24.5012$ & 10.0 & $9.2$ & $9.2$ & $0.32$ & 0.50 & 17.22$\pm$2.42 & 44.0 & 100 & 102$\pm$26 & D,SA \\
NGC1433$^{\bullet}$ & $55.5062$ & $-47.2219$ & 10.9 & $9.3$ & $9.4$ & $0.05$ & $-$0.36 & 18.63$\pm$1.76 & 28.6 & 441 & 185$\pm$48 & D,R,B,C \\
NGC1512 & $60.9756$ & $-43.3487$ & 10.7 & $9.1$ & $9.9$ & $0.11$ & $-$0.21 & 18.83$\pm$1.78 & 42.5 & 270 & 253$\pm$65 & D,R,SA,B,C \\
NGC1566$^{\bullet}$ & $65.0016$ & $-54.9380$ & 10.8 & $9.7$ & $9.8$ & $0.66$ & 0.29 & 17.69$\pm$1.91 & 29.5 & 212 & 216$\pm$56 & D,SA,B \\
NGC1672 & $71.4270$ & $-59.2473$ & 10.7 & $9.9$ & $10.2$ & $0.88$ & 0.56 & 19.4$\pm$2.72 & 42.6 & 255 & 184$\pm$47 & D,R,SA,B,C \\
NGC2835 & $139.4704$ & $-22.3547$ & 10.0 & $8.8$ & $9.5$ & $0.09$ & 0.26 & 12.22$\pm$0.9 & 41.3 & 88 & 192$\pm$49 & D,SA,B \\
NGC3351 & $160.9906$ & $11.7037$ & 10.4 & $9.1$ & $8.9$ & $0.12$ & 0.05 & 9.96$\pm$0.32 & 45.1 & 76 & 216$\pm$56 & D,R,B,C \\
NGC3627 & $170.0625$ & $12.9915$ & 10.8 & $9.8$ & $9.1$ & $0.58$ & 0.19 & 11.32$\pm$0.47 & 57.3 & 87 & 308$\pm$79 & D,SA,B,C \\
NGC4254$^{\bullet}$ & $184.7068$ & $14.4164$ & 10.4 & $9.9$ & $9.5$ & $0.49$ & 0.37 & 13.1$\pm$1.87 & 34.4 & 174 & 151$\pm$39 & D,SA,C \\
NGC4303$^{\bullet}$ & $185.4789$ & $4.4737$ & 10.5 & $9.9$ & $9.7$ & $0.73$ & 0.54 & 16.99$\pm$2.78 & 23.5 & 220 & 206$\pm$53 & D,SA,B,C \\
NGC4321$^{\bullet}$ & $185.7289$ & $15.8223$ & 10.7 & $9.9$ & $9.4$ & $0.55$ & 0.21 & 15.21$\pm$0.49 & 38.5 & 196 & 182$\pm$47 & D,R,SA,B,C \\
NGC4535$^{\bullet}$ & $188.5846$ & $8.1980$ & 10.5 & $9.6$ & $9.6$ & $0.33$ & 0.14 & 15.77$\pm$0.36 & 44.7 & 126 & 244$\pm$63 & D,SA,B,C \\
NGC5068 & $199.7281$ & $-21.0387$ & 9.4 & $8.4$ & $8.8$ & $-0.56$ & 0.02 & 5.2$\pm$0.22 & 35.7 & 23 & 224$\pm$58 & D,B \\
NGC7496$^{\bullet}$ & $347.4470$ & $-43.4278$ & 10.0 & $9.3$ & $9.1$ & $0.35$ & 0.53 & 18.72$\pm$2.63 & 35.9 & 89 & 100$\pm$26 & D,B \\
IC5332 & $353.6145$ & $-36.1011$ & 9.7 & $-$ & $9.3$ & $-0.39$ & 0.01 & 9.01$\pm$0.39 & 26.9 & 34 & 182$\pm$47 & D,C \\

\hline
\end{tabular}
\end{center}
\caption[Summary of the adopted global parameters of our sample galaxies]{Summary of the galactic parameters of our sample adopted through this work. $^{\bullet}$: Galaxies observed with MUSE WFM-AO mode. Values in columns (4), (5), (6), and (7) correspond to those presented in \citet{Leroy2021b}. Column (8) provides the vertical offset of the galaxy from the integrated main sequence of galaxies, as defined in \citet{Leroy2019}. The distances in column (9) are derived through redshift-independent techniques, whose details are provided in \citet{Anand2021}. Inclinations correspond to those reported in \citet{Lang2020}. Uncertainties for the values in columns (4), (5), (6), (7), and (8) are on the order of $0.1$ dex. Column (11) lists the area mapped by MUSE. Column (12) shows the 25th magnitude isophotal $B$-band radius in arcseconds, from LEDA \citep{Makarov2014}. Column (13) indicates the morphological components identified in each galaxy by \citet{Querejeta2021} (see also Sec.~\ref{sec:enviromental_mask}). Different environments are labeled as D (disk), R (ring), SA (sp. arms), B (bar), and C (center).} 
\label{tab:sample_ch3}
\end{table*}

\subsection{Environmental masks}
\label{sec:enviromental_mask}

We employ the environmental masks described in \citet{Querejeta2021} to classify the different environments of each galaxy and label them as disks, spiral arms, rings, bars, and centers (see, e.g., Fig.~\ref{fig:NGC1566_maps}). This classification was done using photometric data, mostly from the Spitzer Survey of Stellar structure in Galaxies \citep[S$^4$G;][]{Sheth2010}. We refer the reader to that paper for a detailed explanation of how the masks are defined.

In summary, disks and centers are identified via 2D photometric decomposition of $3.6~\mu$m images \citep[see, e.g.,][]{Salo2015}. Centers are defined as a central excess of light, independently of its surface brightness profile. If a central excess of light is not identified, the inner galactic region is not labeled as center. Bars and rings properties are defined visually on the NIR images, following \citet{Herrera-Endoqui2015} for S$^4$G galaxies. Spiral arms are defined by fitting a log-spiral function to bright regions along the arms on the NIR images, only when the spiral arms are clear features across the galaxy, and their width is empirically determined based on CO emission. We use these environmental masks to probe the properties of stellar populations separately across different galactic environments.

\subsection{ALMA CO(2-1) Maps}
\label{sec:alma}

We compare our measurements to ALMA CO-based measurements of the velocity dispersion. We use the ``effective width''-based estimates, which estimate the line width based on the observed peak intensity and line-integrated intensity. These effective width measurements are robust in the presence of noise and are expressed as an equivalent rms velocity dispersion assuming a Gaussian line profile. We expect that these provide a reasonable first-order estimate the local molecular gas velocity dispersion across most of the area in our target galaxies. The primary shortcoming of these measurements will arise in the central parts of galaxies where complex line profiles can mean that the interpretation of the measurement in terms of a single Gaussian line profile represents an oversimplication.


\section{Results and discussion}
\label{sec:results}
 \label{sec:stellarpops_analysis_sec}
 
 \begin{figure*}
\centering
 	\includegraphics[width=0.95\textwidth, trim=100 80 100 100, clip]{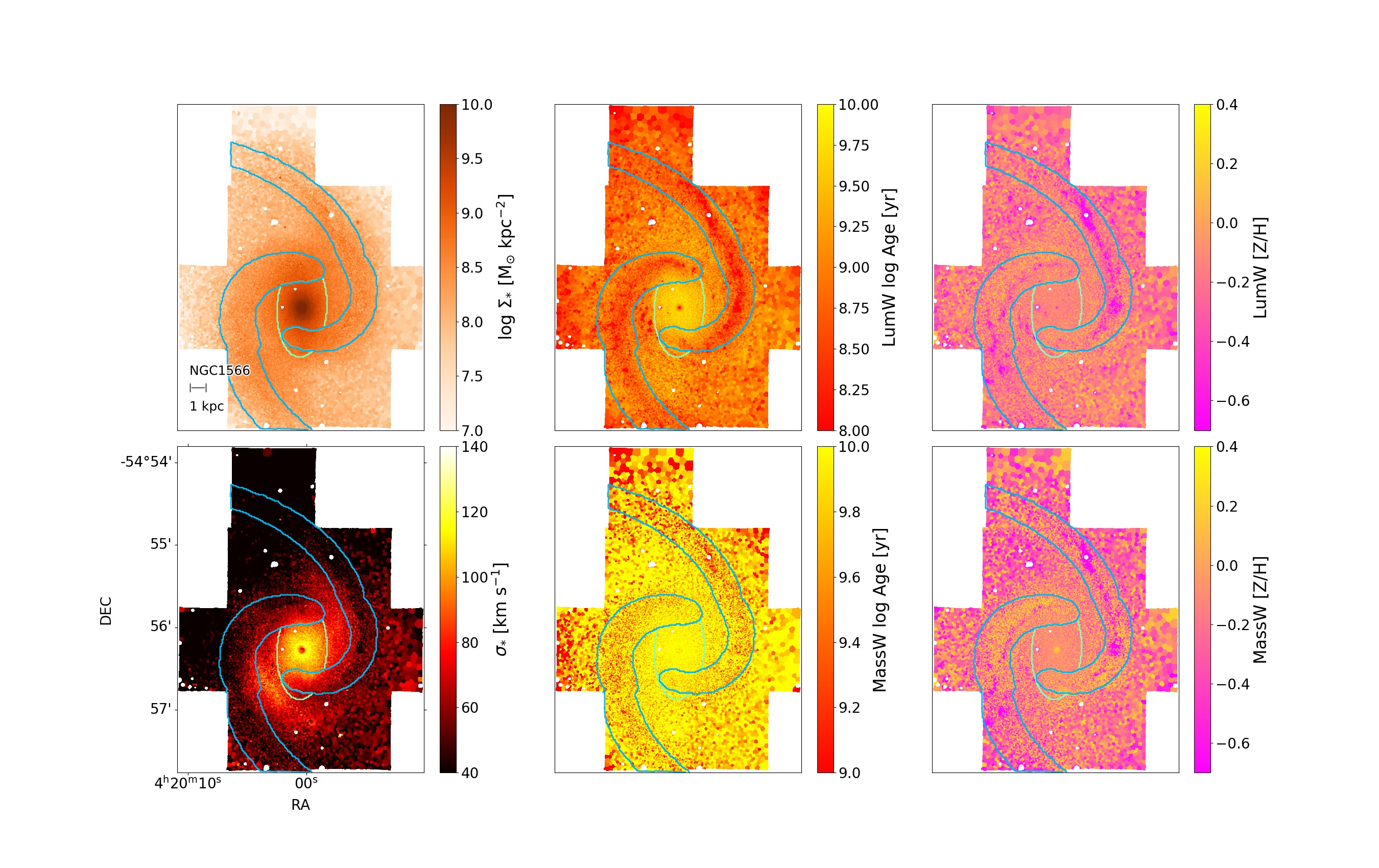}
     \caption[Stellar mass surface density, velocity dispersion, age, and metallicity maps obtained for the mosaic of NGC\,1566]{Example of some of the DAP outputs obtained for the mosaic of NGC\,1566. Stellar mass surface density (top-left), luminosity-weighted age (top-center), luminosity-weighted [Z/H] (top-right), stellar velocity dispersion (bottom-left), mass-weighted age (bottom-center), and mass-weighted [Z/H] (bottom-right). The spiral arms and bar of the galaxy are marked with blue and green contours, respectively.}
     \label{fig:NGC1566_maps}   
\end{figure*}

In Sec.~\ref{sec:stellarpops_maps}, we described in detail the methodology adopted to fit the spectra of galaxies in our sample with a linear combination of SSPs to derive their SFHs, in a spatially resolved manner. In the following sections, we show how we can use this information to gain insights into galaxy assembly and evolution. 
 
\subsection{Radial structure of stellar population properties}
\label{sec:radil_profiles_ag_z}


In this section, we show the radial distribution of the light- and mass-weighted mean age and metallicity of the stellar populations of PHANGS-MUSE galaxies, at a resolution of $\sim100$~pc, thus, resolving their different morphological components. While luminosity-weighted quantities are very sensitive to recent star formation events, because young stars dominate the B-band light emission, and thus, encode information about the ongoing galaxy evolution, mass-weighted quantities trace the properties of the long-lived, low-mass stars that dominate the stellar mass budget, and thus, encode information about the mass growth and secular redistribution history of galaxies. Fig.~\ref{fig:NGC1566_maps} shows the stellar mass surface density, velocity dispersion, mean luminosity-weighted (LumW) and mass-weighted (MassW) age, and mean light- and mass-weighted metallicity for a galaxy in our sample (NGC\,1566), as an example of our mean stellar population properties maps. Overplotted contours enclose the spiral structure and the bar environments \citep[as defined in][]{Querejeta2021}.

We derive the radial variation of these parameters, averaging the quantities azimuthally in concentric rings. Each radial bin has a width of $0.03\,R_{25}$, where $R_{25}$ represents the 25th isophotal magnitude radius in B-band. This radial bin corresponds to a physical size of about $0.4\pm0.2$~kpc for the galaxies in our sample. To define the radial bins consistently, we use the galactic inclinations and position angles reported in \citet{Lang2020}, measured from CO(2-1) data. Only those radial bins in which at least one-third of the ring is covered by the MUSE mosaic are considered valid for the radial profiles.

\subsubsection{Age radial profiles}
\begin{figure*}[h!]
\centering
 	\includegraphics[width=0.9\textwidth]{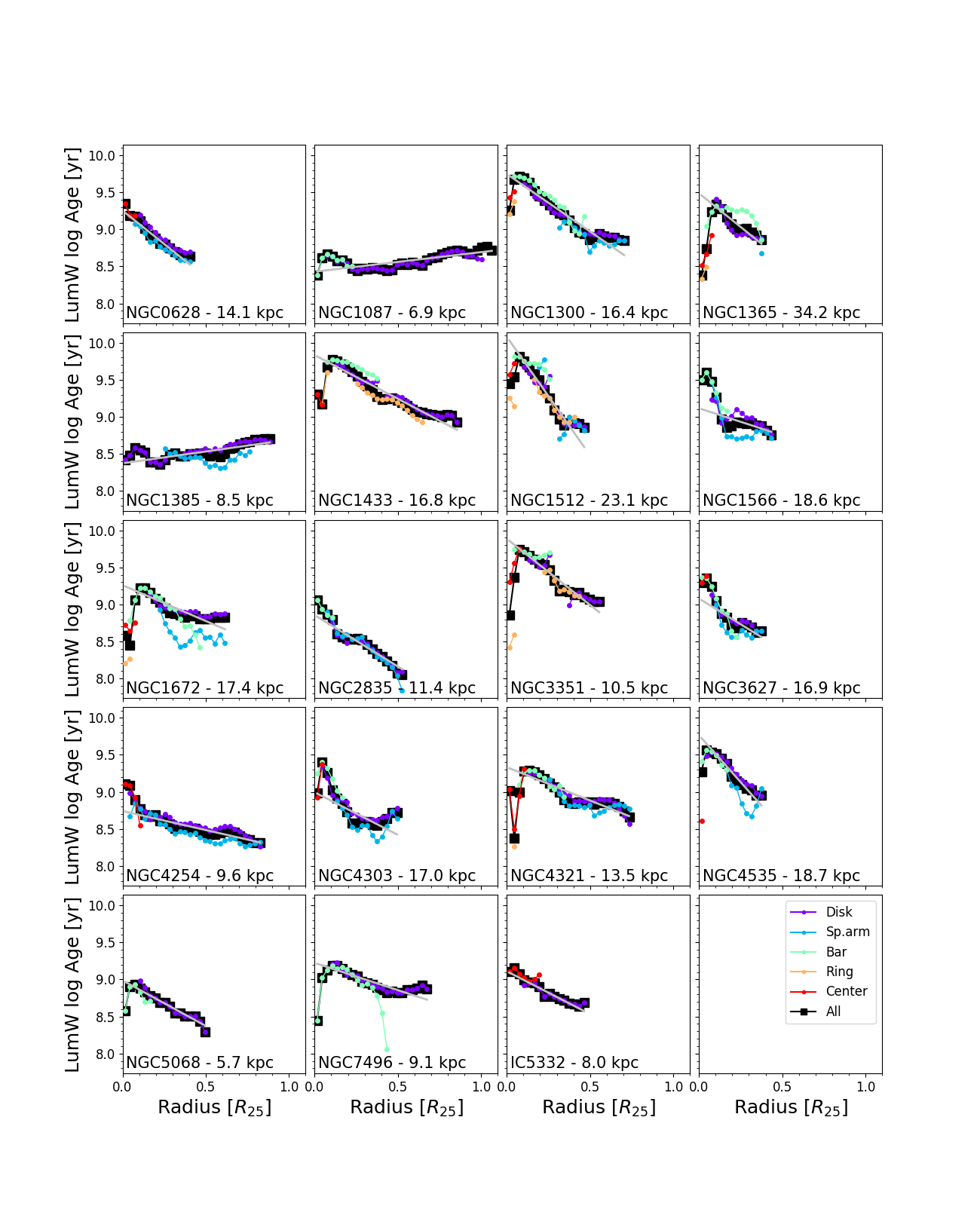}
     \caption[Luminosity-weighted stellar age radial profiles for the galaxies in our sample]{Luminosity-weighted stellar age radial profiles for the galaxies in our sample. Different colors indicate the radial profile measured across different environments, as indicated in the legend of the bottom-right panel. The black line shows the radial profile measured for the entire FoV (i.e., all environments together). The galactocentric distance is measured in units of $R_{25}$, in order to measure radial distance homogeneously across our sample. The value of $R_{25}$ (kpc) of each galaxy is indicated in each corresponding panel. The solid gray line shows the best-fit linear gradient for each galaxy.}
     \label{fig:radial_LW_age}   
\end{figure*}

Figure~\ref{fig:radial_LW_age} shows the LumW age profiles for the 19 galaxies in our sample, for the full field of view (FoV), and each morphological component separately. LumW quantities are strongly biased toward young populations \citep[e.g., ][]{Zibetti2017}. Therefore, sudden declines in these profiles are associated with recent local star formation.

\begin{figure*}[h!]
\centering
 	\includegraphics[width=\textwidth]{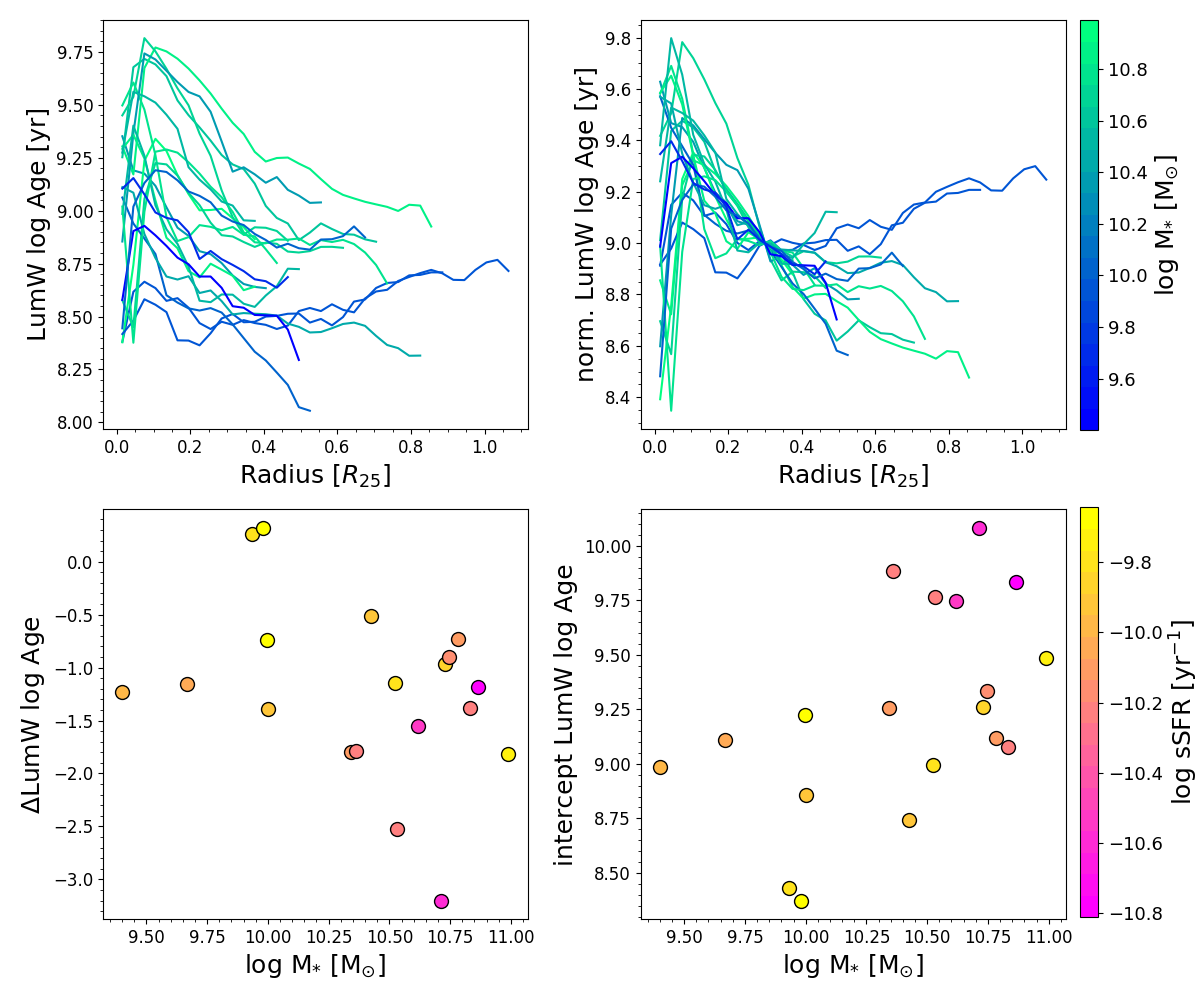}
     \caption{Luminosity-weighted stellar age radial profiles for the galaxies in our sample, considering all environments. Top left panel shows the measured LumW log Age radial profiles, colored by total galaxy stellar mass. Top right panel shows the profiles normalized by their value at $0.3$ R$_{25}$, to better compare their shapes. The bottom panels show the slope ($\Delta$LumW log Age) and intercept of the LumW radial age profile (indicated with a solid gray line in each panel of Fig.~\ref{fig:radial_LW_age}) as a function of the total stellar mass of galaxies, color coded by total specific star formation rate (sSFR).}
     \label{fig:summary_radial_LW_age}   
\end{figure*}

The first obvious feature here is the overall decreasing trend with radius present for most of the galaxies in our sample. Negative age gradients have been reported in previous studies \citep[e.g., ][]{Moustakas2010,SanchezBlazquez2014, GonzalezDelgado2014}. The only exceptions are NGC\,1087 and NGC\,1385, which are both at the low-mass end of our sample (log M$_{*}/M_\odot$ of $9.9$ and $10.0$, respectively). This is consistent with previous findings of a larger diversity in the radial structure of low-mass galaxies \citep{IbarraMedel2016, Smith2022}. This trend can be more clearly appreciated in the top panels of Fig.~\ref{fig:summary_radial_LW_age}, where low-mass galaxies (blue) show either flat or negative gradients, while higher mass galaxies (green) show mostly negative gradients. The figure also reveals a clear dependence of the normalization of the LumW age radial profile with respect to the total stellar mass of the galaxies, where more massive galaxies are in general older than lower mass galaxies. 

The slope of LumW log age radial gradients span a wide range of values. We measure the slope and intercept of the radial profiles using an ordinary least-squares (OLS) fitting routine to fit a linear model to the full FoV profile (i.e., all environments together), but dropping the innermost radial bins (r < 0.06 $R_{25}$), that are often outliers in the radial trend. The linear models are indicated with a gray solid line in Figs.~\ref{fig:radial_LW_age}. We acknowledge the caveat that although the linear models generally capture well the observed radial trend, some galaxies show gradients that are intrinsically hard to capture with a simple linear model (e.g., the log (LumW age) radial profiles of NGC\,1566 or NGC\,3627). In these galaxies, a single slope is not sufficient to describe the observed gradients, and a more appropriate representation of the data would be provided by a compound model with two different slopes. However, a more sophisticated modeling is beyond the scope of the analyses presented here. Fig.~\ref{fig:summary_radial_LW_age} shows explicitly how the normalization and slope of the LumW age radial profile depend on the galaxy stellar mass and specific SFR.

Now we look at specific galactic structures. Some galaxies show evidence of recent star formation in their inner-most region ($R \lesssim 0.2 R_{25}$), leading to a sudden decrease of stellar age (e.g.,\ NGC\,1300, NGC\,1365 among others). Although our sample suggests that this rejuvenation occurs preferentially in the central region of barred galaxies, the small number of nonbarred galaxies in our sample makes it hard to conclude exactly the impact of the bar on this phenomenon. Spiral arms, when present, usually show up as younger structures at any given radius, compared to the disk, indicating that star formation is occurring more often in spiral arms. 
Bars also show a very clear negative profile. Some galaxies show a sharp drop in LumW age at the outer radius of the bar (NGC\,1300, NGC\,1365, NGC\,1672, NGC\,7496), consistent with a scenario where the oldest populations dominate in the inner and rounder part of the bar, while young populations are located on more elongated orbits, populating the bar ends, where they spend a larger fraction of their orbiting period \citep{Beuther2017, Wozniak2007, Neumann2020}. The impact of bars in the orbital configuration of stars of different ages, together with the suppression of star formation along bars \citep[e.g.,][]{Neumann2019, Fraser2020}, caused by the bar dynamics \citep[i.e., velocity shocks and shear, see, e.g.,][]{Zurita2004, DiazGarcia2020}, could be the origin of the consistently higher LumW log Ages in bars, relative to disks, at a fixed galactocentric distance, across most of the radial range covered by both environments (except for the bar ends), in several of our sample galaxies (e.g., NGC\,1365, NGC\,1433, NGC\,1512, NGC\,3351). It is also worth noting that many galaxies exhibit a clear flattening of their radial gradients toward larger radii. This break in the radial gradient roughly coincides with the bar radius, when a bar is present.

There have been other works that have reported differences in the stellar age of different morphological components of late-type galaxies. \citet{MendezAbreu2021} use data from CALIFA \citep{Sanchez2012}, and also find older bulges that formed earlier in cosmic history than disks. In another work of spectro-photometric decomposition of bulges and disks, \citet{Costantin2022} find that bulge ages show a bimodal distribution, for which they propose a two-wave formation scenario. Similarly, \citet{Coccato2018} separate bulge and disk of a single spiral galaxy \citep[log M$_{*} \approx 10.7$; ][]{Leroy2008}, and find that young stars contribute mostly to the disk, with the young contribution increasing with galactocentric radius. They also find a strong negative age gradient in the disk, but no evidence of a gradient in the bulge, consistent with the results that we find in our sample, where some galaxies show a clear age gradient in their center and others do not (although most of them show a clear negative gradient in their disk). Although there is still no consensus about the origin of the stellar age gradients in bulges\citep[See, e.g.,][and references therein]{SanchezBlazquez2016}, theoretical models suggest a link between the radial gradient of stellar population properties of bulges and their formation mechanisms \citep{Matteucci2019}. \citet{Morelli2015} measured the stellar population gradients in the bulges of a sample of spiral galaxies, and found (on average) flat age gradients, albeit with significant dispersion ($\sigma_{\rm gradients} = 1.3$), which could point to a hierarchical merging contribution in the assembly of bulges. On this subject, several authors agree that drops in stellar ages in the central part of galaxies, leading to positive measured age gradients in bulges, are likely caused by the presence of central disks or nuclear rings \citep{Morelli2008, Bittner2020}. Within this framework, the high angular resolution of our data allows us to clearly determine the presence of a nuclear ring, and to show that its presence indeed often leads to an overall positive LumW age gradient in the innermost galactic region (e.g., NGC\,1365 or NGC\,3351, among others). Furthermore, the measurement at high-resolution of gradients in stellar population properties in the center of galaxies, together with stellar kinematics could enable a systematic study of the origin of bulges in late-type galaxies \citep[see, e.g.,][]{Moorthy2006, RosadoBelza2020, Gadotti2019, Gadotti2020}; however, this is beyond the scope of this paper. Lastly, we remind the reader that, unlike other works that compare bulges and disks, defining bulges according to their surface brightness profile or kinematic properties, our definition of ``centers'' implies only a central excess of light, as explained in Sec.~\ref{sec:enviromental_mask}. 

\begin{figure}[h!]
\centering
 	\includegraphics[width=\columnwidth]{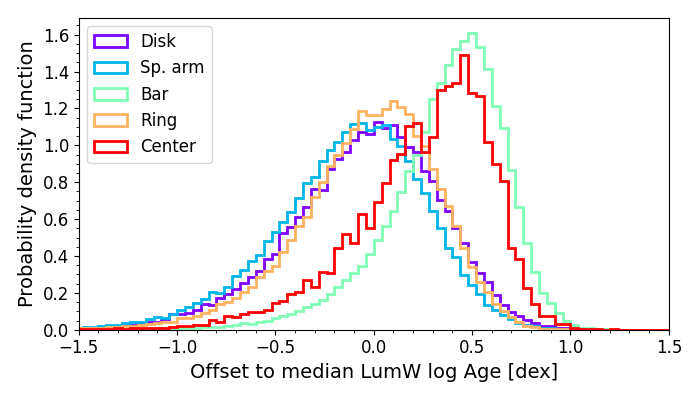}
     \caption{Distribution of the offset of the LumW log Age of the pixels within each environments, with respect to the median galaxy log LumW Age of each galaxy, in bins of $0.04$ dex. Different colors indicate the distribution of different environments, as indicated in the legend.}
     \label{fig:dist_age_envs}   
\end{figure}

Differences between the LumW log Age of different environments are summarized in Fig.~\ref{fig:dist_age_envs}. Bars show a clear peak at old ages ($\sim+0.5$ dex offset with respect to the median age of each galaxy), being the environment with the oldest LumW ages on average. Centers also peak at old ages, but they show a bimodal distribution, with a second peak toward younger ages ($\sim+0.2$ dex), due to the younger stellar populations present in the central region of some galaxies. Rings show a narrower distribution than disks and spiral arms, suggesting a lower dispersion in the age of the stars within this environment (although we acknowledge that this could be partially due to that the Ring distribution is dominated by the outer ring of a few galaxies). Finally, spiral arms show a distribution similar in shape to that of the disk, but displaced toward slightly younger ages. Table~\ref{tab:median_lw_age_per_env} shows the median and standard deviation of the LumW log Age distribution within each environment, and the median offset of each environment with respect to the full galaxy measurement for the full sample in the last row (i.e., median and standard deviation of distributions shown in Fig.~\ref{fig:dist_age_envs}).

The results from \citet{Querejeta2021} indicate that a spatial resolution of $\sim100$ pc is sufficient to separate the different morphological components of galaxies. For comparison, in Appendix~\ref{app:degrade} we explore how these results would change with lower angular resolution data, by degrading our dataset to a fixed angular resolution of 15 arcsec ($1.1\pm0.3$ kpc for the distances of our sample galaxies), at which contamination among different galactic environments is significantly larger.

\begin{table*}
\centering
\begin{tabular}{ccccccc}
\hline
\hline
Target & Disk & Sp. arms & Bar & Ring & Center & All \\
\hline
NGC0628 & 8.86$\pm$0.35 & 8.76$\pm$0.38 & - & - & 9.24$\pm$0.23 & 8.83$\pm$0.37 \\
NGC1087 & 8.60$\pm$0.32 & - & 8.60$\pm$0.29 & - & - & 8.65$\pm$0.33 \\
NGC1300 & 9.00$\pm$0.37 & 8.92$\pm$0.37 & 9.55$\pm$0.30 & 9.22$\pm$0.22 & 9.50$\pm$0.19 & 9.02$\pm$0.40 \\
NGC1365 & 8.87$\pm$0.42 & 8.69$\pm$0.34 & 9.28$\pm$0.29 & 8.44$\pm$0.43 & 8.67$\pm$0.30 & 8.93$\pm$0.43 \\
NGC1385 & 8.61$\pm$0.32 & 8.43$\pm$0.32 & - & - & - & 8.58$\pm$0.33 \\
NGC1433 & 9.14$\pm$0.38 & - & 9.72$\pm$0.20 & 9.24$\pm$0.31 & 9.28$\pm$0.12 & 9.22$\pm$0.38 \\
NGC1512 & 9.02$\pm$0.54 & 8.89$\pm$0.46 & 9.77$\pm$0.16 & 9.15$\pm$0.40 & 9.63$\pm$0.12 & 9.16$\pm$0.51 \\
NGC1566 & 8.91$\pm$0.34 & 8.81$\pm$0.35 & 9.51$\pm$0.25 & - & - & 8.89$\pm$0.37 \\
NGC1672 & 8.90$\pm$0.34 & 8.59$\pm$0.37 & 9.07$\pm$0.35 & 8.25$\pm$0.30 & 8.73$\pm$0.30 & 8.90$\pm$0.37 \\
NGC2835 & 8.30$\pm$0.45 & 8.46$\pm$0.45 & 8.94$\pm$0.25 & - & - & 8.36$\pm$0.46 \\
NGC3351 & 9.16$\pm$0.42 & - & 9.70$\pm$0.17 & 9.28$\pm$0.36 & 9.59$\pm$0.25 & 9.28$\pm$0.41 \\
NGC3627 & 8.77$\pm$0.36 & 8.65$\pm$0.39 & 9.16$\pm$0.28 & - & 9.34$\pm$0.16 & 8.77$\pm$0.38 \\
NGC4254 & 8.53$\pm$0.38 & 8.43$\pm$0.40 & - & - & 9.07$\pm$0.23 & 8.47$\pm$0.44 \\
NGC4303 & 8.76$\pm$0.33 & 8.61$\pm$0.37 & 9.28$\pm$0.23 & - & 8.94$\pm$0.26 & 8.72$\pm$0.37 \\
NGC4321 & 8.90$\pm$0.37 & 8.86$\pm$0.35 & 9.26$\pm$0.27 & 8.33$\pm$0.35 & 8.89$\pm$0.38 & 8.92$\pm$0.37 \\
NGC4535 & 9.14$\pm$0.31 & 9.01$\pm$0.35 & 9.52$\pm$0.22 & - & 8.73$\pm$0.48 & 9.14$\pm$0.34 \\
NGC5068 & 8.52$\pm$0.45 & - & 8.88$\pm$0.34 & - & - & 8.54$\pm$0.45 \\
NGC7496 & 8.91$\pm$0.31 & - & 9.07$\pm$0.32 & - & - & 8.93$\pm$0.31 \\
IC5332 & 8.78$\pm$0.36 & - & - & - & 9.06$\pm$0.29 & 8.81$\pm$0.36 \\
\hline
Offset to median & -0.02$\pm$0.39 & -0.09$\pm$0.38 & 0.41$\pm$0.31 & 0.00$\pm$0.36 & 0.31$\pm$0.35 & - \\
(all galaxies) \\
\hline
\end{tabular}
\caption{Median LumW log Age measured for each environment in each galaxy. The right-most column shows the median LumW log Age across the full MUSE mosaic, considering all environments. Errors correspond to the standard deviation of the LumW log Age distribution within each environment. The bottom row shows the median offset (in dex, for the full sample) of the LumW log Age distribution within each environment of a given galaxy, with respect to the overall distribution (considering all environments) of the same galaxy. }
\label{tab:median_lw_age_per_env}
\end{table*}

\begin{figure*}[h!]
\centering
 	\includegraphics[width=0.9\textwidth]{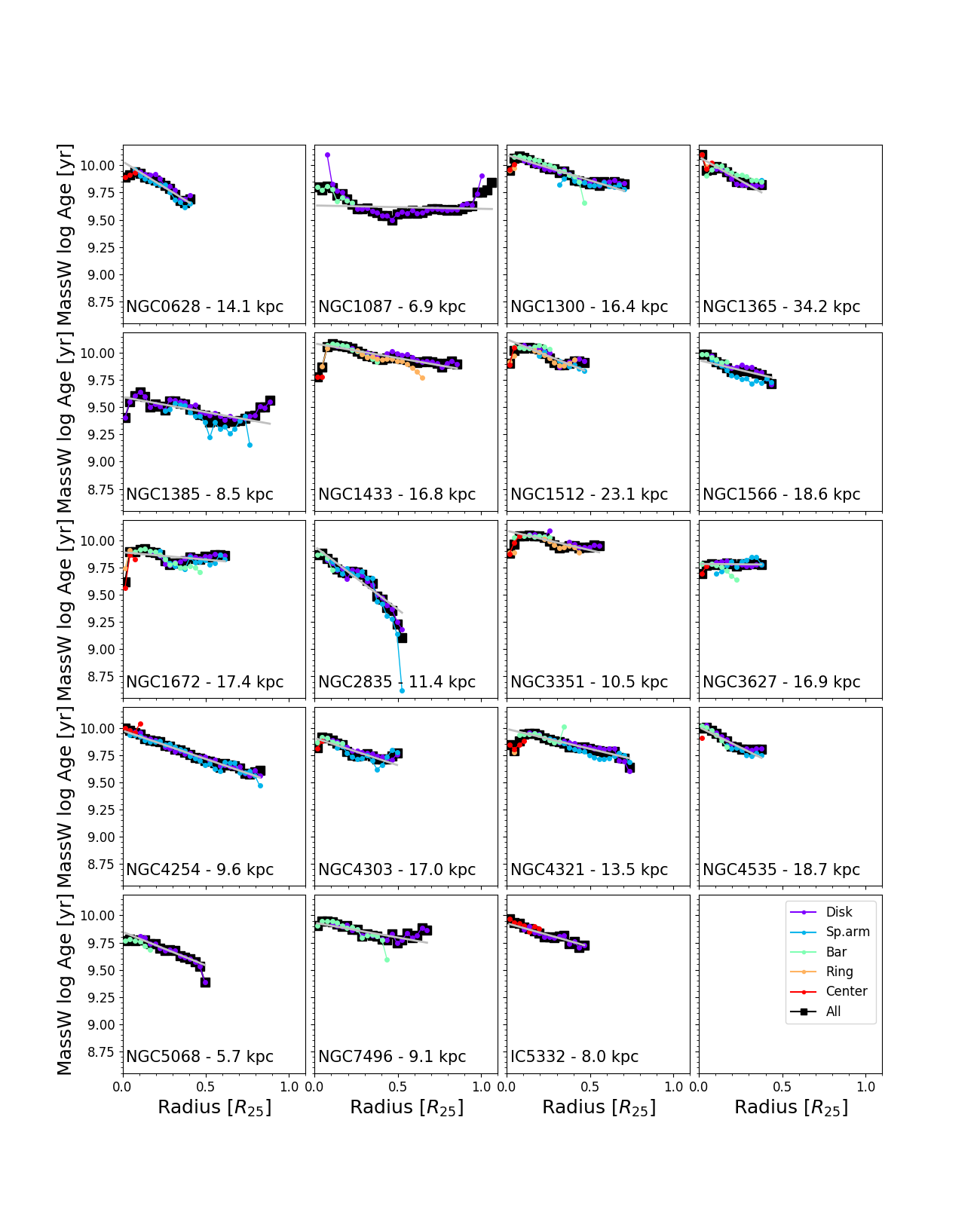}
     \caption[Mass-weighted stellar age radial profiles for the galaxies in our sample]{Mass-weighted stellar age radial profiles for the galaxies in our sample. Different colors indicate the radial profile measured across different environments, as indicated in the legend of the bottom-right panel. The black line shows the radial profile measured for the entire FoV (i.e., all environments together). The galactocentric distance is measured in units of $R_{25}$, in order to measure radial distance homogeneously across our sample. The value of $R_{25}$ (kpc) of each galaxy is indicated in each corresponding panel. The solid gray line shows the best-fit gradient for each galaxy.}
     \label{fig:radial_MW_age}   
\end{figure*}

Figure~\ref{fig:radial_MW_age} shows the mass-weighted (MassW) age profiles for the galaxies in our sample. Since MassW quantities are less biased by recent star formation events, as explained above, MassW ages are higher than their LumW counterpart, as the bulk of the mass of a galaxy is dominated by old stellar populations \citep[see, e.g.,][]{Cook2020}.


\begin{figure*}[h!]
\centering
 	\includegraphics[width=\textwidth]{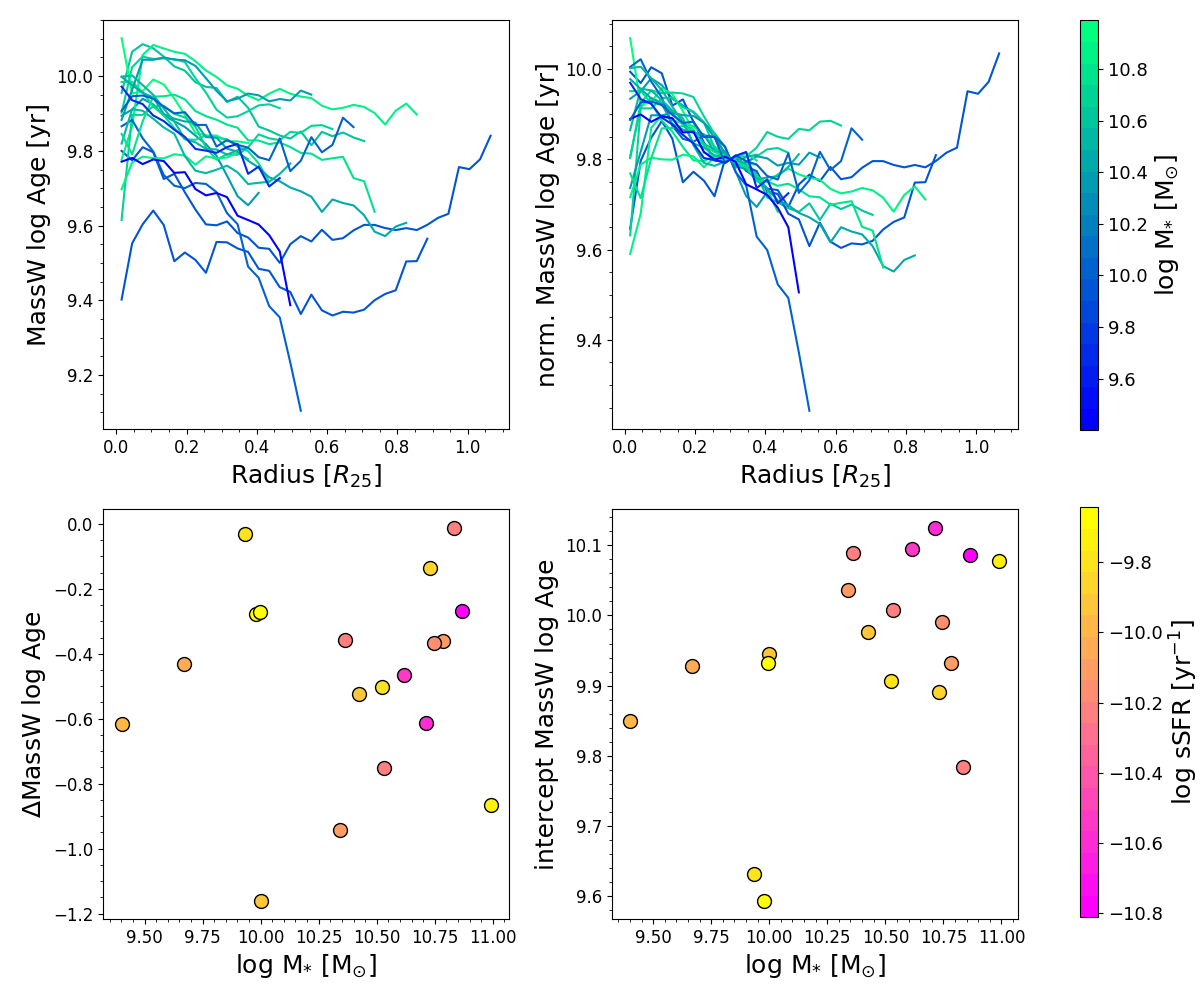}
     \caption{Mass-weighted stellar age radial profiles for the galaxies in our sample, considering all environments. Top left panel shows the measured MassW log Age radial profiles, colored by total galaxy stellar mass. Top right panel shows the profiles normalized by their value at $0.3$ R$_{25}$, to better compare their shapes. The bottom panels show the slope ($\Delta$MassW log Age) and intercept of the MassW radial age profile (indicated with a solid gray line in each panel of Fig.~\ref{fig:radial_MW_age}) as a function of the total stellar mass of galaxies, color coded by total specific star formation rate (sSFR).}
     \label{fig:summary_radial_MW_age}   
\end{figure*}

The most evident feature in the radial MassW age profiles is that environmental differences are basically absent and that the dynamic range in age is drastically reduced, typically confined to values larger than $\log_{10}$ (age/yr) $\approx 9.5$ ($\sim 3$ Gyr). This implies that the distribution of older stellar populations is significantly more homogeneous across the galactic structure than for the younger populations. Clear negative slopes persist in the MassW profile meaning that inner regions assembled the bulk of their stellar mass earlier in cosmic history than outer regions (although they are shallower, consistent with findings from \citet{Parikh2021} and \citet{Neumann2020}). These age gradients have been interpreted as a spatially resolved imprint of the inside-out  \citep{Mo1998} growth of galaxies by several studies \citep[e.g., ][]{GonzalezDelgado2014, Goddard2017, Parikh2021}. Fig.~\ref{fig:summary_radial_MW_age} shows the normalization and slope of the MassW age radial profile as a function of galaxy total stellar mass and specific SFR. There is a clear dependence of the normalization of the MassW age profile with the total stellar mass of galaxies, where more massive galaxies show older MassW ages. This can be interpreted as an imprint of the downsizing in galaxy evolution \citep[][see also Appendix.~\ref{sec:downsizing}]{Lennox1996,Heavens2004, PerezGonalez2008}.  

Finally, it is worth noting that the break of the age gradient at approximately the bar radius discussed above persists for the MassW age gradients, although it is more subtle, due to the smaller dynamic range of the MassW ages. The sharp decrease of age in the central region also persists in the MassW age profile of some galaxies, suggesting that these centers have formed a substantial amount of stellar mass in relatively recent times. In Sec.~\ref{sec:inside_out}, we investigate this further. One possibility is that the MassW age measurement in the center environment is contaminated by the surrounding ring structure in galaxies where both morphological components are present. However, given that we are able to spatially resolve the nuclear rings at the resolution of our dataset \citep{Querejeta2021} and we observe this pattern in the innermost (thus furthest from the ring) radial bin of some centers, we believe that contamination from the ring is not the primary driver of these trends. Furthermore, a careful inspection reveals that, while low LumW ages in the inner region of these galaxies strongly correlate with the location of the ring environment, the decrease in the MassW age correlates spatially with the center environment, indicating the presence of secularly evolved nuclear disks \citep[see also][]{Bittner2020}. Flatter or even slightly positive radial gradients of stellar population properties in the innermost regions of galaxies have been reported before in the literature \citep[e.g., ][for NGC\,0628]{SanchezBlazquez2014b}. In Appendix~\ref{sec:disc_stellar_age} we explore the correlation between the slope of the stellar age radial gradient and a set of global galaxy properties. Table~\ref{tab:median_mw_age_per_env} shows the same information as Table~\ref{tab:median_lw_age_per_env}, but for the MassW log age distribution rather than the LumW log age distribution.

\begin{table*}
\centering
\begin{tabular}{ccccccc}
\hline
\hline
Target & Disk & Sp. arms & Bar & Ring & Center & All \\
\hline
NGC0628 & 9.87$\pm$0.29 & 9.84$\pm$0.33 & - & - & 9.94$\pm$0.15 & 9.86$\pm$0.31 \\
NGC1087 & 9.68$\pm$0.40 & - & 9.79$\pm$0.27 & - & - & 9.74$\pm$0.41 \\
NGC1300 & 9.94$\pm$0.29 & 9.91$\pm$0.27 & 10.05$\pm$0.14 & 9.95$\pm$0.10 & 9.99$\pm$0.10 & 9.95$\pm$0.28 \\
NGC1365 & 9.89$\pm$0.31 & 9.85$\pm$0.29 & 9.95$\pm$0.16 & 10.11$\pm$0.22 & 9.98$\pm$0.14 & 9.91$\pm$0.29 \\
NGC1385 & 9.59$\pm$0.47 & 9.45$\pm$0.48 & - & - & - & 9.61$\pm$0.48 \\
NGC1433 & 10.00$\pm$0.21 & - & 10.06$\pm$0.12 & 9.98$\pm$0.20 & 9.78$\pm$0.08 & 10.00$\pm$0.20 \\
NGC1512 & 10.03$\pm$0.26 & 9.98$\pm$0.29 & 10.05$\pm$0.08 & 9.97$\pm$0.24 & 9.93$\pm$0.09 & 10.01$\pm$0.24 \\
NGC1566 & 9.87$\pm$0.34 & 9.85$\pm$0.30 & 9.98$\pm$0.12 & - & - & 9.88$\pm$0.32 \\
NGC1672 & 9.90$\pm$0.26 & 9.88$\pm$0.32 & 9.89$\pm$0.23 & 9.90$\pm$0.16 & 9.73$\pm$0.21 & 9.90$\pm$0.26 \\
NGC2835 & 9.59$\pm$0.63 & 9.69$\pm$0.60 & 9.92$\pm$0.21 & - & - & 9.64$\pm$0.62 \\
NGC3351 & 10.01$\pm$0.20 & - & 10.04$\pm$0.09 & 9.97$\pm$0.17 & 9.97$\pm$0.08 & 10.00$\pm$0.18 \\
NGC3627 & 9.84$\pm$0.28 & 9.84$\pm$0.28 & 9.79$\pm$0.15 & - & 9.71$\pm$0.09 & 9.83$\pm$0.27 \\
NGC4254 & 9.80$\pm$0.40 & 9.81$\pm$0.44 & - & - & 9.99$\pm$0.11 & 9.79$\pm$0.48 \\
NGC4303 & 9.84$\pm$0.32 & 9.83$\pm$0.36 & 9.88$\pm$0.10 & - & 9.81$\pm$0.08 & 9.84$\pm$0.33 \\
NGC4321 & 9.88$\pm$0.31 & 9.85$\pm$0.29 & 9.96$\pm$0.15 & 9.78$\pm$0.14 & 9.85$\pm$0.14 & 9.88$\pm$0.29 \\
NGC4535 & 9.86$\pm$0.19 & 9.82$\pm$0.19 & 9.96$\pm$0.13 & - & 9.94$\pm$0.18 & 9.86$\pm$0.19 \\
NGC5068 & 9.64$\pm$0.57 & - & 9.80$\pm$0.29 & - & - & 9.65$\pm$0.56 \\
NGC7496 & 9.91$\pm$0.27 & - & 9.94$\pm$0.20 & - & - & 9.92$\pm$0.26 \\
IC5332 & 9.82$\pm$0.33 & - & - & - & 9.95$\pm$0.21 & 9.84$\pm$0.32 \\
\hline
Offset to median  & -0.00$\pm$0.36 & -0.02$\pm$0.36 & 0.05$\pm$0.17 & -0.03$\pm$0.21 & 0.05$\pm$0.18 & - \\
(all galaxies) \\
\hline
\end{tabular}
\caption{Median MassW log Age measured for each environment in each galaxy. The right-most column shows the median MassW log Age across the full MUSE mosaic, considering all environments. Errors correspond to the standard deviation of the MassW log Age distribution within each environment. The bottom row shows the median offset (in dex, for the full sample) of the MassW log Age distribution within each environment of a given galaxy, with respect to the overall distribution (considering all environments) of the same galaxy. }
\label{tab:median_mw_age_per_env}
\end{table*}

\subsubsection{Metallicity radial profiles}

\begin{figure*}[h!]
\centering
 	\includegraphics[width=0.9\textwidth]{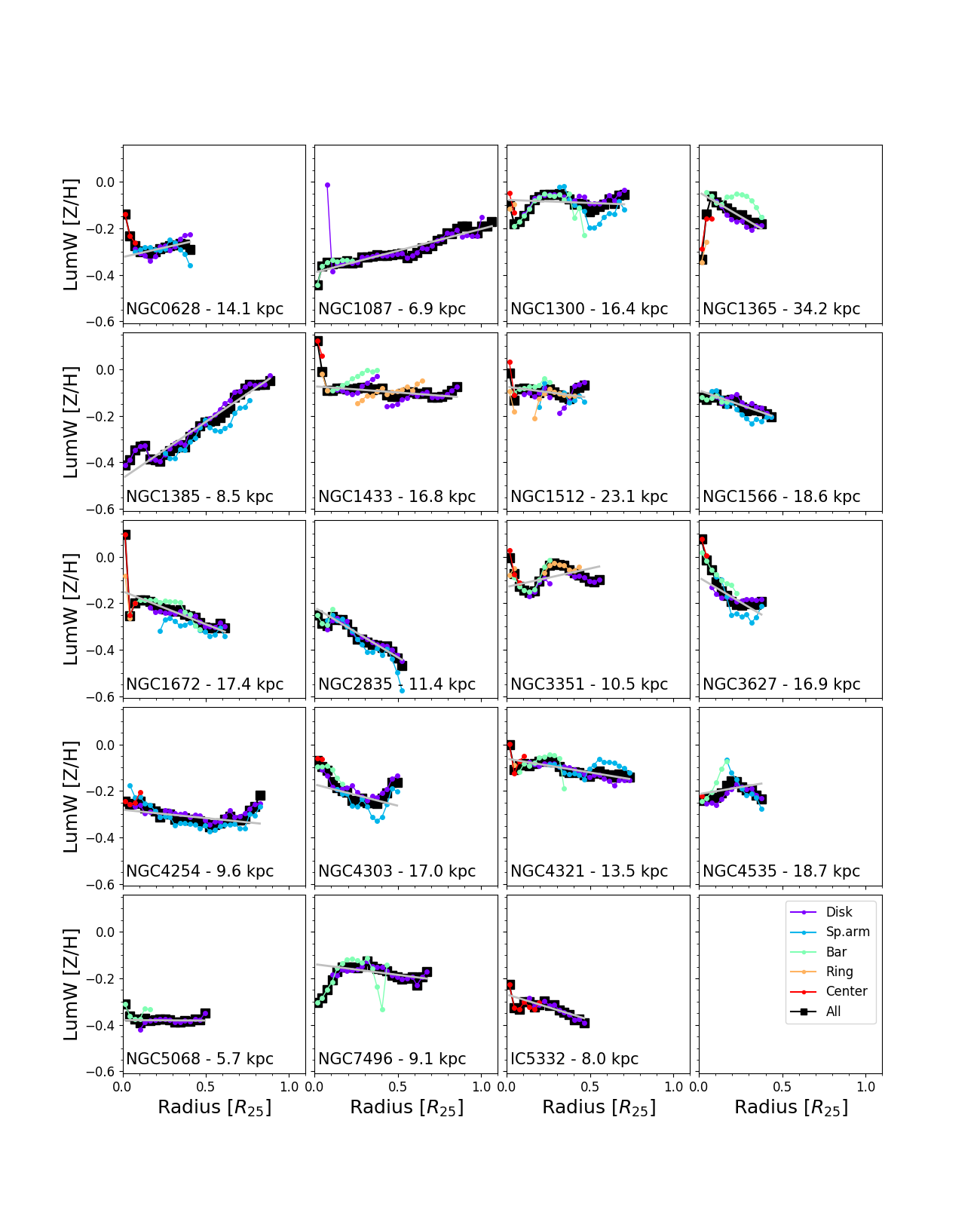}
     \caption[Luminosity-weighted stellar metallicity radial profiles for the galaxies in our sample]{Luminosity-weighted stellar [Z/H] radial profiles for the galaxies in our sample. Different colors indicate the radial profile measured across different environments, as indicated in the legend of the bottom-right panel. The black line shows the radial profile measured for the entire FoV (i.e., all environments together). The galactocentric distance is measured in units of $R_{25}$, in order to measure radial distance homogeneously across our sample. The value of $R_{25}$ (kpc) of each galaxy is indicated in each corresponding panel. The solid gray line shows the best-fit gradient for each galaxy.}
     \label{fig:radial_LW_Z}   
\end{figure*}

\begin{figure*}[h!]
\centering
 	\includegraphics[width=0.9\textwidth]{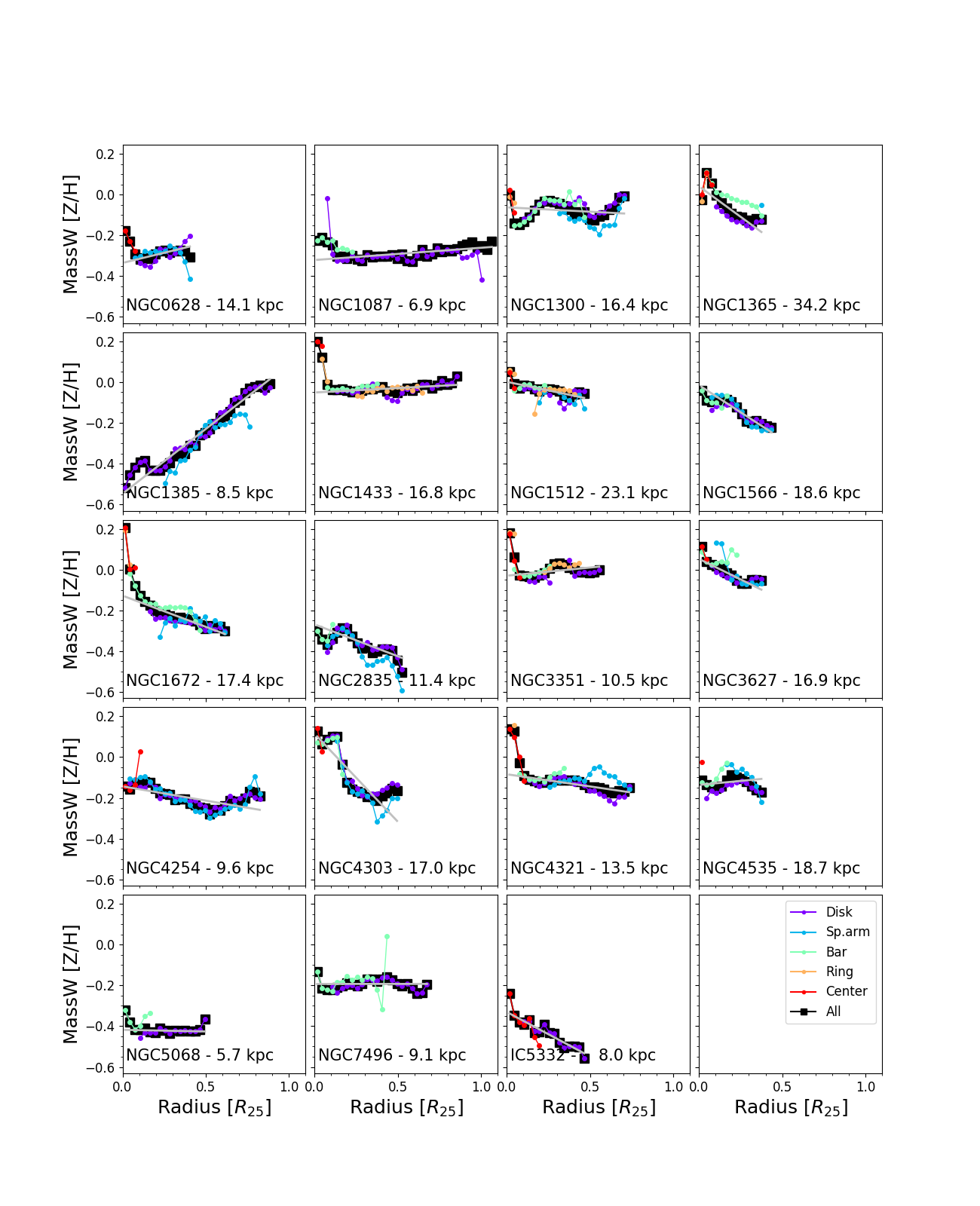}
     \caption[Mass-weighted stellar metallicity radial profiles for the galaxies in our sample]{Mass-weighted stellar [Z/H] radial profiles for the galaxies in our sample. Different colors indicate the radial profile measured across different environments, as indicated in the legend of the bottom-right panel. The black line shows the radial profile measured for the entire FoV (i.e., all environments together). The galactocentric distance is measured in units of $R_{25}$, in order to measure radial distance homogeneously across our sample. The value of $R_{25}$ (kpc) of each galaxy is indicated in each corresponding panel. The solid gray line shows the best-fit gradient for each galaxy.}
     \label{fig:radial_MW_Z}   
\end{figure*}

Regarding stellar metallicity, Figs.~\ref{fig:radial_LW_Z} and \ref{fig:radial_MW_Z} show the LumW and MassW stellar metallicity profiles measured for our sample galaxies. The figures show flatter, or generally slightly negative slopes, with the exception of the same two low-mass galaxies mentioned earlier (NGC\,1087, NGC\,1385). Overall, negative slopes are consistent with results reported previously in the literature \citep{Goddard2017, Coenda2020}.  Moreover, \citet{Morelli2015b} and \citet{Coccato2018} have studied the stellar metallicity gradients in the morphologically decoupled bulges and disks of spiral galaxies, also finding predominantly negative gradients. Interestingly, NGC\,1385 shows a strongly positive metallicity gradient. This is consistent with the findings from \citet{Zhuang2019}, who reported that low-mass and late morphological types commonly show positive metallicity slopes.  
In some galaxies, spiral arms show up with a slightly lower metallicity than the disk at a given radius. This is likely an artifact arising from the SSP fitting, and the imperfect masking of the youngest regions (see Appendix~\ref{sec:improving_young}).

\citet{Zhuang2019} also show that these metallicity gradients are a consequence of the local [Z/H]-stellar mass surface density ($\Sigma_{*}$) relation, and that they are expected to arise primarily due to the insitu local SFH. Indeed, \citet{Neumann2021} report a tight correlation between [Z/H] and $\Sigma_{*}$, with a higher scatter toward larger galactocentric radii and among lower-mass galaxies, implying that in the outer regions of galaxies (as well as within less massive galaxies), additional mechanisms such as gas accretion or outflows might become relevant in determining the chemical enrichment.

\begin{figure*}[h!]
\centering
 	\includegraphics[width=\textwidth]{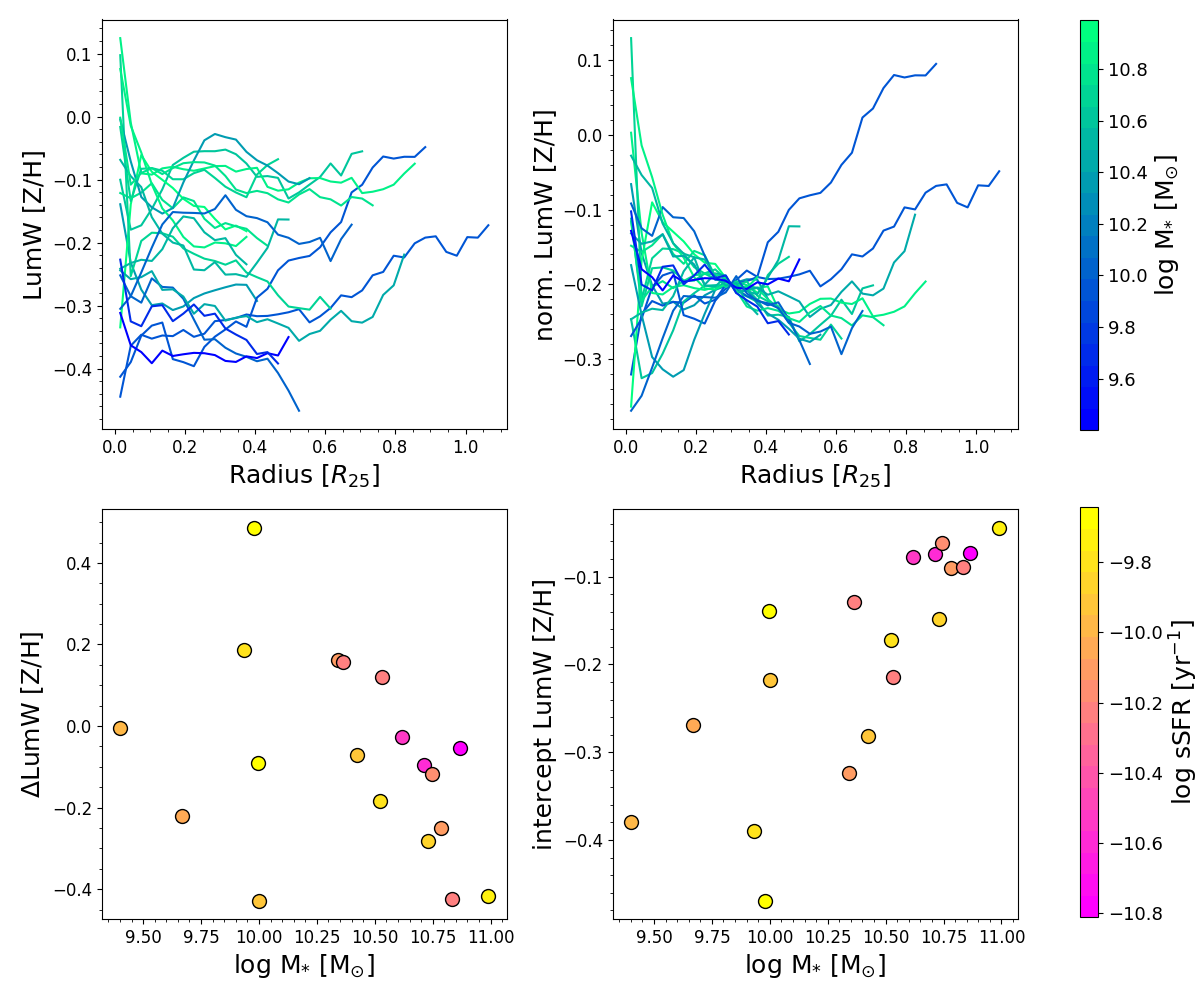}
     \caption{Luminosity-weighted stellar [Z/H] radial profiles for the galaxies in our sample, considering all environments. Top left panel shows the measured LumW [Z/H] radial profiles, colored by total galaxy stellar mass. Top right panel shows the profiles normalized by their value at $0.3$ R$_{25}$, to better compare their shapes. The bottom panels show the slope ($\Delta$LumW [Z/H]) and intercept of the LumW radial [Z/H] profile (indicated with a solid gray line in each panel of Fig.~\ref{fig:radial_LW_Z}) as a function of the total stellar mass of galaxies, color coded by total specific star formation rate (sSFR).}
     \label{fig:summary_radial_LW_Z}   
\end{figure*}

\begin{figure*}[h!]
\centering
 	\includegraphics[width=\textwidth]{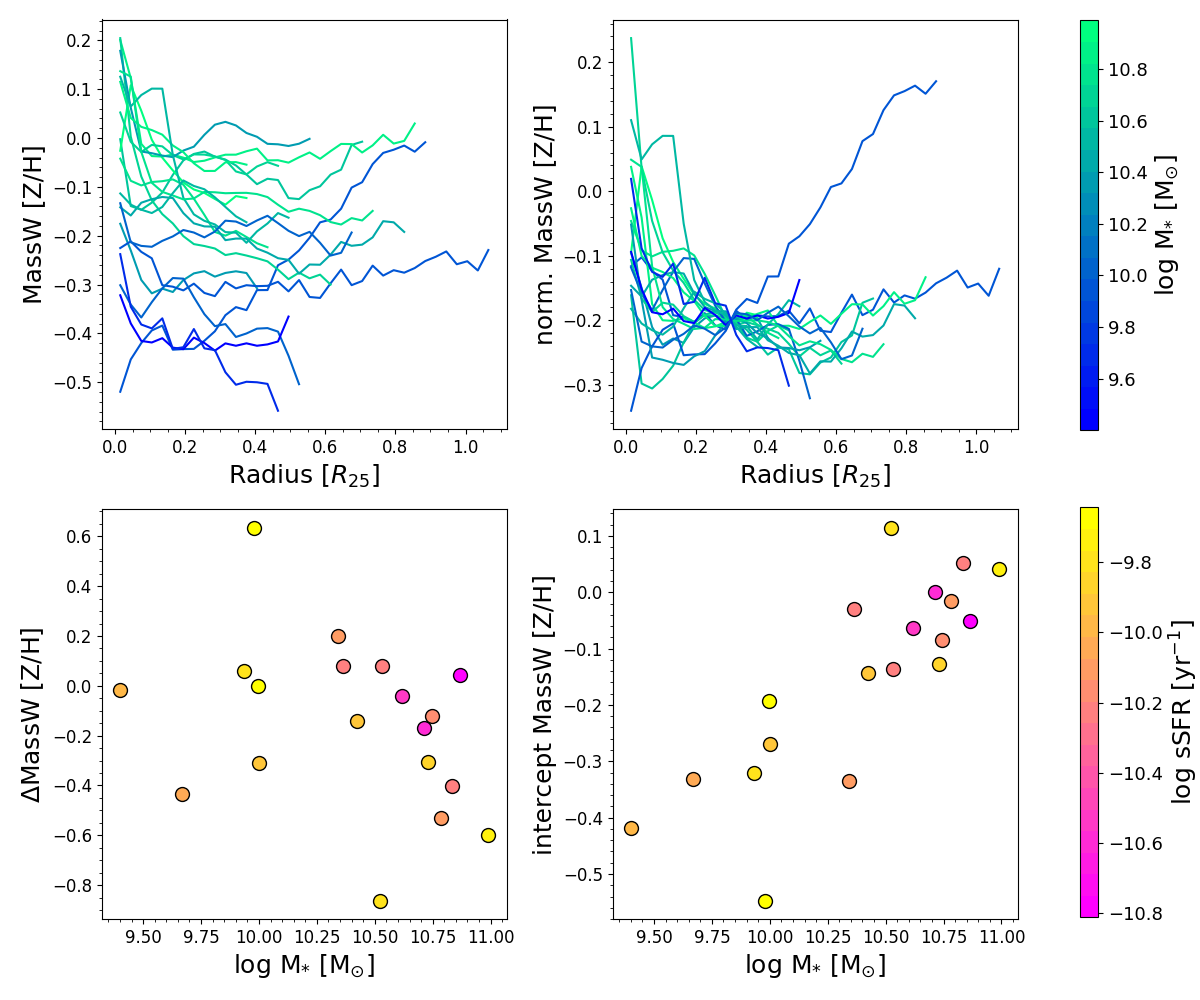}
     \caption{Mass-weighted stellar [Z/H] radial profiles for the galaxies in our sample, considering all environments. Top left panel shows the measured MassW [Z/H] radial profiles, colored by total galaxy stellar mass. Top right panel shows the profiles normalized by their value at $0.3$ R$_{25}$, to better compare their shapes. The bottom panels show the slope ($\Delta$MassW [Z/H]) and intercept of the MassW radial [Z/H] profile (indicated with a solid gray line in each panel of Fig.~\ref{fig:radial_MW_Z}) as a function of the total stellar mass of galaxies, color coded by total specific star formation rate (sSFR).}
     \label{fig:summary_radial_MW_Z}   
\end{figure*}

Figures~\ref{fig:summary_radial_LW_Z} and ~\ref{fig:summary_radial_MW_Z} show explicitly how the slope and intercept of the radial metallicity profiles (LumW and MassW, respectively) depend on total stellar mass and total sSFR of each galaxy. The slope and intercept of the LumW- and MassW-[Z/H] radial profiles are measured in the same fashion as explained above for stellar age profiles. We find a clear trend between the normalization of the stellar metallicity gradients and total stellar mass of each galaxy. This trend is expected from the well studied mass-metallicity relation of galaxies \citep{Foster2012, Xiangcheng2016}, which is thought to arise because more massive galaxies are more likely to retain a larger fraction of their metals due to their stronger gravitational potential, as opposed to low-mass galaxies, in which stellar feedback is able to remove metals from the galactic disk more efficiently \citep[see, e.g., ][]{Xiangcheng2016}.

Finally, there are some peculiarities in the metallicity gradients that are worth mentioning. NGC\,3351 exhibits an ``S'' shape LumW metallicity gradient, which originates from recent star formation occurring in the central region and in the outer ring, leading to enhanced metallicities at these radii. NGC\,4535 also has strong star formation at the ends of its bar and in the spiral arms (hence, lower LumW ages), leading to enhanced metallicities in these environments. NGC\,7496 shows a sharp radial decrease in metallicity in its bar profile, which is likely an artifact from the spectral fitting (see Appendix~\ref{sec:improving_young}). The LumW metallicity gradient of NGC\,1433 shows a strange discontinuous pattern that is likely an artifact due to subtle systematic differences between the MUSE pointings of the mosaic. In Appendix~\ref{sec:disc_stellar_z} we explore the correlation between the slope of the stellar metallicity radial gradient and a set of global properties of galaxies. Tables~\ref{tab:median_lw_z_per_env} and ~\ref{tab:median_mw_z_per_env} show the same information as Table~\ref{tab:median_lw_age_per_env}, for the LumW [Z/H] and MassW [Z/H] distributions, respectively.

\begin{table*}
\centering
\begin{tabular}{ccccccc}
\hline
\hline
Target & Disk & Sp. arms & Bar & Ring & Center & All \\
\hline
NGC0628 & -0.28$\pm$0.16 & -0.29$\pm$0.16 & - & - & -0.23$\pm$0.10 & -0.28$\pm$0.16 \\
NGC1087 & -0.29$\pm$0.19 & - & -0.35$\pm$0.15 & - & - & -0.26$\pm$0.21 \\
NGC1300 & -0.06$\pm$0.18 & -0.11$\pm$0.17 & -0.10$\pm$0.12 & -0.12$\pm$0.10 & -0.11$\pm$0.11 & -0.08$\pm$0.17 \\
NGC1365 & -0.18$\pm$0.20 & -0.22$\pm$0.19 & -0.07$\pm$0.12 & -0.28$\pm$0.08 & -0.16$\pm$0.09 & -0.15$\pm$0.19 \\
NGC1385 & -0.16$\pm$0.20 & -0.26$\pm$0.19 & - & - & - & -0.16$\pm$0.20 \\
NGC1433 & -0.09$\pm$0.17 & - & -0.06$\pm$0.11 & -0.09$\pm$0.16 & 0.12$\pm$0.06 & -0.09$\pm$0.16 \\
NGC1512 & -0.08$\pm$0.20 & -0.13$\pm$0.20 & -0.08$\pm$0.08 & -0.10$\pm$0.17 & -0.01$\pm$0.08 & -0.09$\pm$0.18 \\
NGC1566 & -0.16$\pm$0.15 & -0.17$\pm$0.17 & -0.12$\pm$0.09 & - & - & -0.16$\pm$0.15 \\
NGC1672 & -0.28$\pm$0.17 & -0.32$\pm$0.18 & -0.19$\pm$0.13 & -0.26$\pm$0.07 & -0.21$\pm$0.18 & -0.26$\pm$0.17 \\
NGC2835 & -0.40$\pm$0.19 & -0.36$\pm$0.19 & -0.28$\pm$0.10 & - & - & -0.38$\pm$0.19 \\
NGC3351 & -0.11$\pm$0.15 & - & -0.13$\pm$0.08 & -0.04$\pm$0.13 & -0.07$\pm$0.06 & -0.09$\pm$0.14 \\
NGC3627 & -0.18$\pm$0.15 & -0.23$\pm$0.16 & -0.07$\pm$0.09 & - & 0.07$\pm$0.06 & -0.17$\pm$0.15 \\
NGC4254 & -0.30$\pm$0.18 & -0.33$\pm$0.18 & - & - & -0.25$\pm$0.07 & -0.30$\pm$0.21 \\
NGC4303 & -0.18$\pm$0.16 & -0.24$\pm$0.17 & -0.10$\pm$0.09 & - & -0.08$\pm$0.08 & -0.20$\pm$0.16 \\
NGC4321 & -0.12$\pm$0.16 & -0.11$\pm$0.16 & -0.07$\pm$0.11 & -0.10$\pm$0.11 & -0.07$\pm$0.11 & -0.11$\pm$0.16 \\
NGC4535 & -0.21$\pm$0.12 & -0.15$\pm$0.13 & -0.19$\pm$0.10 & - & -0.22$\pm$0.09 & -0.20$\pm$0.12 \\
NGC5068 & -0.40$\pm$0.20 & - & -0.37$\pm$0.14 & - & - & -0.39$\pm$0.20 \\
NGC7496 & -0.15$\pm$0.17 & - & -0.17$\pm$0.14 & - & - & -0.15$\pm$0.16 \\
IC5332 & -0.36$\pm$0.17 & - & - & - & -0.33$\pm$0.14 & -0.35$\pm$0.17 \\
\hline
Offset to median  & -0.01$\pm$0.18 & -0.02$\pm$0.17 & 0.04$\pm$0.12 & 0.01$\pm$0.15 & 0.04$\pm$0.12 & - \\
(all galaxies) \\
\hline
\end{tabular}
\caption{Median LumW [Z/H] measured for each environment in each galaxy. The right-most column shows the median LumW [Z/H] across the full MUSE mosaic, considering all environments. Errors correspond to the standard deviation of the LumW [Z/H] distribution within each environment. The bottom row shows the median offset (in dex, for the full sample) of the LumW [Z/H] distribution within each environment of a given galaxy, with respect to the overall distribution (considering all environments) of the same galaxy. }
\label{tab:median_lw_z_per_env}
\end{table*}

\begin{table*}
\centering
\begin{tabular}{ccccccc}
\hline
\hline
Target & Disk & Sp. arms & Bar & Ring & Center & All \\
\hline
NGC0628 & -0.30$\pm$0.23 & -0.31$\pm$0.25 & - & - & -0.24$\pm$0.14 & -0.30$\pm$0.24 \\
NGC1087 & -0.33$\pm$0.27 & - & -0.23$\pm$0.22 & - & - & -0.30$\pm$0.28 \\
NGC1300 & -0.00$\pm$0.22 & -0.09$\pm$0.24 & -0.07$\pm$0.13 & -0.01$\pm$0.09 & -0.05$\pm$0.11 & -0.03$\pm$0.22 \\
NGC1365 & -0.12$\pm$0.24 & -0.11$\pm$0.24 & -0.00$\pm$0.13 & 0.07$\pm$0.11 & 0.13$\pm$0.10 & -0.07$\pm$0.23 \\
NGC1385 & -0.13$\pm$0.28 & -0.27$\pm$0.26 & - & - & - & -0.13$\pm$0.28 \\
NGC1433 & 0.00$\pm$0.18 & - & -0.03$\pm$0.11 & -0.01$\pm$0.17 & 0.22$\pm$0.05 & -0.00$\pm$0.17 \\
NGC1512 & -0.00$\pm$0.24 & -0.04$\pm$0.25 & -0.02$\pm$0.08 & -0.01$\pm$0.19 & 0.02$\pm$0.05 & -0.01$\pm$0.21 \\
NGC1566 & -0.18$\pm$0.21 & -0.13$\pm$0.23 & -0.09$\pm$0.11 & - & - & -0.16$\pm$0.21 \\
NGC1672 & -0.30$\pm$0.23 & -0.30$\pm$0.26 & -0.16$\pm$0.16 & 0.01$\pm$0.11 & 0.05$\pm$0.12 & -0.26$\pm$0.22 \\
NGC2835 & -0.49$\pm$0.29 & -0.47$\pm$0.27 & -0.35$\pm$0.19 & - & - & -0.48$\pm$0.28 \\
NGC3351 & -0.00$\pm$0.17 & - & -0.02$\pm$0.08 & 0.02$\pm$0.14 & 0.05$\pm$0.08 & 0.00$\pm$0.15 \\
NGC3627 & -0.02$\pm$0.20 & -0.02$\pm$0.21 & 0.04$\pm$0.10 & - & 0.10$\pm$0.07 & -0.01$\pm$0.20 \\
NGC4254 & -0.19$\pm$0.27 & -0.19$\pm$0.28 & - & - & -0.15$\pm$0.10 & -0.16$\pm$0.28 \\
NGC4303 & -0.09$\pm$0.24 & -0.17$\pm$0.26 & 0.09$\pm$0.10 & - & 0.15$\pm$0.07 & -0.10$\pm$0.25 \\
NGC4321 & -0.15$\pm$0.23 & -0.11$\pm$0.22 & -0.10$\pm$0.15 & 0.20$\pm$0.09 & 0.07$\pm$0.14 & -0.13$\pm$0.22 \\
NGC4535 & -0.14$\pm$0.15 & -0.06$\pm$0.15 & -0.11$\pm$0.10 & - & -0.01$\pm$0.13 & -0.13$\pm$0.15 \\
NGC5068 & -0.48$\pm$0.28 & - & -0.41$\pm$0.21 & - & - & -0.48$\pm$0.28 \\
NGC7496 & -0.20$\pm$0.22 & - & -0.19$\pm$0.18 & - & - & -0.20$\pm$0.21 \\
IC5332 & -0.56$\pm$0.23 & - & - & - & -0.40$\pm$0.20 & -0.54$\pm$0.23 \\
\hline
Offset to median  & -0.01$\pm$0.24 & -0.01$\pm$0.25 & 0.03$\pm$0.14 & 0.00$\pm$0.17 & 0.10$\pm$0.16 & - \\
(all galaxies) \\
\hline
\end{tabular}
\caption{Median MassW [Z/H] measured for each environment in each galaxy. The right-most column shows the median MassW [Z/H] across the full MUSE mosaic, considering all environments. Errors correspond to the standard deviation of the MassW [Z/H] distribution within each environment. The bottom row shows the median offset (in dex, for the full sample) of the MassW [Z/H] distribution within each environment of a given galaxy, with respect to the overall distribution (considering all environments) of the same galaxy. }
\label{tab:median_mw_z_per_env}
\end{table*}

\subsection{Further insights on the inside-out growth of galaxies}
\label{sec:inside_out}

In this section, we explore in more detail what we can learn about the radial structure of the stellar mass assembly of PHANGS galaxies, beyond the mean age radial gradients.

\begin{figure*}
\centering
 	\includegraphics[width=0.9\textwidth]{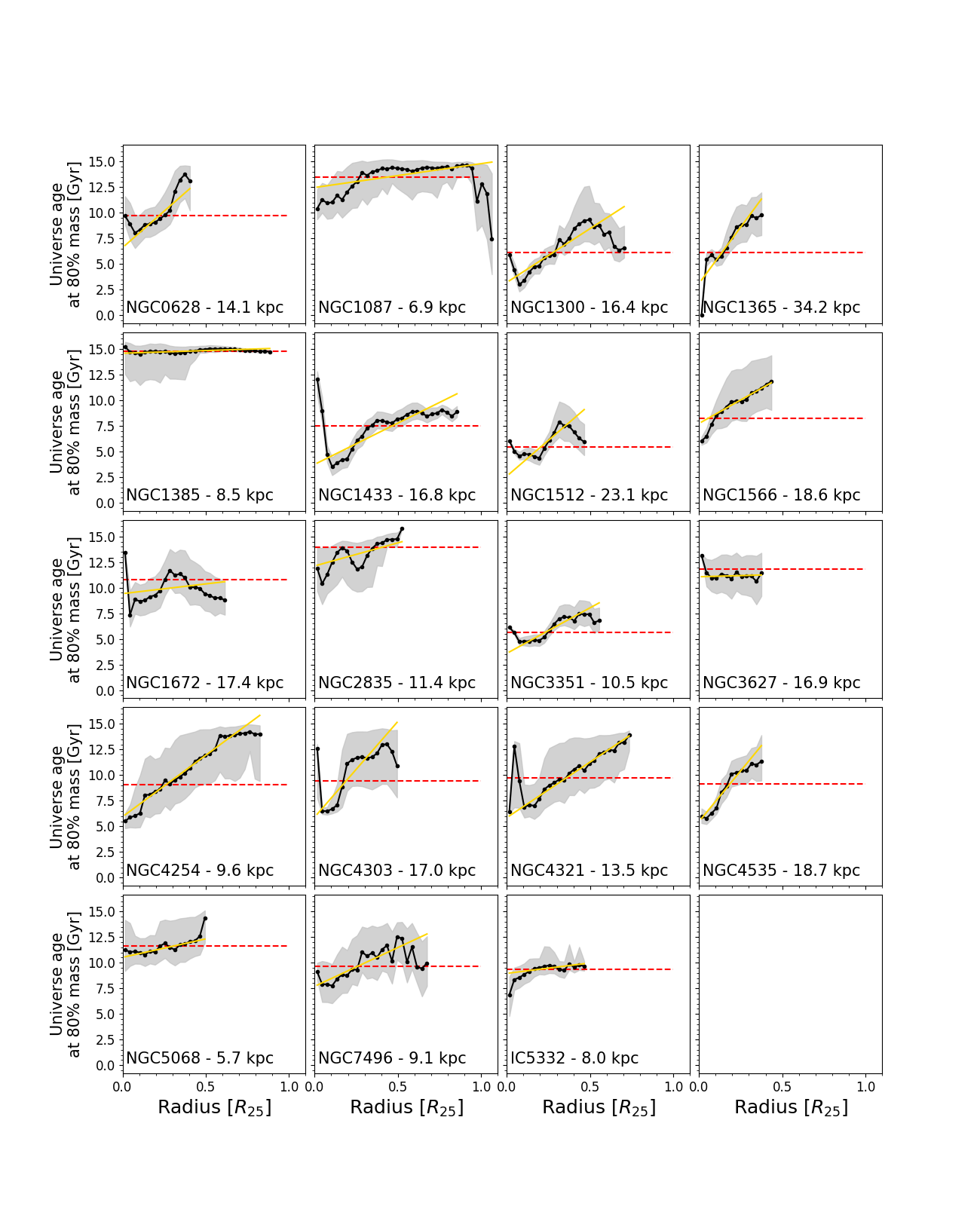}
     \caption[Radial assembly history of PHANGS galaxies]{Radial assembly history of our sample galaxies. The $y$-axis shows the age of the Universe at which each radial bin assembled $80\%$ of its total current stellar mass. A positive trend in these plots implies that outer regions assembled their stellar mass later in cosmic history than inner regions. The yellow solid lines show the best-fit linear model for each galaxy. The red dashed line show the time at which the full galaxy formed $80\%$ of its total current stellar mass. The gray shaded area represents the uncertainty in our measurement, estimated by propagating the stellar mass uncertainty in the integration of the SFH of each radial bin. The value of $R_{25}$ (kpc) of each galaxy is indicated in each corresponding panel. The bottom-right panel displays the trends for all galaxies, color-coded by total stellar mass. By construction, the highest possible value for the $y$-axis is given by the age of our oldest age bin in the SSP fitting ($\sim15$ Gyr, see Sec.~\ref{sec:stellarpops_maps}).}
     \label{fig:radial_inside_out}   
\end{figure*}

Figure~\ref{fig:radial_inside_out} shows the age of the Universe at which each radial bin formed $80\%$ of its current total stellar mass \citep[neglecting the impact of stellar migration, whose contribution in shaping the radial structure of stellar populations is estimated to be of second order; see e.g., ][]{Zhuang2019}. A positive trend implies that outer regions assembled their stellar mass later in cosmic time than inner regions. The figure shows that most galaxies follow an overall well-defined positive trend, a direct imprint of the inside-out growth of galaxies. The most remarkable exception is NGC\,1385, that shows an almost perfectly flat trend. This, together with its positive age and metallicity radial trends make NGC\,1385 a very peculiar system. Another exception is NGC\,3627, which also shows a nearly flat MassW age radial profile. NGC\,3627 is a member of the interacting group Leo Triplet \citep{Zhang1993}, interactions with nearby galaxies could have triggered episodes of star formation \citep[e.g., ][]{Renaud2019} that might explain its peculiar radial assembly history of stellar mass.

Other galaxies, such as NGC\,1672 or NGC\,1300 show a positive trend across parts of their stellar disk, and then a negative trend in the outer radii. Such a negative trend suggests that in these galaxies, the outer regions have remained relatively quiescent for long periods of time, compared to the inner radii. This does not imply that they lack current star formation, but it means they have not formed a substantial amount of stellar mass in the last $\sim5$ Gyr. Indeed, Fig.~\ref{fig:radial_LW_age} shows that these galaxies do not show higher LumW ages in their outer radii.

\begin{figure*}[h!]
\centering
 	\includegraphics[width=\textwidth]{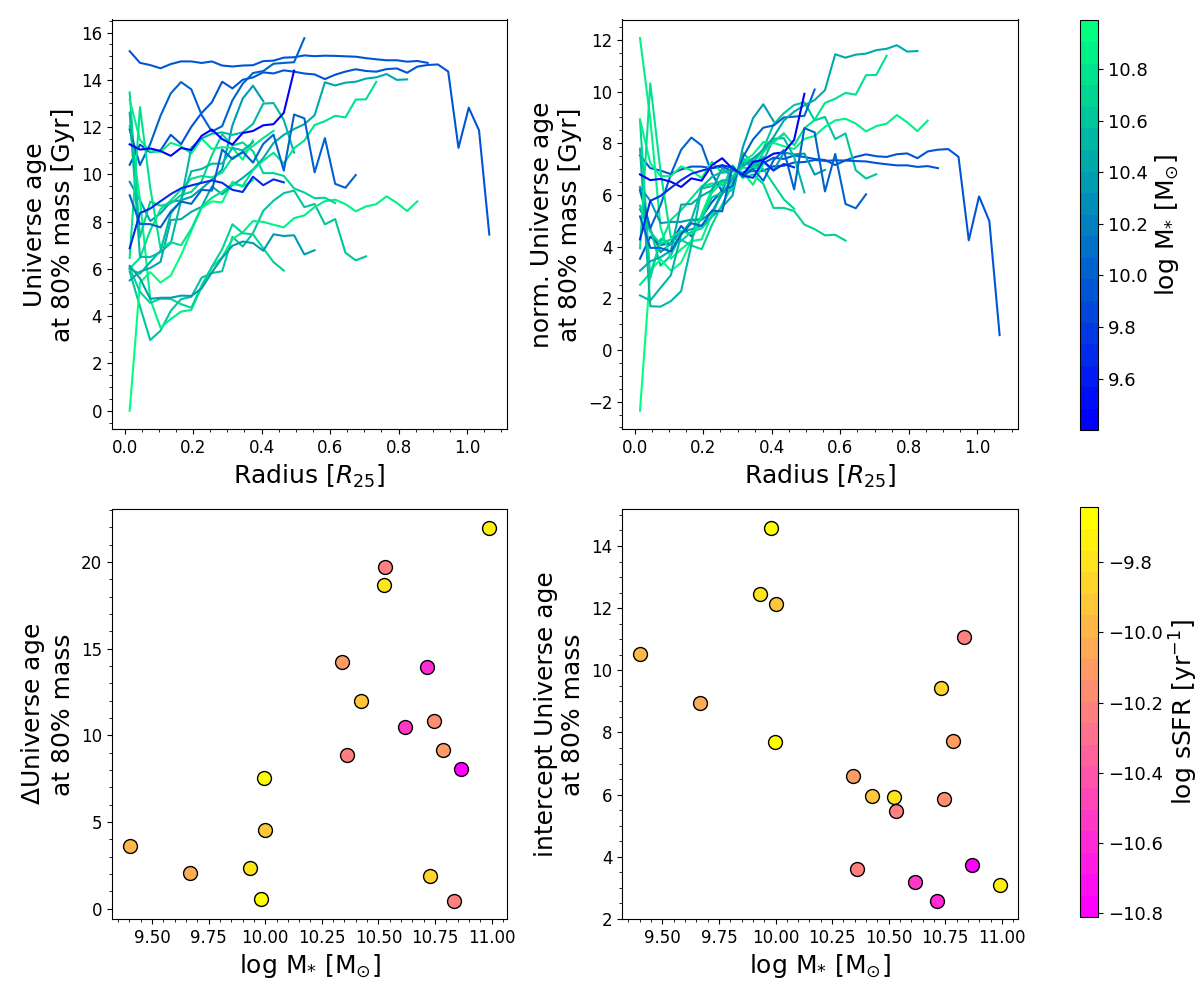}
     \caption{Radial assembly history profiles for measured for our sample galaxies, considering all environments. Top left panel shows the measured profiles, colored by total galaxy stellar mass. Top right panel shows the profiles normalized by their value at $0.3$ R$_{25}$, to better compare their shapes. The bottom panels show the slope and intercept of the radial profiles (indicated with a solid yellow line in each panel of Fig.~\ref{fig:radial_inside_out}) as a function of the total stellar mass of galaxies, color coded by total specific star formation rate (sSFR).}
     \label{fig:summary_radial_assembly}   
\end{figure*}

Fig.~\ref{fig:summary_radial_assembly} shows clearly how the normalization and slope of the of the radial assembly profiles depend on the galaxy stellar mass and specific SFR. We find a trend such that the radial assembly profiles of more massive galaxies (green) are steeper, with values at smaller radii being preferentially located at younger ages (i.e., in the earlier Universe). Less massive galaxies (blue) show flatter gradients, with their inner regions preferentially showing formation times at higher cosmic times. 

\subsection{Searching for evidence of radial mixing in bars}
\label{sec:radial_mix_bar}
Interaction of stars or gas with nonaxisymmetric structures can lead to a change in angular momentum, ultimately causing their radial migration toward different orbits. Different mechanisms, such as the exchange of angular momentum at the corotation resonance of spiral arms \citep*{Sellwood2002}, or induced by the bar-spiral arms resonance overlap \citep{Minchev2011} are thought to drive the migration of stars. Radial migration of stars is expected to produce a flattening of radial gradients of the stellar population properties, especially for older stellar populations \citep[e.g., ][]{DiMateo2013}.

Simulations show that bars provide very efficient mechanisms for the radial redistribution of gas and angular momentum \citep{Brunetti2011, Sormani2015, Spinoso2017, Perry2021}; hence, radial gradients of stellar population properties along the bar are expected to be flatter than along the galactic disk due to orbital mixing and stellar radial migration \citep[although the latter is mostly visible from the corotation radius outward, e.g., ][]{Friedli1994, DiMateo2013}. 

A number of studies have tested this prediction using nearby galaxies \citep{SanchezBlazquez2014, Seidel2016, Fraser2019, Neumann2020}, obtaining a variety of results. \citet{SanchezBlazquez2014} see no significant differences between the radial profiles of stellar age and metallicity of barred and unbarred galaxies from the CALIFA sample. \citet{Seidel2016} use data from the BaLROG project \citep{Seidel2015}, and find that stellar metallicity gradients along the bar major axis are considerably flatter than along its minor axis. \citet{Fraser2019} measure flatter stellar age and metallicity gradients along bars, compared to disks in galaxies from the MaNGA sample. \citet{Neumann2020} used data from the TIMER project \citep{Gadotti2019}, and find flatter MassW age and metallicity radial profiles along bars, compared to the disk, but differences are only mild with significant scatter, and no clear trend with bar strength.

\begin{figure}[h!]
\centering
 	\includegraphics[width=\columnwidth]{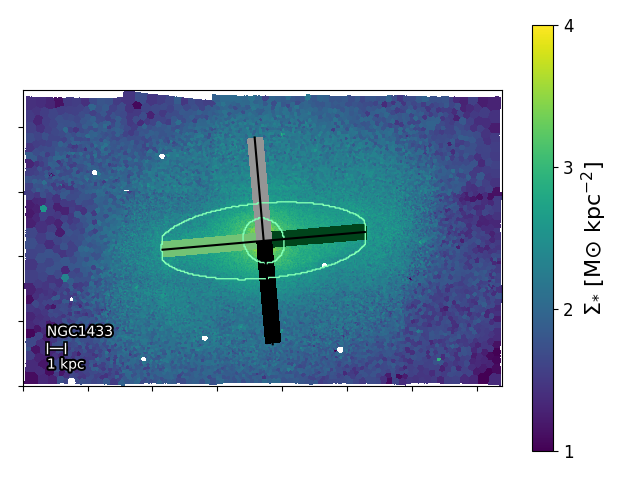}
     \caption[Representation of the location of the pseudo-slits along the bar of NGC\,1433, and perpendicular to the bar axis]{Representation of the location of the pseudo-slits along the bar of NGC\,1433 (pale and dark green rectangles), and perpendicular to the bar axis (gray and black rectangles). The position of the bar is indicated by the light-green ellipses. The background shows the stellar mass surface density map of the galaxy.}
     \label{fig:NGC1433_bar_mask}   
\end{figure}

\begin{figure*}[h!]
\centering
 	\includegraphics[width=0.7\textwidth]{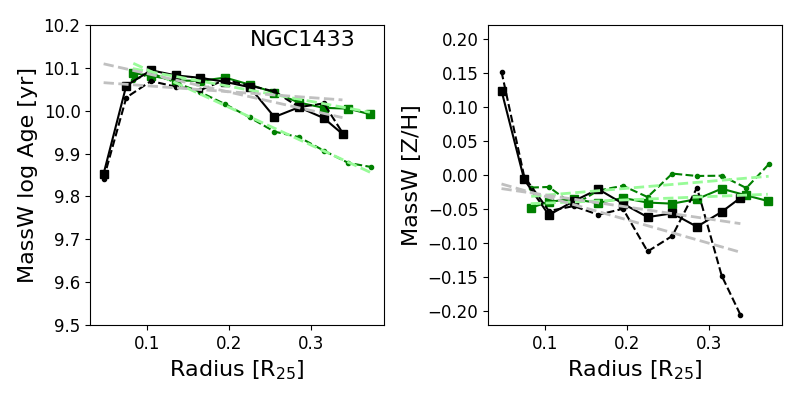}
     \caption[Mass-weighted age and metallicity radial profiles along the bar major axis, and along its perpendicular direction, for NGC\,1433]{Mass-weighted age (left) and [Z/H] (right) radial profiles along the bar major axis (dark green), and along its perpendicular direction (black), for NGC\,1433. The solid and dashed dark green and black lines represent the gradient measured toward each possible direction from the center. The pale green dashed lines show the best-fitting linear model to the bar major axis gradient (dark green lines), and the silver dashed lines show the best-fitting to the gradients measured in the direction perpendicular to the bar.}
     \label{fig:NGC1433_bar_profiles}   
\end{figure*}

Here, we measure MassW age and metallicity radial gradients along pseudo-slits placed across the bar major axis and perpendicular to it. Figure~\ref{fig:NGC1433_bar_mask} shows the position of the pseudo-slits overplotted on the stellar mass surface density map of NGC\,1433. 
We have measured radial gradients along each of these two pseudo-slits (excluding centers). The radial profiles (and the best-fit gradients, measured using an OLS fitting routine) of MassW age and metallicity measured along each one of these pseudo-slits for NGC\,1433 are shown in Fig.~\ref{fig:NGC1433_bar_profiles}. We excluded the innermost and outermost radial bins when measuring the radial gradients, in order to avoid potential deviations from the radial trend driven by these extreme values. The gradients along the bar major axis, and their perpendicular counterparts, are calculated as the mean gradient of the two possible directions along each axis from the center (i.e., mean gradient between dark-green (black) and pale-green (gray) pseudo-slits in Fig.~\ref{fig:NGC1433_bar_mask}). The difference between both measurements span a wide range of values, from $\sim0$, to more than 1 dex, and it is interpreted as the uncertainty in the galaxy gradient (along a given axis).

\begin{figure*}
\centering
 	\includegraphics[width=0.95\textwidth]{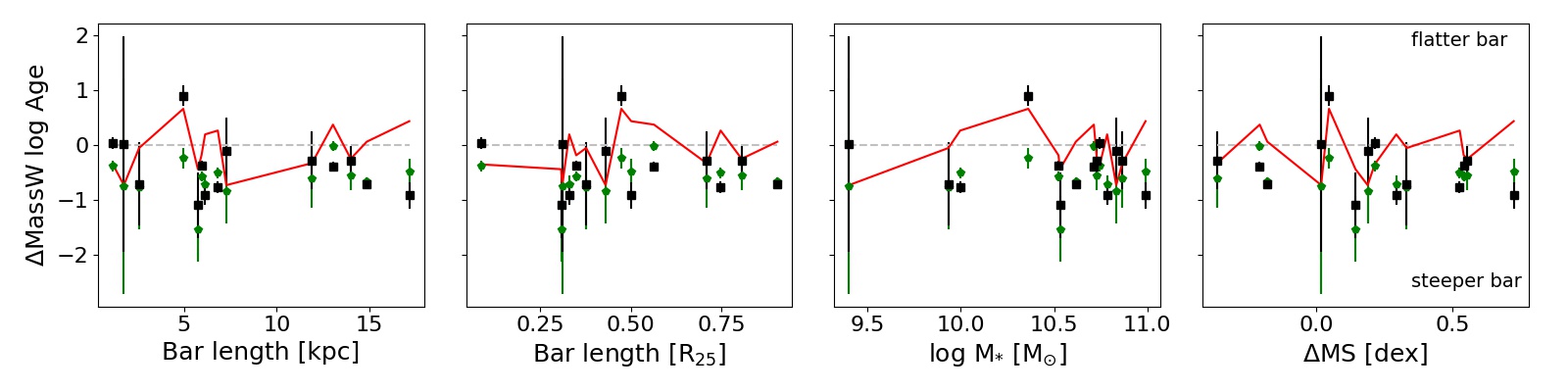}
     \caption[Comparison of mass-weighted age gradient along bars, and along their minor axis, as a function of bar size and global galaxy properties]{Mass-weighted age gradients measured along bar major axis (green pentagons), and along its perpendicular direction (black squares), as a function of the bar length (in units kpc and $R_{25}$), total stellar mass, and offset from the global main sequence of galaxies ($\Delta$MS), for 14 barred galaxies in our sample. The red line is calculated following Eq.~\ref{eq:red_line_age}. It is defined to be positive in galaxies that show a flatter gradient along their bar major axes, compared to the bar perpendicular direction, and negative values otherwise.}
     \label{fig:mixing_evidence_age}   
\end{figure*}

\begin{figure*}
\centering
 	\includegraphics[width=0.95\textwidth]{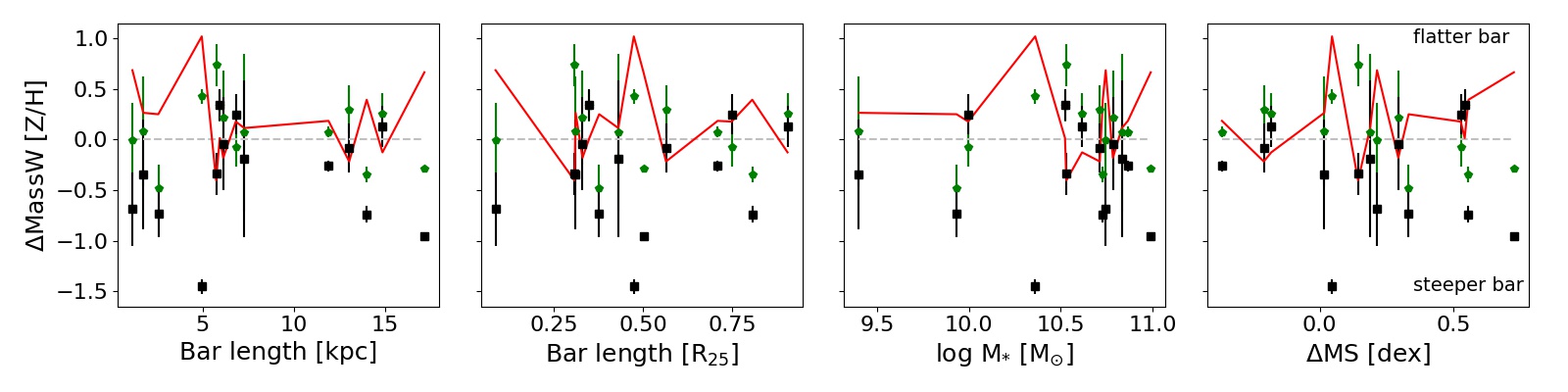}
     \caption[Comparison of mass-weighted metallicity gradient along bars, and along their minor axis, as a function of bar size and global galaxy properties]{Mass-weighted [Z/H] gradients measured along bar major axis (green pentagons), and along its perpendicular direction (black squares), as a function of the bar length (in units kpc and $R_{25}$), total stellar mass, and offset from the global main sequence of galaxies ($\Delta$MS), for 14 barred galaxies in our sample. The red line is calculated following Eq.~\ref{eq:red_line_z}. It is defined to be positive in galaxies that show a flatter gradient along their bar major axis, compared to the bar perpendicular direction, and negative values otherwise.}
     \label{fig:mixing_evidence_z}   
\end{figure*}

Figures~\ref{fig:mixing_evidence_age}  and \ref{fig:mixing_evidence_z} show the slopes of the MassW age and metallicity gradients measured along the bar axis (green), and its perpendicular direction (black), for the 14 barred galaxies in our sample, as a function of bar length (in units of kpc and $R_{25}$), total stellar mass, and offset from the main sequence of galaxies ($\Delta$MS). NGC\,2835 was excluded because the small size of its bar makes the measurement of the gradients unreliable.

The red line is defined as the difference of the absolute value of the gradient measured along the bar and its perpendicular direction, that is,
\begin{equation}
    \label{eq:red_line_age}
    \begin{aligned}
    \Delta \mathrm{MassW}\,\mathrm{log Age_{disk-bar}} = 
    | \Delta \mathrm{MassW}\,\mathrm{log Age}_{\mathrm{disk}} | \\
    - | \Delta \mathrm{MassW}\,\mathrm{log Age}_{\mathrm{bar}} |
    \end{aligned}
\end{equation}
and
\begin{equation}
    \label{eq:red_line_z}
    \begin{aligned}
    \Delta \mathrm{MassW}\,\mathrm{[Z/H]_{disk-bar}} = | \Delta \mathrm{MassW}\,\mathrm{[Z/H]}_{\mathrm{disk}} | \\
    - | \Delta \mathrm{MassW}\,\mathrm{[Z/H]}_{\mathrm{bar}} |
    \end{aligned}
\end{equation}
implying that positive values correspond to a flatter profile along the bar, compared to the perpendicular direction, and negative values otherwise.

For $\Delta$MassW age, we do not see evidence of flatter profiles along the bar. On the contrary, we find that, on average, MassW age gradients are steeper along the bar ($-0.59\pm0.33$), than along its perpendicular direction ($-0.38\pm0.50$), and less than half of the galaxies (6 out of 14) show a flatter MassW age gradient along the bar. We do not find any trend between flatter profiles, and bar length, total stellar mass, or $\Delta$MS.

On the other hand, we observe flatter MassW [Z/H] gradients on average along the bar. We calculate a mean gradient of $0.08\pm0.31$ along bars, and $-0.30\pm0.48$ in the perpendicular direction. In terms of individual galaxies, 10 out of the 14 barred galaxies in our sample show a flatter stellar metallicity profile along the bar, without any significant trend with bar length, total stellar mass, or $\Delta$MS.

Finding more negative age profiles along the bar is not surprising. \citet{Seidel2016} also find that age profiles are more negative along the bar semi-major axis, than along its semi-minor axis. This is consistent with chemodynamic simulations \citep{Wozniak2007} that predict that younger stellar populations will be confined to more elongated orbits, populating the edges of the bar. \citet{Neumann2020} report a similar spatial distribution of stellar populations within the bar of nearby galaxies, with old (>8 Gyr) stars shaping the inner and rounder part of the bar, intermediate age (2-6 Gyr) stars trapped on more elongated orbits, and an accumulation of young (<2 Gyr) stars at the bar edges. 

In the case of metallicity, we find considerably flatter profiles along bars, compared to their perpendicular direction. This result is consistent with simulations that investigate the bar-driven secular evolution of galaxies, and predict flatter metallicity gradients in barred galaxies \citep{DiMateo2013}, and that have been observationally confirmed \citep[e.g., ][]{Seidel2016}. 

In conclusion, we find steeper age gradients, but flatter metallicity gradients along bars, compared to their semi-minor axis. The natural explanation for flatter [Z/H] gradients is orbital mixing of gas along the bar, as gas is more susceptible to nonaxisymmetric structure than stars. Hence, gas from which stars form is homogenized along the bar. Another explanation for flatter [Z/H] gradients along bars could be a flatter surface mass density profile along bars as compared to the disk, such that the local environment in the bar could be more efficient in recycling the gas. However, we acknowledge that due to the limited size of our sample, our results are also consistent with similar age and metallicity gradients along bars, and perpendicular to them.

\subsection{Kinematic imprints of different stellar populations across the galactic disk}
\label{sec:dynamical_heating}
\subsubsection{Previous measurements of the age -- velocity dispersion relation}
The dynamical evolution of galactic disks is encoded in present-day observations of the kinematic properties of its different stellar populations. It has long been known that older populations of solar neighborhood stars have larger velocity dispersions, compared to younger stars \citep{Roman1954, Wielen1977, Freeman1987}. The trend of increasing velocity dispersions with age is known as the age-velocity  relation (AVR). \citet{Binney2000} used \texttt{Hipparcos} \citep{Perryman1997} data to quantify the rate at which the velocity dispersion of a coeval group of stars increases with time, and found that it scales with age as $\tau^{0.33}$, with $\tau$ in Gyr. Recent works \citep{Yu2018,Mackereth2019,Tarricq2021} have performed similar measurements for open clusters and field stars in our Galaxy, using data from Gaia DR2 \citep{GaiaCollaboration2018}. 

An alternative explanation to the AVR is that the interstellar medium at high redshift was more turbulent, and thus, older stars retain larger velocity dispersion to present day \citep{Stott2016}. Along this line, \citet{Leaman2017} find that observations of galaxies in the Local Group are consistent with a model in which stars are born with a velocity dispersion close to that of the gas from which they formed, and are then dynamically heated with an efficiency that depends on the galaxy mass, among other factors \citep[see, e.g., ][]{Pinna2018}. 

 Different mechanisms can contribute to the progressive dynamical heating of stars that lead to the AVR, such as interaction with giant molecular clouds \citep*{Spitzer1953}, interaction with spiral arms \citep{Barbanis1967, Sellwood1984, Mackereth2019}, bars \citep{Grand2016}, or even external sources of perturbation \citep{Grand2016, Pinna2018}. However, measuring the relative importance of these different mechanisms is not straightforward. Studies of the AVR are limited to the Local Group, where individual stars can be resolved. This limitation means that the parameter space of factors that could contribute to the diffusion of stars (e.g., galactic structure) is poorly covered. 

 \subsubsection{Age -- Velocity dispersion relation in our data}
 Here, we present an exploration of the AVR measured in nearby galaxies from the PHANGS sample, using the LumW age and velocity dispersion maps derived as explained in Sec.~\ref{sec:data}. Although the spatial resolution of $\sim 100$~pc is far from being sufficient to resolve individual stars, it is possible to resolve young star-forming regions, and more generally, regions dominated by stellar populations of distinguishable ages. Furthermore, since at this spatial scale the structural components of the galaxy (namely bars, spiral arms, centers, rings, and disks) are clearly resolved, we can also search for changes in dynamical heating of stars as a function of local environment.
 
\begin{figure*}
\centering
 	\includegraphics[width=0.9\textwidth]{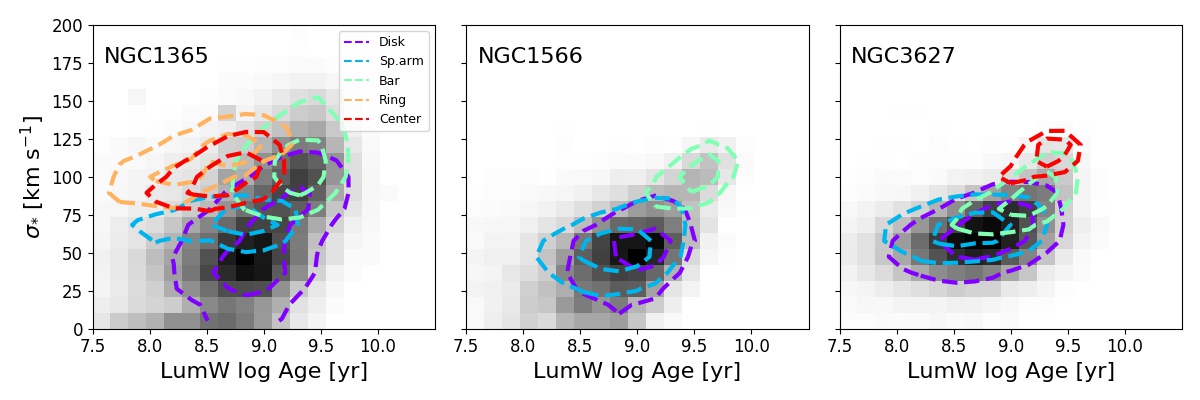}
     \caption[2D histogram that shows the distribution of stellar velocity dispersion versus log age (luminosity-weighted) for three sample galaxies]{2D histogram that shows the distribution of stellar velocity dispersion ($\sigma_{*}$) and log age (luminosity-weighted), for NGC\,1365 (left), NGC\,1566 (center), and NGC\,3627 (right). The color of each [LumW log Age, $\sigma_{*}$] bin scales with the number of pixels within that bin. The contours show the one- and two-sigma limits of the distribution of each individual galactic environment, following the color code indicated in the top-right corner of the right of the panel.}
     \label{fig:NGC1566_NGC3627_age_sigma}   
\end{figure*}

 Figure~\ref{fig:NGC1566_NGC3627_age_sigma} shows the AVR (using LumW age) as 2D histograms for three galaxies (NGC\,1365, NGC\,1566 and NGC\,3627) of our sample. These three galaxies are good examples of the general trend that we see in our sample; spiral arms and disk share similar stellar velocity dispersion ($\sigma_{*}$) values, with the former located preferentially at lower ages, and centers and bars dominated by older stellar populations, with higher $\sigma_{*}$ values (although some galaxies, such as NGC\,1365, can also host young stellar populations in their centers). These positive correlations could, at first glance, be naively interpreted as the AVR. However, in order to properly interpret these trends, it is important to take into account the overall negative $\sigma_{*}$ radial gradient set by the underlying stellar mass distribution, following the virial theorem \citep[see, e.g., ][]{Bittner2020}. The overall negative $\sigma_{*}$ radial profile is clearly visible in the bottom left panel of Fig.~\ref{fig:NGC1566_maps}. The $\sigma_{*}$ trend, in combination with the negative LumW age profiles discussed in Sec.~\ref{sec:radil_profiles_ag_z}, result in the positive trends seen in Fig.~\ref{fig:NGC1566_NGC3627_age_sigma}.

Therefore, in order to explore the trend between $\sigma_{*}$ and LumW age in a meaningful way, we investigate the radial gradients of $\sigma_{*}$ for three different LumW age bins; young (LumW age $<100$ Myr), intermediate ($100$ Myr $ <$ LumW age $< 600$ Myr), and old (LumW age $> 3$ Gyr). This choice is motivated by the results from \citet{Tarricq2021}, who find that the increase in $\sigma_{*}$ occurs more rapidly within the first Gyr after stars form, following a much slower increase from this point. To do this measurement, we take the square root of the azimuthally averaged square of the $\sigma_{*}$ values in a given radial bin, considering separately spaxels with LumW ages in each one of the three age bins defined.

\begin{figure*}
\centering
 	\includegraphics[width=0.9\textwidth, trim=0 0 0 0, clip]{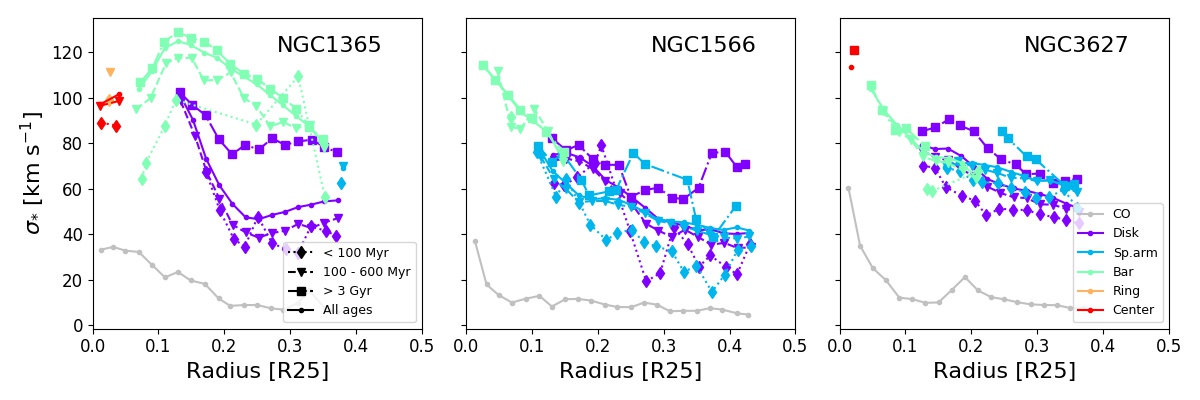}
     \caption[Radial profile of stellar velocity dispersion for three sample galaxies, plotted as a function of galactic structure and LumW age of the underlying stellar populations.]{Radial profile of stellar velocity dispersion ($\sigma_{*}$) for NGC\,1365 (left), NGC\,1566 (center), and NGC\,3627 (right), plotted as a function of galactic structure (colors) and LumW age of the underlying stellar populations. Stellar populations younger than 100 Myr are indicated by diamonds (dotted line). Stellar populations of ages in the range of $100$ Myr $<$ log LumW age $< 600$ Myr are marked with triangles (dashed line). Stars older than 3~Gyr are represented by squares (dashdotted line). Circles (solid line) show the radial profile considering all stellar populations. The gray line shows the radial velocity dispersion gradient measured for the molecular gas, from the PHANGS-ALMA CO(2-1) data.}
     \label{fig:NGC1365_NGC1566_NGC3627_sigma_prof}   
\end{figure*}

Figure~\ref{fig:NGC1365_NGC1566_NGC3627_sigma_prof} shows the $\sigma_{*}$ radial profiles for the same three galaxies, separately for the three age bins defined above, and across different galactic environments. For reference, we also display the molecular gas $\sigma_{\mathrm{mol}}$ profile, measured from the CO(2-1) PHANGS-ALMA data.

The figures show a clear age gradient in the $\sigma_{*}$ value at a given galactocentric distance, with younger populations having lower velocity dispersions than older ones. This effect is very clear across disk and spiral arms in general, but the age gradient is generally nonexistent in bars. Out of the 14 barred galaxies in our sample, the only noticeable exceptions are NGC\,1365 and NGC\,5068, with a very clear age gradient present also in the bar. It is interesting to note that these two galaxies represent the high- and low-stellar mass extremes of our sample ($\log $M$_{*}/M_\odot = $ 10.99 and 9.40, respectively). Therefore, this similarity is likely not connected to their total stellar mass. On the other hand, we do not observe significant differences between spiral arms and disks, at any given radius. Finally, the molecular gas from which young stars form shows a nearly flat radial $\sigma_{\mathrm{mol}}$ profile, with very low ($\lesssim 20$~km s$^{-1}$) velocity dispersion values, and a rapid increase toward the inner radii. Overall, these findings suggest that (i) the dynamical heating of stellar populations occurs at any given radius, increasing the velocity dispersion from $\lesssim 20$ km s$^{-1}$, up to $\gtrsim 50$~km s$^{-1}$ on timescales of hundreds of Myr; (ii) assuming that molecular gas traces the velocity dispersion of stars at birth \citep{Leaman2017}, the increase of velocity dispersion takes place in shorter timescales within the bar, as we do not see clear difference in the $\sigma_{*}$ of the different populations in bars for most barred galaxies. However, we acknowledge that this result could be driven by the lack of young stellar populations in the bar hindering the measurement of a clear age gradient. (iii) Spiral arms do not play a primary role in the heating of stellar populations, as we observe this phenomenon across the entire disks, including galaxies that do not contain distinct spiral arms. Alternatively, spiral arms could heat stellar populations in the radial and azimuthal directions, contributing only mildly to the line-of-sight velocity dispersion on galaxies with relatively low inclinations (as those in our sample)

\begin{figure*}[h!]
\centering
 	\includegraphics[width=0.9\textwidth]{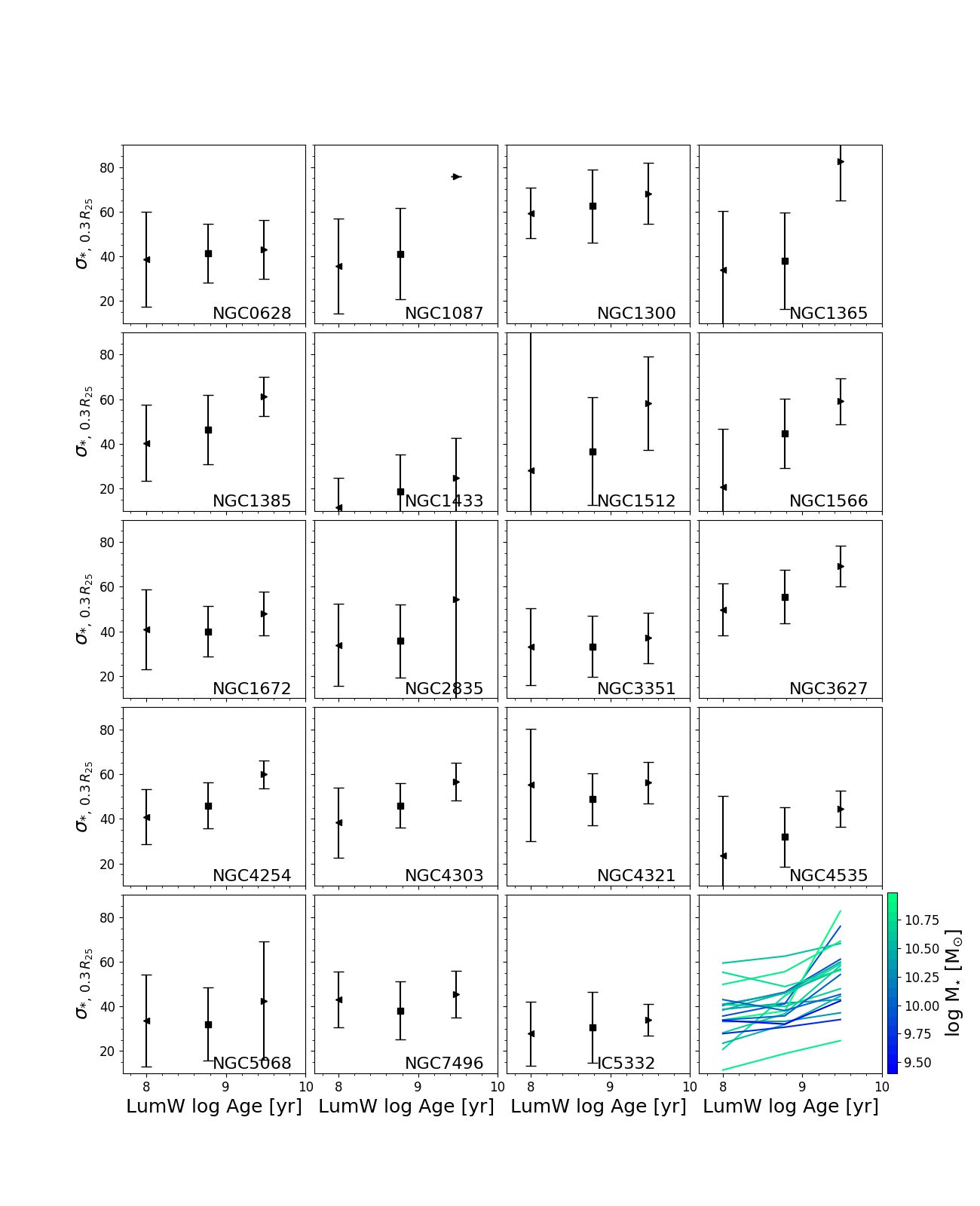}
     \caption[Mean stellar velocity dispersion measured at a galactocentric radius of 0.25-0.30 $R_{25}$ in the disk environment of all our sample galaxies for stellar populations with different luminosity-weighted ages.]{Mean stellar velocity dispersion ($\sigma_{*}$) measured at a galactocentric radius of 0.25-0.30 $R_{25}$ in the disk environment of all our sample galaxies for stellar populations with luminosity-weighted ages within the three different age bins defined in the main text. We note that for visualization purposes, the $x$-axis position of each data point is located at the edge of the age bin it represents. In each panel, the leftmost triangle represents the mean (area-weighted) velocity dispersion of all populations younger than $100$ Myr. The central square marks the mean velocity dispersion of populations with 100 Myr < LumW age < 300 Myr, and the rightmost triangle shows the same quantity for populations older than 3 Gyr. The errorbar indicates the standard deviation of the $\sigma_{*}$ values of the pixels within a given age bin. The bottom right panel shows the trend for all galaxies, colored by total stellar mass.}
     \label{fig:sigma_age_fixed_radii}   
\end{figure*}

To illustrate these trends across our full sample, Fig.~\ref{fig:sigma_age_fixed_radii} shows the $\sigma_{*}$ values measured for stellar populations within the three different age bins defined at a fixed radius (0.25 - 0.30 $R_{25}$) for the disk environment. We choose this specific range of radii because it is covered by the disk of most of our sample galaxies. For those few galaxies that do not show these three stellar ages present within the previously defined radius range (NGC\,1087, NGC\,1433, NGC\,1512, NGC\,3351, NGC\,4303), we opted by choose a different radial bin of the same length (0.05 $R_{25}$), located at the minimum possible radius in which the three populations are present in the disk environment. The figure reveals a positive trend between $\sigma_{*}$ and LumW age of the underlying stellar populations that exists for most of the galaxies in our sample. However, the differences in $\sigma_{*}$ between different age bins are often comparable or even smaller than the standard deviation of the distribution of $\sigma_{*}$ values within a given age bin (shown as the error bars in Fig.~\ref{fig:sigma_age_fixed_radii}). Although this could, in principle, imply that the measured $\sigma_{*}$ differences are not statistically significant, the fact that we find the positive slope of $\sigma_{*}$ with LumW age to be nearly ubiquitous in our sample and that error bars within a given age bin are intrinsically overestimated due to the underlying gradient (introducing an additional spread of $\sim5 - 10$ km s$^{-1}$ in a 0.05 $R_{25}$ bin), points to the fact that these differences in  $\sigma_{*}$ across stellar populations of different ages are indeed a real feature in our data. Furthermore, while it is true that the light (and thus, measurement of $\sigma_{*}$) in the pixels with young LumW ages is dominated by the spectra of young stars, these pixels also likely have a subdominant (in light) old component with higher velocity dispersion, that would bias the $\sigma_{*}$ measurements toward higher values, reducing the measured differences. This effect would become stronger in pixels with relatively higher light contributions from old stars.  

It is also worth noting that the finite spectral resolution of VLT/MUSE limits our ability to measure low ($<40$ km s$^{-1}$) $\sigma_{*}$ values depending on the S/N in each Voronoi bin. As described in Sec.~\ref{sec:stellarpops_maps}, we bin our data to a S/N of at least 35, though single spaxels can have higher S/N, especially in the high surface brightness innermost ($<0.3 R_{25}$) regions of galaxies. This suggest that the results in Fig.~\ref{fig:sigma_age_fixed_radii} are robust, but the low $\sigma_{*}$ values in Fig.~\ref{fig:NGC1566_NGC3627_age_sigma} should be treated with cauthion, and likely explains some of the scatter at low $\sigma_{*}$ values in Fig.~\ref{fig:NGC1365_NGC1566_NGC3627_sigma_prof}, especially at large radii in the young disk.

Finally, the bottom right panel of Fig.~\ref{fig:sigma_age_fixed_radii} shows that there is not any clear trend between the change of velocity dispersion with stellar age and total stellar mass of the galaxies.

The scenario in which young stellar populations get heated progressively toward higher vertical heights (i.e., increasing $\sigma_{*,z}$), producing a continuous gradient between age and $\sigma_{*,z}$ (at a given radius), is also consistent with the results from \citet{Bovy2012}, where the authors report that the vertical mass distribution of the disk of our Galaxy is consistent with a single structural component, rather than a combination of a `t`hin'' and a ``thick'' disk. This diffusion would also naturally lead to higher galactic latitudes being populated exclusively by old stars, while galactic latitudes closer to the galactic plane could be populated by a wider variety of stellar populations, a trend that has been indeed observed in our Galaxy \citep{Casagrande2016}.

It is worth noting that because of the generally low inclination of the galaxies in our sample ($<60^{\circ}$), our setting allows us to measure the velocity dispersion in the vertical direction primarily (i.e., $\sigma_{*,z}$). Although different works have shown that the diffusion of stars occurs similarly in all directions \citep{Yu2018, Mackereth2019, Tarricq2021}, there is debate in the literature about the relative strength of the diffusion toward different directions. \citet{Yu2018}, \citet{Mackereth2019} and \citet{Sharma2021} find that the diffusion in the vertical direction is stronger than in the azimuthal and radial directions. More recently, \citet{Tarricq2021} found that the diffusion in the radial and angular directions are stronger than in the vertical direction. In these works, the rate at which the velocity dispersion of young stars increases with time is quantified by the index of the power law ($\beta$) between age and $\sigma_{*}$, such that $\sigma_{*} \propto \tau^{\beta}$, following \citet{Binney2000}. A similar quantification scheme is not applicable to our data. This is because we define age bins significantly broader than the time-resolution of the measured SFH in order to maximize the radial coverage at a given age, and decrease the statistical uncertainty in our measurements. The exact offset between the young, intermediate, and old populations depends on the definition of the bins, the galactocentric distance, local galactic environment, as well as systematic uncertainties associated with our spectral fitting. Nevertheless, our analysis provides useful insights on the impact of galactic structure on the progressive dynamical heating of young stellar populations, and the speed at which it occurs in star-forming galaxies beyond our Local Group.

Finally, these findings suggest it could be worthwhile to consider a refinement of our spectral fitting methodology currently implemented in the PHANGS-MUSE DAP, to include two independent kinematic components rather than just one. Zhang et al. (in prep.) will present an exploration of this dimension in more detail, characterizing the bias in the derived stellar population parameters yielded by the different kinematic conditions of young stellar populations.

\subsection{Star formation histories across the galactic disks}
\label{sec:SFH_Disk}

In this section, we present how different galactic regions have formed stars through the lifetime of a galaxy. Interpreting the present-day spatially resolved SFHs as the real assembly history of a given galactic region neglects the impact of mechanisms that could cause stars to travel far away from their birthplace. However, the negative age and metallicity profiles, and the general inside-out formation found for these galaxies point to an evolution dominated primarily by insitu star formation, with radial migration being a second-order effect  \citep{IbarraMedel2016, Zhuang2019, Neumann2021}.

\begin{figure*}
\centering
 	\includegraphics[width=0.8\textwidth]{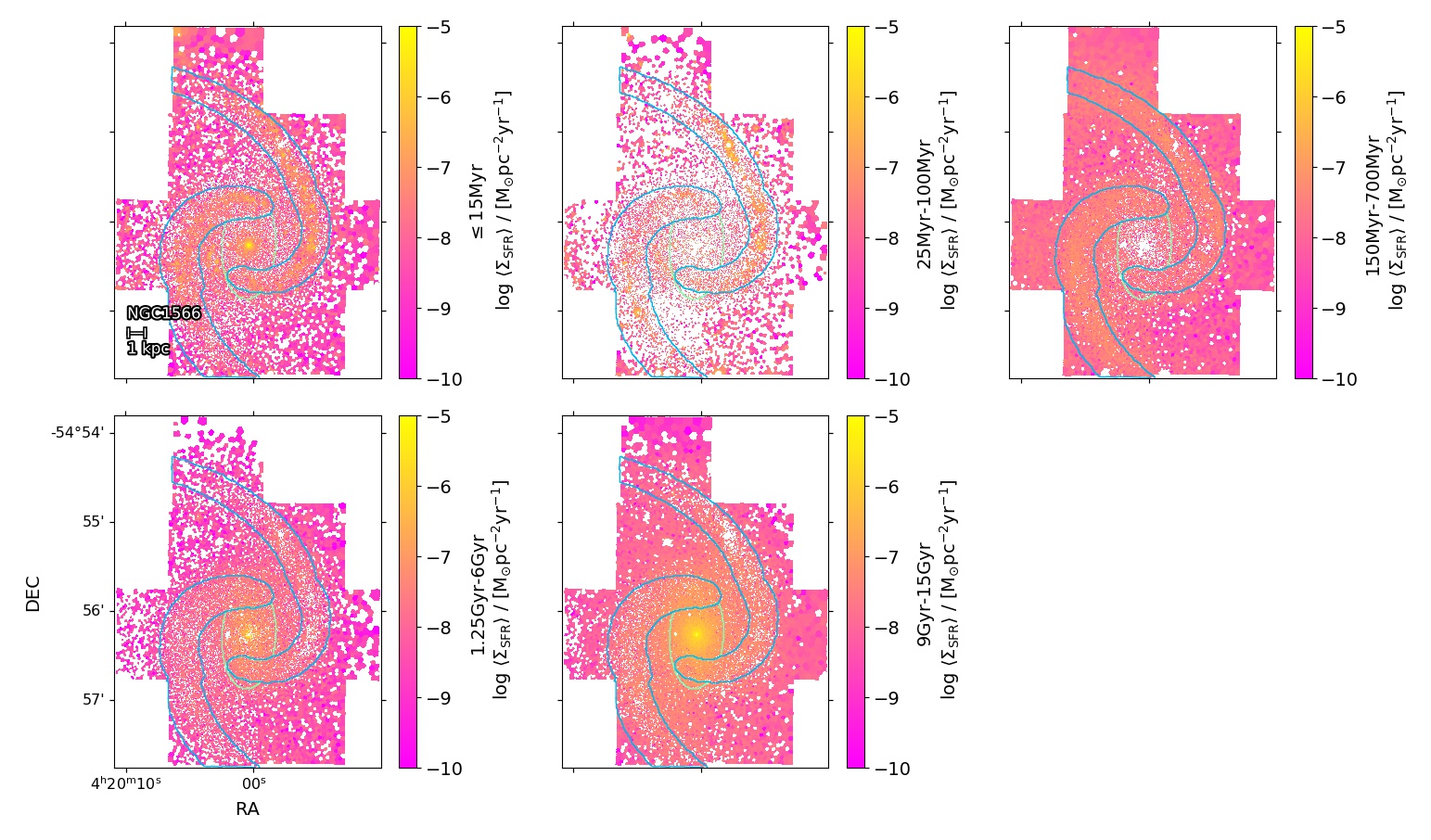}
     \caption[Time averaged SFR maps for NGC\,1566, calculated from the spatially resolved SFH.]{Time-averaged SFR in five different time bins, calculated from the spatially resolved SFH, for NGC\,1566. The panels show how different regions of the galactic disk have assembled their stellar mass in different moments of cosmic history. The blue and green contours mark the position of the spiral arms and the bar, respectively.}
     \label{fig:NGC1566_resolvedSFH}   
\end{figure*}

\begin{figure*}
\centering
 	\includegraphics[width=0.8\textwidth]{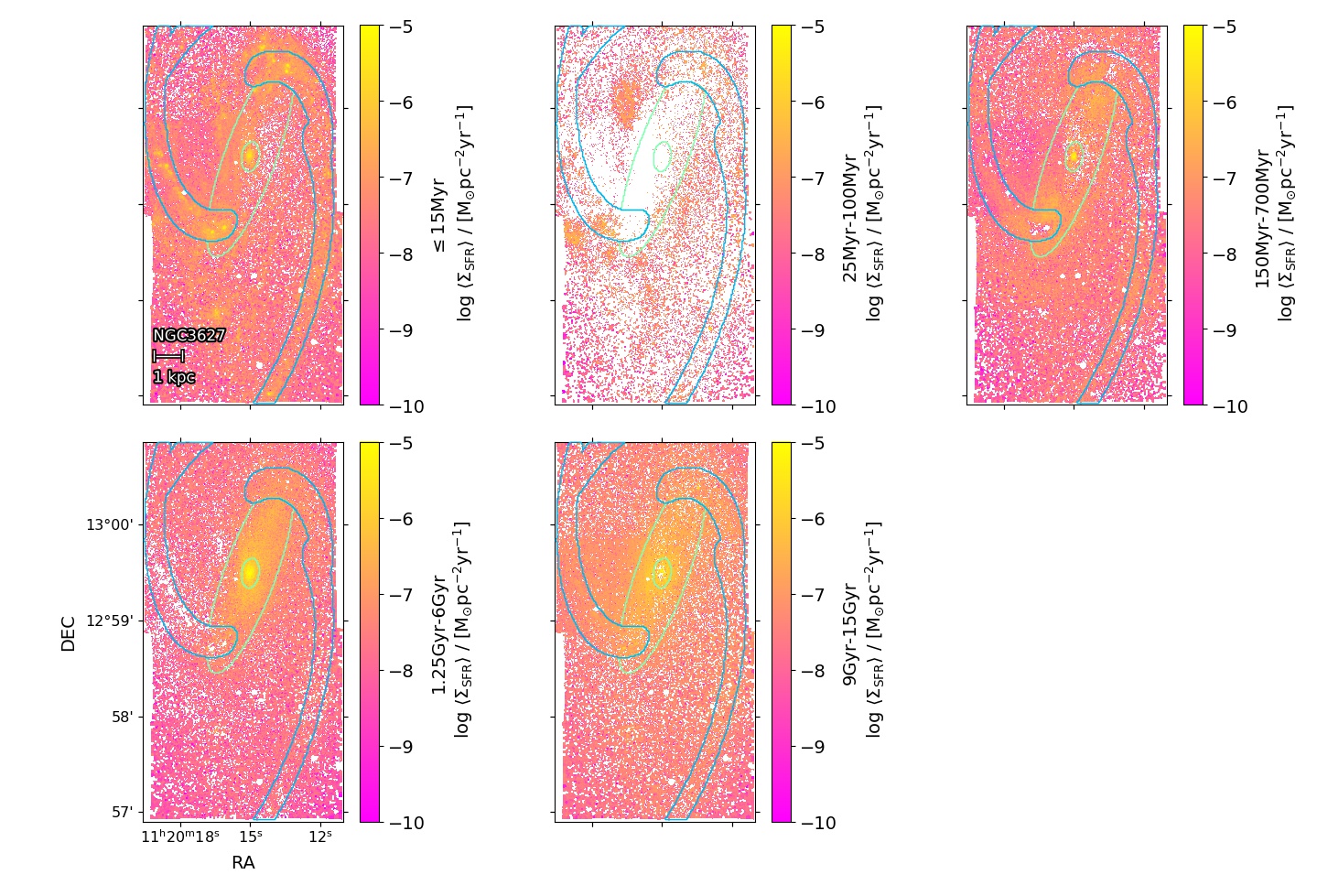}
     \caption[Time averaged SFR maps for NGC\,3627, calculated from the spatially resolved SFH.]{Time-averaged SFR in five different time bins, calculated fro the spatially resolved SFH, for NGC\,3627. The panels show how different regions of the galactic disk have assembled their stellar mass in different moments of cosmic history. The blue and green contours mark the position of the spiral arms and the bar, respectively.}
     \label{fig:NGC3627_resolvedSFH}   
\end{figure*}

Figures~\ref{fig:NGC1566_resolvedSFH} and \ref{fig:NGC3627_resolvedSFH} show the time-averaged SFR in five time bins, calculated as the total stellar mass formed within that time window (according to the derived SFH), divided by the length of the time-bin, for NGC\,1566, and NGC\,3627. There is a number of interesting features worth noting. The most obvious one is the lack of SFR in the second-youngest age bin (25 Myr - 100 Myr). The low SFR in this time bin is a consequence of the artifact in the SFHs described in Appendix.~\ref{sec:downsizing}, i.e., the dip in the SFH at $\log$ age $\sim 7.7$ Myr. Hence, we avoid physically interpreting this feature.

The central region of NGC\,1566 seems to have been quenched (i.e., stopped forming stars) after the oldest age bin, and restarting star formation activity only very recently. A possible scenario could be that bar-driven secular evolution has enabled star formation in the center, by supplying gas to the innermost region. However, as discussed in Sec.~\ref{sec:inside_out}, the lack of nonbarred galaxies in our sample makes it difficult to quantify the role of the bar in the reactivation of star formation in centers. On the other hand, NGC\,3627 seems to have been forming stars in its center continuously (this is also reflected in its inner radial bin, in Fig.~\ref{fig:radial_inside_out}). In our sample, we see a variety of behaviors, such as a drastic decrease of star formation in the inner regions, from early times to present day (e.g., NGC4254), a continuous star formation activity, or a reactivation of star formation in recent times. These findings are consistent with \citet{Querejeta2021}, who also report a wide range of depletion times for centers, possibly due to bars fuelling gas episodically to the center, cycling between star forming and quiescent phases.

Another clear feature is how galactic structure (sp. arms) is evident in the youngest age bins, but it is progressively less prominent toward older ages, to become essentially nonexistent for old stars. This could be interpreted as a consequence of the dynamical heating discussed in Sec.~\ref{sec:dynamical_heating} disrupting the galactic substructure, leading to a smooth distribution. Along this line,  \citet{Neumann2020} use magneto-hydrodynamical cosmological simulations from the Auriga project \citep{Grand2017} and find that while older stars (> 8 Gyr) follow a smooth distribution with a central concentration, younger stars (< 4 Gyr) are strongly coupled to the galactic structure. However, we can not rule out the scenario in which the AVR is driven by a more turbulent interstellar medium in the past, rather than a progressive increase of velocity dispersion with time.

The diffusion of galactic structure can be clearly appreciated in the time-averaged SFR maps of NGC\,1566, particularly for its spiral arms. While in the youngest bin, stars are located preferentially along the spiral arms (marked in blue), in the third age bin (150 - 700 Myr), the spiral pattern is still clearly visible, although due to the diffusion of stars, the spiral arms are considerably broader. It is worth mentioning that while in Sec.~\ref{sec:dynamical_heating} we presented evidence for dynamical heating in the vertical direction (i.e., increase of $\sigma_{*,z}$ with time), it is the dynamical heating in the perpendicular directions (i.e., radial and angular directions) that is responsible of the broader spiral arms we see in the third age bin, compared to the youngest bin. We also note that the diffusion of stars occurs everywhere across the galactic disk, not only across the spiral arms. However, the existence of a clear spiral pattern provides the opportunity of observing the process of stellar diffusion, and the subsequent ``washing out'' of galactic structure across discrete age bins. A quantification of this diffusion from these SFR maps is challenging, due to the generally low spiral arm-disk contrast in the third age bin map, and due to artifacts in the maps (pointing-to-pointing jumps, or ``holes'' in the maps). A sophisticated approach to quantify the time-scales of this broadening is thus, beyond the scope of this paper.

In order to visually identify this diffusion phenomenon in the time-averaged SFR maps, we require the galaxy to have clear spiral arms, and that the spiral arms had been persisting in time, not formed recently by, for instance, a local instability sheared by differential rotation \citep{Goldreich1965}. Thus, it is not surprising that this effect is more prominent in some galaxies than in other. In the particular case of NGC\,3627, the spiral arms-disk contrast is very low in the third age bin, which could be indicating a more recent origin of its spiral arms, possibly connected to a disruption driven by interactions with other galaxies from the Leo triplet (NGC\,3623, and NGC\,3628). Thus, this approach can provide a powerful tool to study the nature and prevalence of spiral arms across time.

\section{Summary}
\label{sec:summary_stpops}


In this paper, we have presented a detailed analysis of the stellar populations' properties of PHANGS-MUSE galaxies, including investigating the radial structure of age and metallicity, radial assembly history, dynamical heating of young stellar populations, and time-averaged SFRs across the galactic disks of $19$ massive star-forming galaxies. Our main conclusions are the following:

\begin{itemize}
    \item We find generally negative mass-weighted (MassW) and luminosity-weighted (LumW) age profiles. Low-mass galaxies NGC\,1385 and NGC\,1087 are exceptions, showing nearly flat profiles. We find that the slope of the radial gradients correlates reasonably well with global quantities, such that galaxies with higher offset from the main sequence or higher molecular gas content show flatter LumW age profiles. Spiral arms host younger (lower LumW ages) populations than other galactic environments at any given radius.
    
    \item Galaxies have mostly negative metallicity profiles and show some degree of correlation of the slope of the metallicity gradient with total stellar mass and SFR, such that more massive (and star-forming) galaxies exhibit steeper metallicity gradients. The intercept of the metallicity radial profiles also correlates with total mass, such that higher mass galaxies show overall higher metallicites.

    \item We measure a generally well-defined radial assembly history, in which inner regions assembled their stellar mass earlier than outer regions, pointing to an ``inside-out'' growth. This is consistent with previous findings in the literature, where late-type galaxies have been found to show a clear inside-out growth mode.
    
    \item We find flatter metallicity gradients along galactic bars, compared to their minor axis. This is consistent with the expectation from bar-driven secular evolution inducing radial migration within bars. However, we cannot robustly establish that metallicity gradients are flatter along bars due to our small sample size.
    
    
    \item We investigated the progressive increase of velocity dispersion of young stellar populations, and we find a clear gradient in velocity dispersion with age, such that at any given radius, younger stellar populations (< 100 Myr) have in general lower velocity dispersion than intermediate age [100 Myr, 600 Myr] and old (> 3 Gyr) populations. This gradient is can be interpreted as an imprint of the mechanisms dynamically heating young stellar populations, such as interactions with molecular clouds or nonaxisymmetric galactic substructure. We find this process to occur regardless of the presence of spiral arms. The age gradient in velocity dispersion is not present in bars, and we interpret this as dynamical heating being more efficient in the bar region.
    \item We have used the spatially resolved SFHs to reconstruct time-averaged SFR maps and study how different regions of galaxies assembled their mass. We find a wide variety of features, such as centers continuously forming stars, or quenched early in the galaxy lifetime. Younger populations are coupled to galactic structural components, but older stars follow a much more homogeneous and  centrally concentrated distribution. Finally, in some galaxies we see a clear increase in the width of spiral arms with increasing age of the stellar populations probed, and interpret this as a sign of the diffusion of young stellar populations in the radial and angular directions.
\end{itemize}

Overall, the derived SFHs tell us a story in which galaxies assembled their stellar mass in an inside-out mode, leading to negative age and metallicity profiles, and where more massive galaxies assembled their stellar mass earlier in cosmic history than less massive galaxies, consistent with the expectation from previous studies. Furthermore, thanks to the high spatial resolution achieved by PHANGS-MUSE, for the first time, we have been able to investigate the properties of stellar populations across multiple galactic environments separately, covering a significant part of the stellar disk of nearby star-forming galaxies, being also able the measure distinct kinematic properties for the different stellar populations present in our sample galaxies.

\begin{acknowledgements}

We thank the anonymous referee for their insightful comments which helped to improve the paper. This work was carried out as part of the PHANGS collaboration. Based on observations collected at the European Southern Observatory under ESO programmes 094.C-0623 (PI: Kreckel), 095.C-0473,  098.C-0484 (PI: Blanc), 1100.B-0651 (PHANGS-MUSE; PI: Schinnerer), as well as 094.B-0321 (MAGNUM; PI: Marconi), 099.B-0242, 0100.B-0116, 098.B-0551 (MAD; PI: Carollo) and 097.B-0640 (TIMER; PI: Gadotti). This paper also makes use of the following ALMA data: ADS/JAO.ALMA\#2013.1.01161.S, ADS/JAO.ALMA\#2015.1.00925.S, ADS/JAO.ALMA\#2015.1.00956.S and ADS/JAO.ALMA\#2017.1.00886.L. ALMA is a partnership of ESO (representing its member states), NSF (USA) and NINS (Japan), together with NRC (Canada), MOST and ASIAA (Taiwan), and  KASI (Republic of Korea), in cooperation with the Republic of Chile. The Joint ALMA Observatory is operated by ESO, AUI/NRAO and NAOJ. The National Radio Astronomy Observatory is a facility of the National Science Foundation operated under cooperative agreement by Associated Universities, Inc.
This research made use of \texttt{Astropy}, a community-developed core Python package for Astronomy \citep{Astropy2013, Astropy2018}.
IP acknowledges funding by the European Research Council through ERC-AdG SPECMAP-CGM, GA 101020943.
TGW acknowledges funding from the European Research Council (ERC) under the European Union’s Horizon 2020 research and innovation programme (grant agreement No. 694343).
JMDK gratefully acknowledges funding from the ERC under the European Union's Horizon 2020 research and innovation programme via the ERC Starting Grant MUSTANG (grant agreement number 714907).
COOL Research DAO is a Decentralized Autonomous Organization supporting research in astrophysics aimed at uncovering our cosmic origins.
RSK and SCOG acknowledge support from  DFG via  the collaborative research center ``The Milky Way System'' (SFB 881; project ID 138713538; sub-projects B1, B2 and B8), from the Heidelberg cluster of excellence EXC 2181``STRUCTURES'' (project ID 390900948), funded by the German Excellence Strategy, from ERC via the Synergy Grant ``ECOGAL'' (grant 855130), and from the German Ministry for Economic Affairs and Climate Action for funding in  project ``MAINN'' (funding ID 50OO2206). 
KK gratefully acknowledges funding from the Deutsche Forschungsgemeinschaft (DFG, German Research Foundation) in the form of an Emmy Noether Research Group (grant number KR4598/2-1, PI Kreckel). PSB acknowledge support through the
 project  grants  PID2019-107427GB-C31 from the Spanish Ministry of Science, Innovation and Universities (MCIU).
FB acknowledges funding from the European Research Council (ERC) under the European Union’s Horizon 2020 research and innovation programme (grant agreement No.726384/Empire).
MQ acknowledges support from the Spanish grant PID2019-106027GA-C44, funded by MCIN/AEI/10.13039/501100011033.
JN acknowledges funding from the European Research Council (ERC) under the European Union’s Horizon 2020 research and innovation programme (grant agreement No. 694343) and the Science and Technology Facilities for support through the Consolidated Grant Cosmology and Astrophysics at Portsmouth, ST/S000550/1.

\end{acknowledgements}

\bibliographystyle{aa}
\bibliography{biblio} 

\begin{appendix}
\section{Slope of radial gradients of stellar population properties versus global galactic parameters}

\subsubsection{Stellar age}
\label{sec:disc_stellar_age}
\begin{figure*}[h!]
\centering
 	\includegraphics[width=0.7\textwidth]{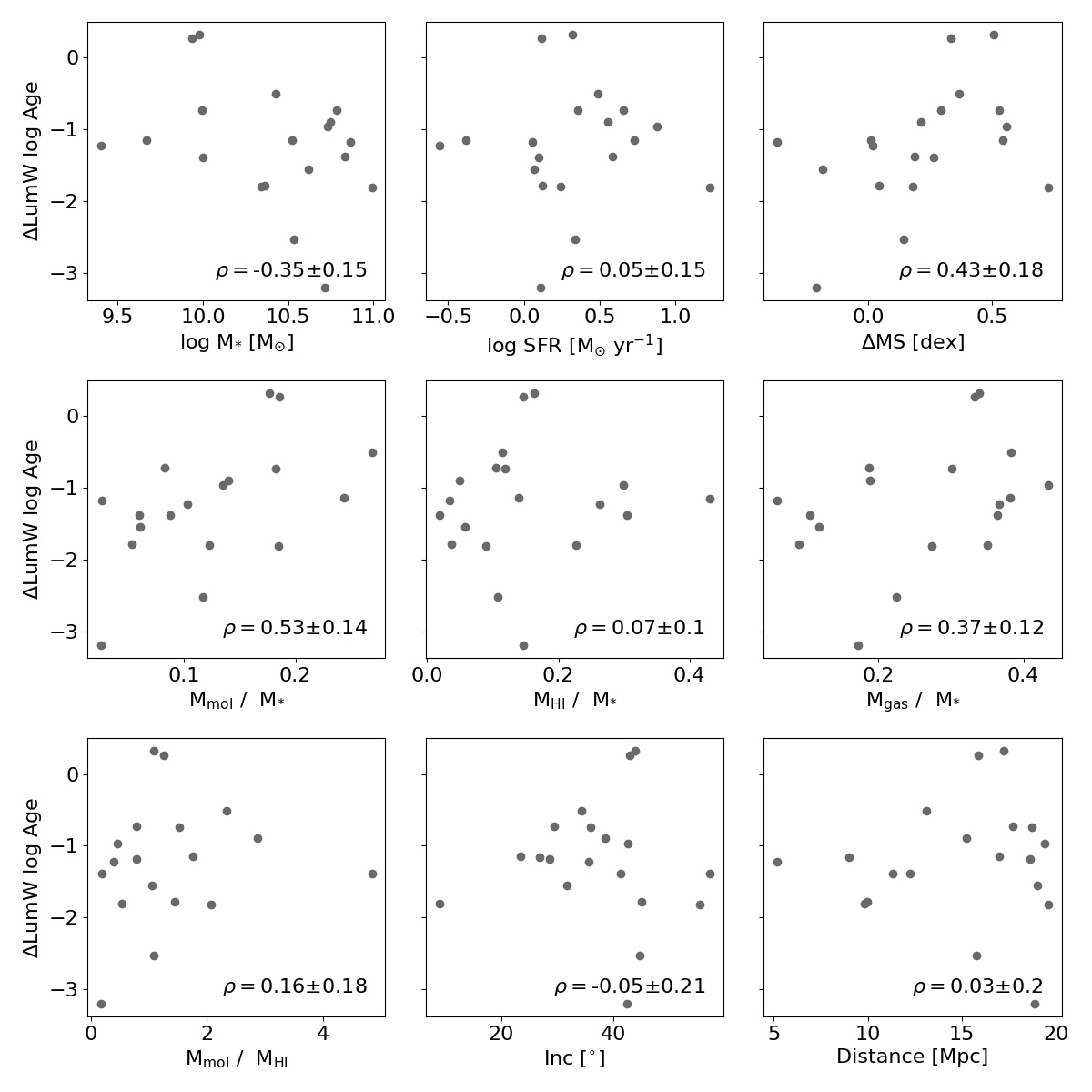}
     \caption[Slope of the LumW-age gradients as a function of global galactic parameters]{Slope of the LumW-age gradient, as a function of (from left to right, and from top to bottom) total stellar mass (log M$_{*}$), total SFR (log SFR), offset from the global main sequence of galaxies ($\Delta$MS), total molecular gas mass-to-stellar mass ratio (M$_\mathrm{mol}$ / M$_{*}$), total atomic gas mass-to-stellar mass ratio (M$_{\mathrm{HI}}$), total gas (molecular + atomic)-to-stellar mass ratio M$_{\mathrm{gas}}$ / M$_{*}$, and molecular-to-atomic gas mass ratio (M$_{\mathrm{mol}}$ / M$_{\mathrm{HI}}$), galaxy inclination and distance. The Pearson correlation coefficient for each parameter is indicated in the bottom-right corner of each panel.}
     \label{fig:trend_LW_age}   
\end{figure*}

\begin{figure*}[h!]
\centering
 	\includegraphics[width=0.7\textwidth]{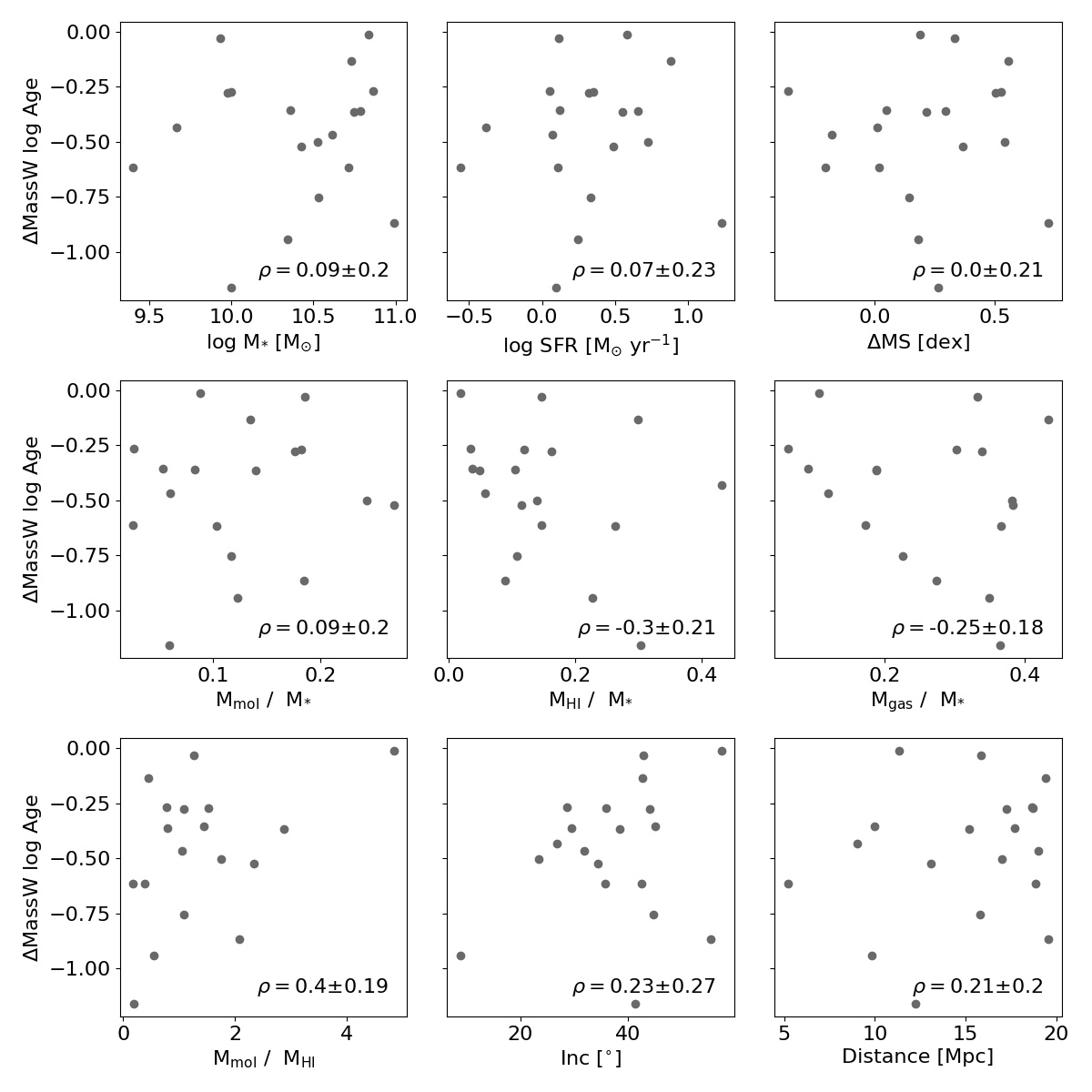}
     \caption[Slope of the MassW-age gradients as a function of global galactic parameters]{Slope of the MassW-age gradient, as a function of (from left to right, and from top to bottom) total stellar mass (log M$_{*}$), total SFR (log SFR), offset from the global main sequence of galaxies ($\Delta$MS), total molecular gas mass-to-stellar mass ratio (M$_\mathrm{mol}$ / M$_{*}$), total atomic gas mass-to-stellar mass ratio (M$_{\mathrm{HI}}$), total gas (molecular + atomic)-to-stellar mass ratio M$_{\mathrm{gas}}$ / M$_{*}$, and molecular-to-atomic gas mass ratio (M$_{\mathrm{mol}}$ / M$_{\mathrm{HI}}$), galaxy inclination and distance. The Pearson correlation coefficient for each parameter is indicated in the bottom-right corner of each panel.}
     \label{fig:trend_MW_age}   
\end{figure*}

Here we investigate if the gradients measured for the LumW log Age radial profiles correlate with global galaxy properties. Figure~\ref{fig:trend_LW_age} shows the slope of the LumW-age gradients, as a function of several global galaxy parameters: total stellar mass (log M$_{*}$), total SFR (log SFR), offset from the global main sequence of galaxies as defined in \citet{Leroy2019} ($\Delta$MS), total molecular gas mass-to-stellar mass ratio (M$_\mathrm{mol}$ / M$_{*}$), total atomic gas mass-to-stellar mass ratio (M$_{\mathrm{HI}}$), total gas (molecular + atomic)-to-stellar mass ratio M$_{\mathrm{gas}}$ / M$_{*}$, and molecular-to-atomic gas mass ratio (M$_{\mathrm{mol}}$ / M$_{\mathrm{HI}}$), with their corresponding Pearson correlation coefficient ($\rho$).  We also include inclination and distance, to ensure that the different gradients are not related to these nonintrinsic properties. The uncertainty in the Pearson coefficient has been calculated by bootstrapping; for each correlation, we repeatedly chose 100 subsamples of datapoints (allowing for repetitions), each subsample with the same size than the original sample, and calculated their $\rho$ coefficient. We adopt the standard deviation of the set of $100$ coefficients as the uncertainty for $\rho$.

The highest $\rho$ correlations are with M$_\mathrm{mol}$ / M$_{*}$ ($\rho = 0.53$) and $\Delta$MS ($\rho = 0.43$) implying that galaxies with higher molecular gas content and a global enhancement of SFR also have a less negative (i.e., flatter) LumW-age radial gradient. This is consistent with the findings from \citet{Ellison2018} and \citet{Pessa2021}, where the authors find that a global enhancement of SFR is reflected primarily on the SFR of the inner galactic regions. In this case, since the LumW-age is very sensitive to recent star-formation, a more negative gradient implies an older and less star-forming central region, with respect to the outer radii.

Figure~\ref{fig:trend_MW_age} provides plots of the MassW-age gradient against the same quantities as Fig.~\ref{fig:trend_LW_age}. The slopes of the MassW-age gradients do not correlate strongly neither with $\Delta$MS, M$_\mathrm{mol}$ / M$_{*}$, nor with any of the parameters explored. However, we see a trend in which galaxies with a high total gas content (M$_{\mathrm{gas}}$ / M$_{*}$) span a wide range of gradients, while galaxies with a low gas content exhibit exclusively flatter gradients. This trend could indicate that galaxies with a low gas content have consumed their gas supply for fueling star formation, leading to younger populations in their centers, hence, flatter MassW-age profiles. Even though the level of correlation of  M$_\mathrm{mol}$ / M$_{*}$ with LumW-age is not particularly high ($\rho = 0.37$), the trend seems to be opposite to the one with MassW-age. This is not in contradiction with the scenario suggested for the origin of the trend between the MassW-age gradient and M$_\mathrm{mol}$ / M$_{*}$, since a low total gas content could certainly imply a lower total present-day SFR (and hence, less SFR in the central region), but it does not imply a lack of SFR on longer time-scales, to which MassW-age is sensitive (as opposed to LumW-age, that is strongly biased toward recent SFR).

In summary, we find generally negative age radial gradients, with LumW profiles being steeper than the MassW ones. Galaxies with higher molecular gas content and that are experiencing a global enhancement of SFR exhibit flatter LumW age profiles.
\subsubsection{Stellar metallicity}
\label{sec:disc_stellar_z}

\begin{figure*}[h!]
\centering
 	\includegraphics[width=0.7\textwidth]{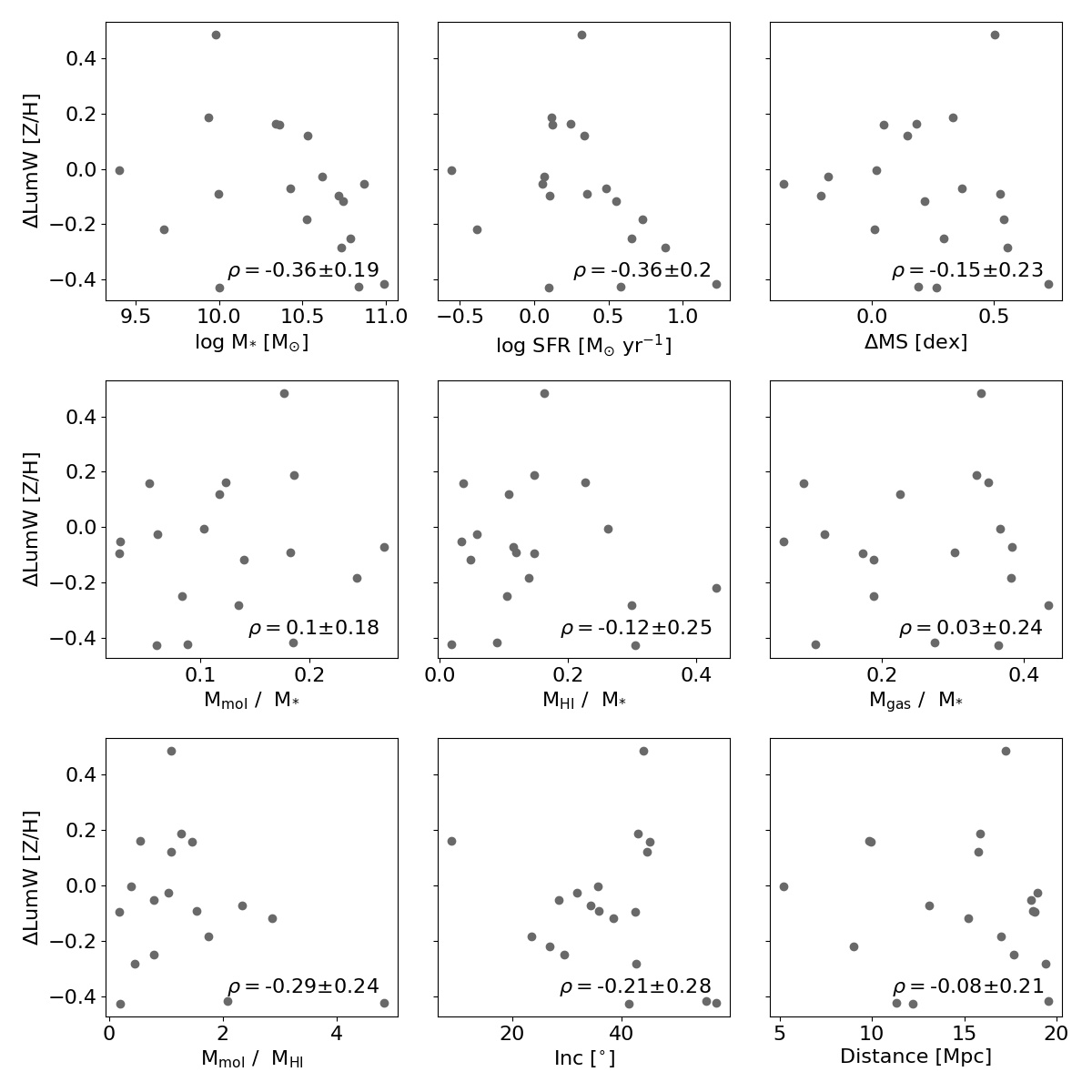}
     \caption[Slope of the LumW-metallicity gradients as a function of global galactic parameters]{Slope of the LumW-[Z/H] gradient, as a function of (from left to right, and from top to bottom) total stellar mass (log M$_{*}$), total SFR (log SFR), offset from the global main sequence of galaxies ($\Delta$MS), total molecular gas mass-to-stellar mass ratio (M$_\mathrm{mol}$ / M$_{*}$), total atomic gas mass-to-stellar mass ratio (M$_{\mathrm{HI}}$), total gas (molecular + atomic)-to-stellar mass ratio M$_{\mathrm{gas}}$ / M$_{*}$, and molecular-to-atomic gas mass ratio (M$_{\mathrm{mol}}$ / M$_{\mathrm{HI}}$), galaxy inclination and distance. The Pearson correlation coefficient for each parameter is indicated in the bottom-right corner of each panel.} 
     \label{fig:trend_LW_z}   
\end{figure*}

\begin{figure*}[h!]
\centering
 	\includegraphics[width=0.7\textwidth]{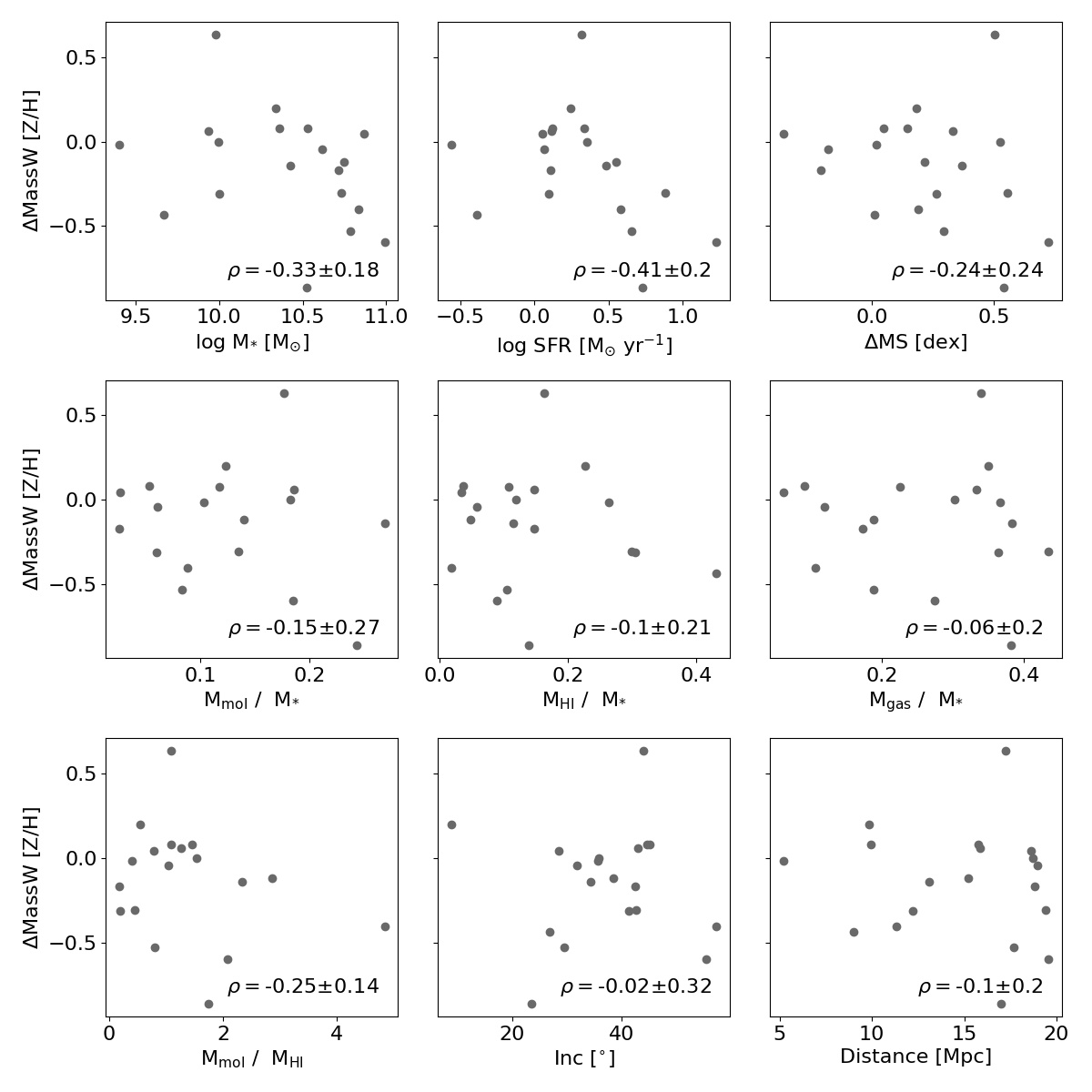}
     \caption[Slope of the mass MassW-metallicity gradients as a function of global galactic parameters]{Slope of the MassW-[Z/H] gradient, as a function of (from left to right, and from top to bottom) total stellar mass (log M$_{*}$), total SFR (log SFR), offset from the global main sequence of galaxies ($\Delta$MS), total molecular gas mass-to-stellar mass ratio (M$_\mathrm{mol}$ / M$_{*}$), total atomic gas mass-to-stellar mass ratio (M$_{\mathrm{HI}}$), total gas (molecular + atomic)-to-stellar mass ratio M$_{\mathrm{gas}}$ / M$_{*}$, and molecular-to-atomic gas mass ratio (M$_{\mathrm{mol}}$ / M$_{\mathrm{HI}}$), galaxy inclination and distance. The Pearson correlation coefficient for each parameter is indicated in the bottom-right corner of each panel.}
     \label{fig:trend_MW_z}   
\end{figure*}

Figures~\ref{fig:trend_LW_z} and~\ref{fig:trend_MW_z} show the LumW- and MassW- stellar metallicity gradients, as a function of the same global properties that we explored for the age gradients. We do not see any obvious trends between metallicity gradients and these global properties. \citet{Goddard2017} report steeper negative metallicity profiles for more massive galaxies. In our sample, we find that the highest $\rho$ values are indeed obtained with total stellar mass and total SFR, however, possibly due to our low-number statistics, we cannot robustly confirm such a trend. In general, it seems that low-mass galaxies (log (M$_{*}$/M$_{\odot}$) $\lesssim 10$) span a more diverse range of gradients, while higher mass galaxies exhibit stellar metallicity gradients that correlate better with total mass. In summary, we find negative stellar metallicity radial profiles, with more massive galaxies exhibiting overall higher metallicities and steeper gradients. 

\section{Uncertainties in age, metallicity, and stellar mass surface density}
\label{app:uncertainties}

In this section we show the distributions of the random uncertainties for the physical quantities derived through our full spectrum fitting approach, as described in Sec.~\ref{sec:stellarpops_maps}, for the full sample. Additionally, we show that performing 20 Monte Carlo iterations to estimate the uncertainties in these quantities for each Voronoi bin is a reasonable compromise in terms of characterizing the dispersion of the distribution of values produced by the Monte Carlo realizations.

\begin{figure}
\centering
 	\includegraphics[width=\columnwidth]{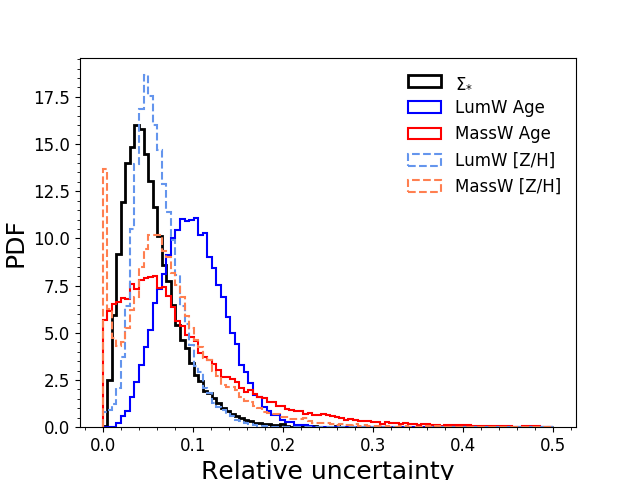}
     \caption{Probability density function of the relative uncertainties estimated for stellar mass surface density ($\Sigma_{*}$), LumW Age, MassW Age, LumW [Z/H], and MassW [Z/H], for our full sample, that is, all available pixels.}
     \label{fig:all_dist_errors}   
\end{figure}

Figure~\ref{fig:all_dist_errors} shows the distribution of the relative errors of $\Sigma_{*}$, LumW Age, MassW Age, LumW [Z/H] and MassW [Z/H]. The relative uncertainty in stellar mass surface density is typically below $10\%$, while the uncertainties in ages and metallicities are typically below $20\%$. The mass-weighted quantities exhibit a long wing toward larger values, although this is not surprising, as mass-weighted quantities are dominated by fainter stars that are often outshone by younger and brighter stars.

To test how many Monte Carlo realizations are required to appropriately estimate the width of the produced distribution, we have chosen five representative Voronoi bins (one for each galactic environment), and we have computed the dispersion of the results of the Monte Carlo iteration, for different numbers of iterations, from 5 to 100.

\begin{figure}
\centering
 	\includegraphics[width=\columnwidth]{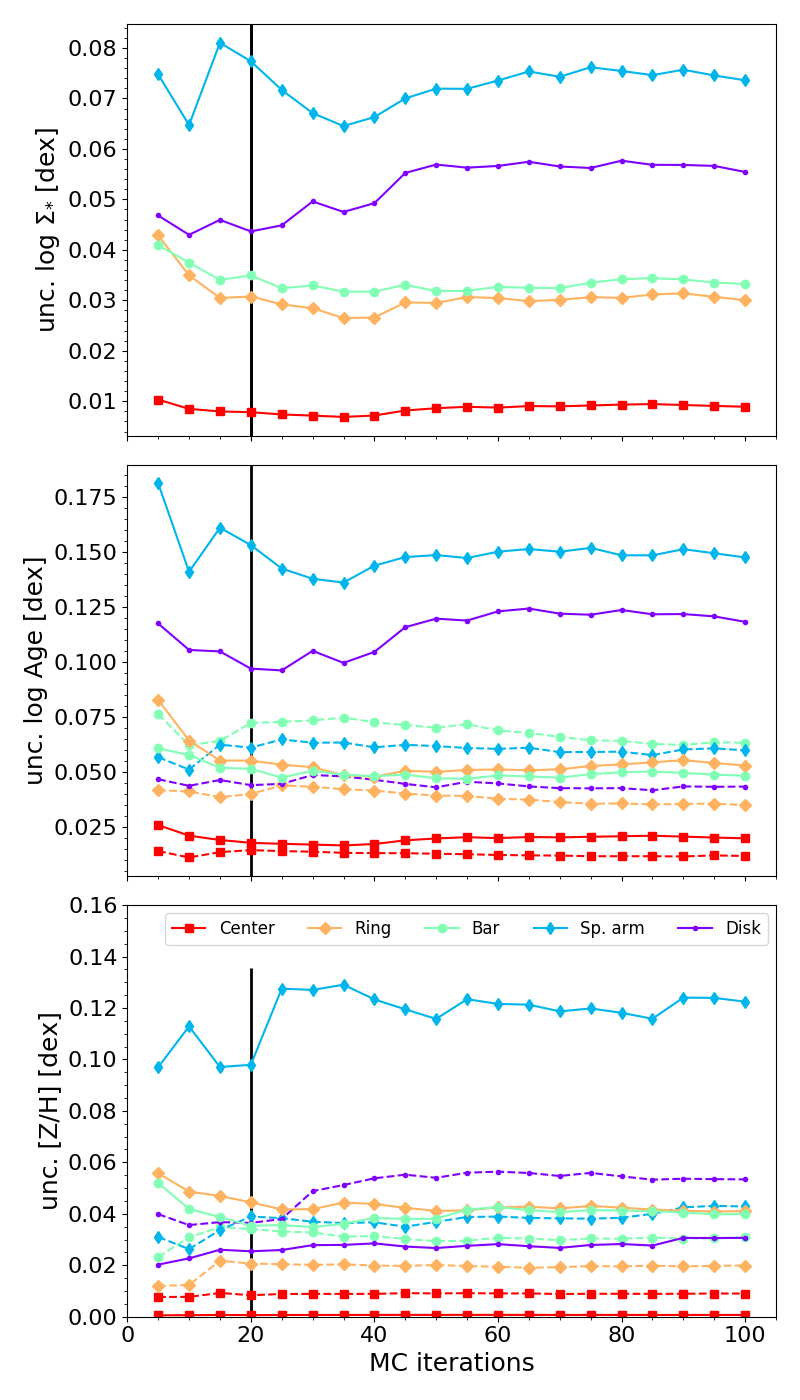}
     \caption{Dispersion of the resulting posterior distribution of stellar mass velocity dispersions ($\Sigma_{*}$), ages and metallicity from the Monte Carlo realizations, as a function of the number of Monte Carlo iterations, for $5$ representative Voronoi bins. The color code indicates the galactic environment of each Voronoi bin. Solid lines correspond to mass-weighted quantities, and dashed lines to their light-weighted counterparts. The vertical black line indicates the adopted number (20) of Monte Carlo iterations to estimate the uncertainties in the analyses presented in this paper.}
     \label{fig:dispersion_per_nMC}   
\end{figure}

Figure~\ref{fig:dispersion_per_nMC} shows the dispersion of the distribution as a function of the number of Monte Carlo iterations, for stellar mass surface density, age, and metallicity, for these five Voronoi bins. In most cases, the width of the dispersion does not change significantly beyond 20 iterations, although in some cases there is some level of variability until $\sim 50$ iterations. Nevertheless, 20 realizations is a reasonable number to estimate the dispersion of the results of the Monte Carlo iterations, and hence, the magnitude of the random uncertainties of our measurements. The figure also shows clearly that the uncertainty in the MassW quantities is significantly larger than that for LumW quantities when young stars are present (e.g., in spiral arms)

\section{Downsizing in galaxy evolution}
\label{sec:downsizing}
Another interesting topic that we can probe within our sample is the downsizing paradigm of galaxy evolution \citep[see, e.g., ][]{Cowie1996, Heavens2004, Thomas2005, PerezGonalez2008}. This phenomenon describes how more massive galaxies assembled their stellar mass earlier in cosmic history than less massive galaxies.

\begin{figure*}
\centering
 	\includegraphics[width=0.8\textwidth]{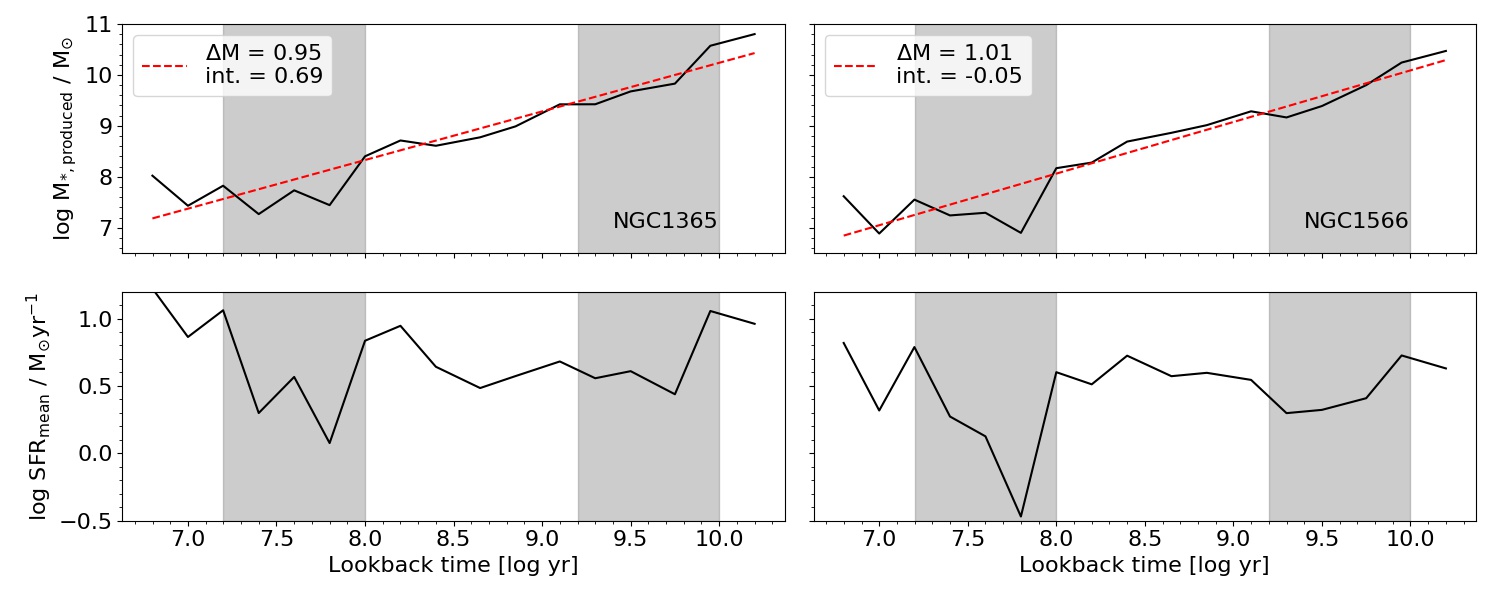}
     \caption[Star formation history and star formation rate history derived for NGC\,1365 and NGC\,1566]{Star formation history (top) and star formation rate history (bottom) derived for NGC\,1365 (left) and NGC\,1566 (right). The best-fitting linear model for the SFH curve of each galaxy is shown as a red dashed line, and its slope and intercept are shown in the top-left corner of the top panels. The shaded areas mark sudden decreases in the SFR at ages $\log$ (age/yr) $\sim7.7$ and $\log$ (age/yr) $\sim9.5$, found to exist in all galaxies in some degree, deemed as an artifact of the computed SFHs.}
     \label{fig:down_NGC1365_NGC1566}   
\end{figure*}

Figure~\ref{fig:down_NGC1365_NGC1566} shows, in the top panels, the total stellar mass formed in a given age bin, and in the bottom panels, the average SFR in a given age bin, defined as the total stellar mass formed, divided by the time-separation between each bin, and the previous one, for two galaxies in our sample (NGC\,1365 and NGC\,1566). In the case of the first bin, the time-separation is calculated with respect to the present time (lookback time = 0). 

The SFHs (and time-averaged SFRs) of most of the galaxies in our sample show a sudden decrease in the formation of stars at ages $\log$ (age/yr) $\sim7.7$ and $\log$ (age/yr) $\sim9.5$ (indicated as the shaded area in Fig.~\ref{fig:down_NGC1365_NGC1566}). We acknowledge that this is very likely an artifact of the derivation of the SFH. This decrease occurs everywhere across the galaxy disk, and we have been not able to identify a clear cause for it. Similar systematic problems at these ages have been identified in the literature. \citet{Peterken2020} report a correlation between the weights assigned for two similar age bins, finding evidence that it is related to the ``UV upturn'' in old stellar populations \citep{Yi2008}. Although we do not report here an identical problem (we do not see a sudden increment in the SFH, as they do), SFH derivations are subject to systematics that depend on the spectral library used \citep[e.g, ][]{Walcher2011, Martins2021}, as well as the code used to derive it \citep[e.g., ][]{SanRoman2019}. Finding the exact cause of this artifact is beyond the scope of this paper, but we exclude it from the interpretation of our results.

The SFHs derived are well represented by almost perfectly linear power-laws (see top panels of Fig.~\ref{fig:down_NGC1365_NGC1566}). This is partially by construction, as we designed our age grid to be log-spaced, and the time-averaged SFR is nearly flat across cosmic time (neglecting the artifacts described earlier). However, deviations from this linearity reflect real features of the mass assembly process of these galaxies. Galaxies that assembled most of their stellar mass earlier in cosmic history show steeper power-laws (i.e., higher $\Delta$M). On the other hand, galaxies that have assembled a substantial amount of their current stellar mass later in cosmic history are characterized by flatter SFHs (i.e., lower $\Delta$M).

\begin{figure*}
\centering
 	\includegraphics[width=0.8\textwidth, trim=0 0 0 0, clip]{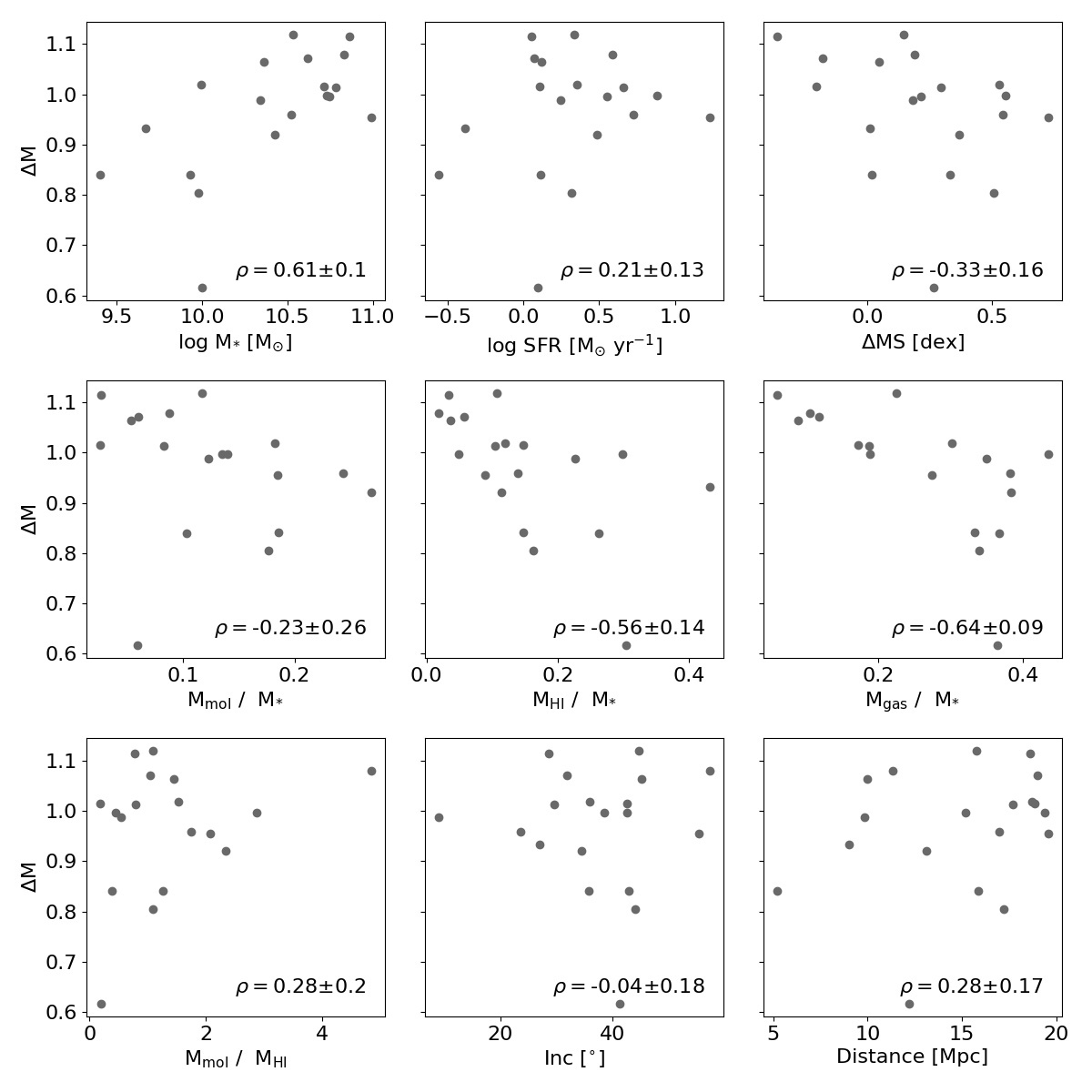}
     \caption[Slope of the SFH of PHANGS galaxies as a function of global galactic parameters]{Slope of the SFH of PHANGS galaxies, as a function of (from left to right, and from top to bottom) total stellar mass (log M$_{*}$), total SFR (log SFR), offset from the global main sequence of galaxies ($\Delta$MS), total molecular gas mass-to-stellar mass ratio (M$_\mathrm{mol}$ / M$_{*}$), total atomic gas mass-to-stellar mass ratio (M$_{\mathrm{HI}}$), total gas (molecular + atomic)-to-stellar mass ratio M$_{\mathrm{gas}}$ / M$_{*}$, and molecular-to-atomic gas mass ratio (M$_{\mathrm{mol}}$ / M$_{\mathrm{HI}}$), galaxy inclination and distance. The Pearson correlation coefficient for each parameter is indicated in the bottom-right corner of each panel.}
     \label{fig:trend_DownSizing}   
\end{figure*}

Figure~\ref{fig:trend_DownSizing} shows the slope of the SFH ($\Delta$M), measured using an OLS fitting routine, as a function of the same global parameters explored in the previous sections. We find clear correlations between $\Delta$M and several of the global properties, including total stellar mass ($\rho = 0.61$), atomic gas fraction ($\rho = -0.56$), and total gas fraction ($\rho = -0.64$).

Despite the artifacts present in our SFHs, we believe that trends in the assembly history of galaxies are robust, where more massive galaxies, with lower gas content assembled their stellar mass earlier in cosmic history, than galaxies with lower stellar masses, and higher gas content.

This reflects the downsizing effect in galaxy evolution, and is consistent with findings from previous works \citep[e.g., ][]{IbarraMedel2016, Peterken2020}. An alternative explanation for this trend is that the ratio of present-day to past SFR is higher for low mass galaxies than for high mass galaxies
Thus, we conclude that the derived SFHs tell a consistent story of the stellar mass assembly of the galaxies in our sample, with strong imprints of an inside-out formation, and with higher mass galaxies assembling their stellar mass on average earlier in cosmic history than lower mass galaxies.

\subsection{Downsizing in individual galactic environments}

Recent works have studied the assembly history of galaxies across their different galactic environments, by performing a photometric decomposition to isolate bulges and disks, and investigating their stellar populations' properties separately \citep[e.g.,][]{Costantin2021,Costantin2022, Johnston2022}. They find that bulges are generally older than their surrounding disks, pointing to an inside-out growth of galaxies. \citet{Costantin2022} also finds that the downsizing paradigm is present in individual galactic environments, such that older disks and older bulges are also more massive than their younger counterparts, while \citep{Johnston2022} find that bulges in more massive galaxies ($\gtrsim 10^{10}$ M$_{\odot}$) form earlier than those in lower mass galaxies. 

\begin{figure*}
\centering
 	\includegraphics[width=\textwidth, trim=0 0 0 0, clip]{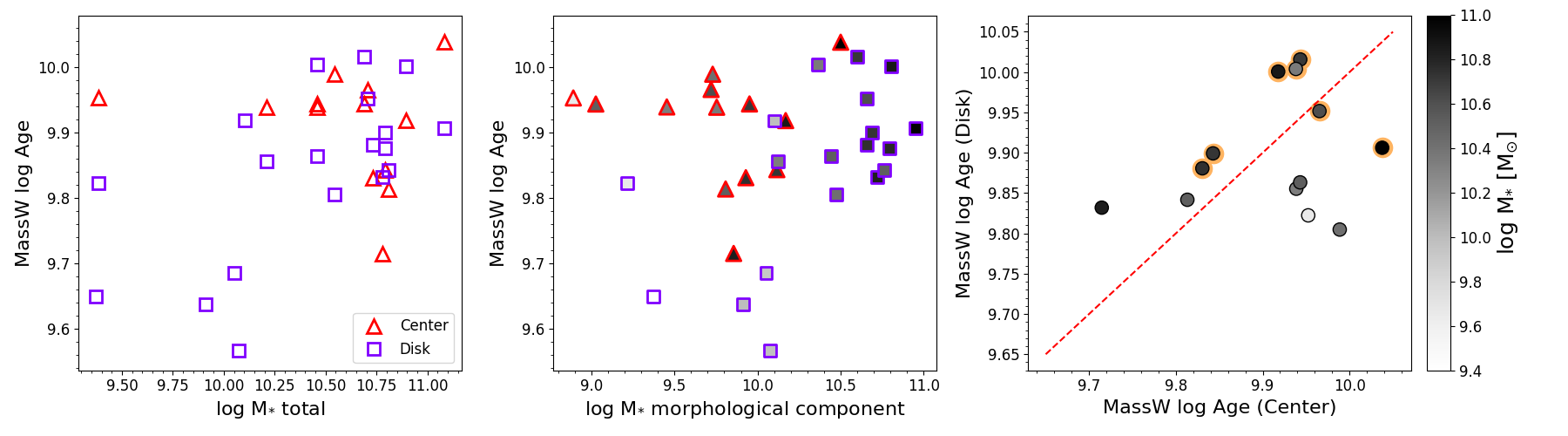}
     \caption{Downsizing paradigm in separated galactic environments. \textit{Left:} correlation between the median MassW log Age and the total galaxy stellar mass for the (combined) disk and center environments, for each galaxy in our sample. The total galaxy stellar mass corresponds to the sum of the masses of the (combined) disk and center environments. \textit{Middle:} median MassW log Age of the (combined) disk and center environments as a function of total stellar mass in each galactic environment. \textit{Right:} Median MassW log Age of the (combined) disk environment as a function of the median MassW log Age age of the (combined) center environment for our sample galaxies. Galaxies that present a nuclear ring are indicated with an additional orange marker.}
     \label{fig:DownSizing_environments}   
\end{figure*}

Motivated by these results, we explore whether we can recover the downsizing paradigm in individual environments. To do so, we have simplified our environmental label, grouping spiral arms, outer rings, and stellar bars together with the ``Disk'' component, and inner rings together with the ``Center'' component. Hereafter, we refer to these combination of morphological structures as combined disk and combined center, respectively. The left and middle panel of Fig.~\ref{fig:DownSizing_environments} show the median MassW age of these combined morphological components as a function of total stellar mass, and stellar mass of each combined morphological component, respectively, for all galaxies in our sample. The figure shows that while the downsizing paradigm is clearly preserved in disks, this is not the case for centers. Furthermore, we do not see a correlation between the age of bulges with the total stellar mass either. 

Finding such different behavior of centers, compared to the properties of bulges reported in \citet{Costantin2022} and \citet{Johnston2022} is not surprising. There are a number of reasons that explain this apparent disagreement between this work and these seemingly similar studies. On one hand, there are inherent differences in the sample selection. For instance, \citet{Johnston2022} uses a sample of mostly unbarred S0 galaxies, while \citet{Costantin2022} uses galaxies at redshift $0.14 < z \leq 1$, and it is known that the abundance of galaxies with stellar bars steadily decreases toward redshift $1$ \citep{Simmons2014, Melvin2014}. This implies that their sample is considerably less affected by the presence of bars than our sample, composed of mostly barred galaxies (14 out of the 19 PHANGS-MUSE galaxies exhibit bars). This is a key difference, as bars can efficiently transport gas from the outer galactic regions to the innermost regions of galaxies \citep{Shiying2022}. Hence, galactic centers in the \citep{Johnston2022} and \citet{Costantin2022} samples are less likely to present bar-driven secularly built nuclear disks/rings. Moreover, if bars are present, a pure bulge-disk decomposition is most likely too simplified to capture and isolate the properties of the bulge \citep[e.g.,][]{Gadotti2009}.

Beyond differences in the sample and in the methodology to separate morphological components, the spatial resolution of the analyses also plays a relevant role. For example, at the spatial resolution of MaNGA \citep[in the case of][]{Johnston2022}, even if a small secularly built central substructure exists, it would probably be unresolved and mixed with the stellar disk in the decomposition. Finally, the definition of bulges is by itself ambiguous. In this work, we define centers as a central excess of light, independently of their exact surface brightness profile or dynamical properties (see Sec.~\ref{sec:enviromental_mask}). This definition differs from that used in other works, which implies a disparity in the comparison.

Finally, the right panel of Fig.~\ref{fig:DownSizing_environments} shows the median MassW age of the (combined) disk as a function of the median MassW age of the (combined) center component. We do not find any clear systematic difference between them, contrasting with previous works that find systematically older bulges. However, this is not unexpected considering the fundamental differences between this and the recent studies discussed, described above. In fact, 5 out of the 7 galaxies where the disk is older than the center show clear bars and inner rings, which further highlights how relevant the sample differences are. This result is consistent with the radial profiles shown in Sec.~\ref{sec:radil_profiles_ag_z}, where some galaxies exhibit clear evidence of young stellar populations in their innermost regions.

To summarize, a direct comparison of our results with recent studies that perform photometric decompositions to investigate the assembly history of bulges and disks is difficult due to the underlying sample and resolution differences. We recover the downsizing paradigm in disks, but not in centers, and we do not find any evidence of centers being systematically older than disks. These apparent discrepancies with previous works can be explained by the fundamental dissimilitude between these  different studies.

\section{Decreasing the angular resolution of our dataset}
\label{app:degrade}
In this section, we show how our results depend on the angular resolution of our data. To do this, we have convolved the MUSE mosaics to a fixed angular resolution of $15$ arcsecs, and derived the stellar populations' property maps following the same methodology described in Sec.~\ref{sec:stellarpops_maps}. Then, we recomputed radial profiles and overall distributions of the stellar population properties for the convolved dataset, across different galactic environments, using the same environmental masks as for our fiducial analysis (see Sec.~\ref{sec:enviromental_mask}). An angular resolution of $15$ arcsecs corresponds to physical scales from $0.4$ kpc to $1.4$ kpc at the distances of our sample galaxies, with a mean value and dispersion of $1.1 \pm 0.3$ kpc.

\begin{figure*}[h!]
\centering
 	\includegraphics[width=0.9\textwidth]{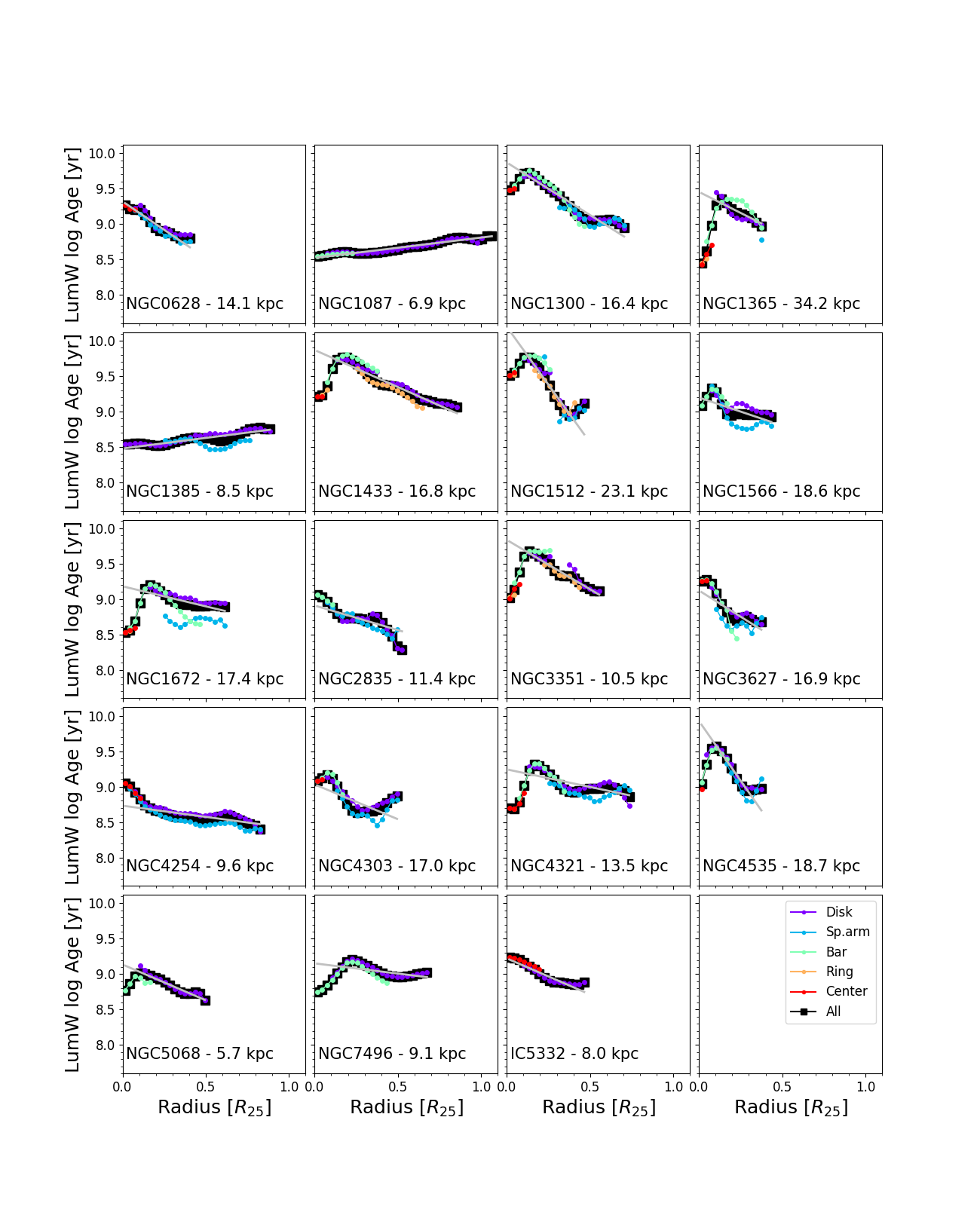}
     \caption{Luminosity-weighted stellar age radial profiles for the galaxies in our sample, measured after convolving our MUSE mosaics to a fixed angular resolution of $15$ arcsecs, and computing the stellar population properties for these convolved cubes. Different colors indicate the radial profile measured across different environments, as indicated in the legend of the bottom-right panel. The black line shows the radial profile measured for the entire FoV (i.e., all environments together). The galactocentric distance is measured in units of $R_{25}$, in order to measure radial distance homogeneously across our sample. The value of $R_{25}$ (kpc) of each galaxy is indicated in each corresponding panel. The solid gray line shows the best-fit gradient for each galaxy. The bottom-right panel displays the entire FoV trends for all galaxies, color-coded by total stellar mass.}
     \label{fig:radial_LW_age_15asec}   
\end{figure*}

\begin{figure}[h!]
\centering
 	\includegraphics[width=\columnwidth]{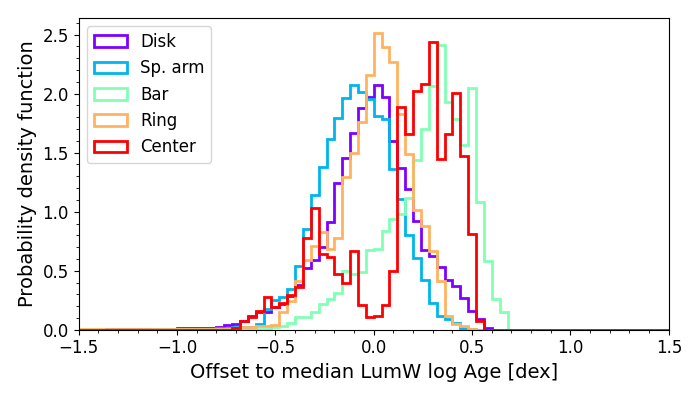}
     \caption{Distribution of the offset of the LumW log Age of the pixels within each environments, with respect to the median galaxy log LumW Age of each galaxy, in bins of $0.04$ dex, obtained for the MUSE mosaics convolved to a fixed angular resolution of $15$ arcsecs. Different colors indicate the distribution of different environments, as indicated in the legend.}
     \label{fig:dist_age_envs_15asec}   
\end{figure}

Figures~\ref{fig:radial_LW_age_15asec} and ~\ref{fig:dist_age_envs_15asec} are identical to Figs.~\ref{fig:radial_LW_age} and ~\ref{fig:dist_age_envs}, but using now the convolved dataset. Fig.~\ref{fig:radial_LW_age_15asec} shows smoother radial profiles. Differences among environments are in general, more subtle than at higher resolution, with rings becoming essentially impossible to distinguish from the center anymore. On the other hand, spiral arms are overall still distinguishable from the stellar disk. This is not unexpected, as \citet{Querejeta2021} found that spiral arms of galaxies from the PHANGS sample exhibit typical width of $\sim 1.5$ kpc. Thus, we can marginally resolve them at a resolution of $\sim 1$ kpc. Other works at $\sim 1$ kpc resolution have reported differences between arms and their surrounding disk. For instance, \citet{SanchezMenguiano2017} find subtle differences in the oxygen abundances of spiral arms with respect to the interarm region using a sample of galaxies from the CALIFA dataset. 

However, beyond environmental differences, the ability to distinguish particularly old or young regions with respect to their surroundings is drastically reduced. This is also illustrated by Fig.~\ref{fig:dist_age_envs_15asec}. In contrast with the fiducial version of the figure, which shows extended tails that reach very young ($<-1.0$ dex with respect to the median galaxy age) and old ($\sim 1.0$ dex with respect to the median galaxy age) ages, the age distributions in the convolved dataset are confined to values much closer to the median galaxy age, mostly within $-0.5$ dex and $+0.5$ dex from this value, implying that although environments might be still distinguishable (as in the case of spiral arms), contamination due to the lower resolution reduces drastically the dynamical range of ages we can measure, in addition of making environments harder to differentiate (or totally washing out their stellar population features, as is the case for inner rings)

\section{Improving the fitting of young stellar populations}
\label{sec:improving_young}
In \citet{Emsellem2021} is reported the presence of low metallicity values ($\mathrm{LumW~[Z/H]} < -1.3$) in a few regions encompassing very young stellar clusters ($\mathrm{LumW~age} < 400$~Myr). Such low metallicity values would be inconsistent with an internal and progressive chemical enrichment of the interstellar medium \citep[e.g.,][]{Ho2017}. This suggests that the fitting process converges toward a misleading local minimum, the bluest available stellar template, constrained by the youngest age bin (30~Myr) of the implemented template library. This issue has already been reported in several studies \citep[e.g.,][]{Carrillo2020, Bittner2020}, who similarly reported unexpected low metallicities in young regions.

In this section, we perform a broader exploration of the parameter space, aiming to define a better approach to fit the spectra of these problematic regions and reduce the artifacts in their derived stellar population properties. This exploration includes the usage of different libraries of templates (Sec.\ref{sec:libraries_test}), the subtraction of the stellar continuum before fitting the observed spectra (Sec.~\ref{sec:stellarcont_test}), accounting for nebular emission (Sec.~\ref{sec:nebcont_test}), and varying the metallicity grid of the templates (Sec.~\ref{sec:metal_test}).

\subsection{Impact of different sets of templates}
\label{sec:libraries_test}

The stellar library, together with the initial mass function, and stellar evolution models (isochrones) are the basic ingredients to generate the spectra of SSPs. Variations in these three ingredients will lead to differences in the resulting SSPs, that ultimately can produce systematic differences in the measured properties of a galaxy, or galactic region, from its integrated spectrum \citep[see, e.g., ][]{Ge2019}. In this section, we explore how our results vary as a function of the set of SSP templates used to fit our data. In addition to the fiducial E-MILES \citep{Vazdekis2016} templates used in \citet{Emsellem2021}, which include 13 log-spaced age bins of [0.03, 0.05, 0.08, 0.15, 0.25, 0.40, 0.60, 1.0, 1.75, 3.0, 5.0, 8.5, 13.5] Gyr, and 6 metallicity bins of [Z/H] = [-1.49, -0.96, -0.35, +0.06, +0.26, +0.4], for these tests we consider the following set of templates:
\begin{itemize}
    \item Charlot \& Bruzual (2007, priv. comm.) - updated version 2016, to which we will refer as ``CB07''. These templates are constructed using the empirical stellar library STELIB \citep[][]{LeBorgne2003}, and use the stellar evolution prescription of \citet[][]{Marigo2007} to model the TP-AGB evolution of low- and intermediate mass stars. We employ five [Z/H] bins: [-1.7, -0.7, -0.4, 0, 0.4 ], and 25 age bins ranging from 1 Myr to 20 Gyr, of which 9 are younger than our fiducial age limit of 30 Myr.
    \item E-MILES + young extension presented in \citet{GonzalezDelgado2005, GonzalezDelgado2014}, to which we will refer as ``E-MILES + G-D''. This combined set of templates consists of our fiducial grid, plus six extra age bins, ranging from 1 Myr to 20 Myr, with four [Z/H] bins:  [-0.71, -0.4, 0, 0.22]. These younger templates are only available at a spectral resolution of $\sim 6 \,\AA$, lower than our fiducial E-MILES templates (FWHM $\sim 2.51 \AA$)
    \item E-MILES + newly computed young MILES SSP models, presented in \citet{Asad2017}, to which we will refer as ``E-MILES young''. These templates are an extension of the MILES library toward younger ages, available for a Padova isochrone. The combined library consist of 17 age bins, ranging from 6.3 Myr to 15 Gyr, and five [Z/H] bins, ranging from -1.33 to 0.22 for templates older than 60 Myr, and from -1.33 to 0.41 for templates younger than 60 Myr.

\end{itemize}
Finally, the tests are performed in the central region of NGC\,3351, which shows a prominent star-forming ring, and provides a perfect testing sample to evaluate the impact of the different templates/setting on the resultant fitting of young stellar populations. The quantities that we consider to evaluate the improvement of the quality of the solution are LumW age, LumW metallicity, E(B-V), and stellar mass surface density ($\Sigma_{*}$).

\begin{figure*}
 	\includegraphics[width=0.9\textwidth]{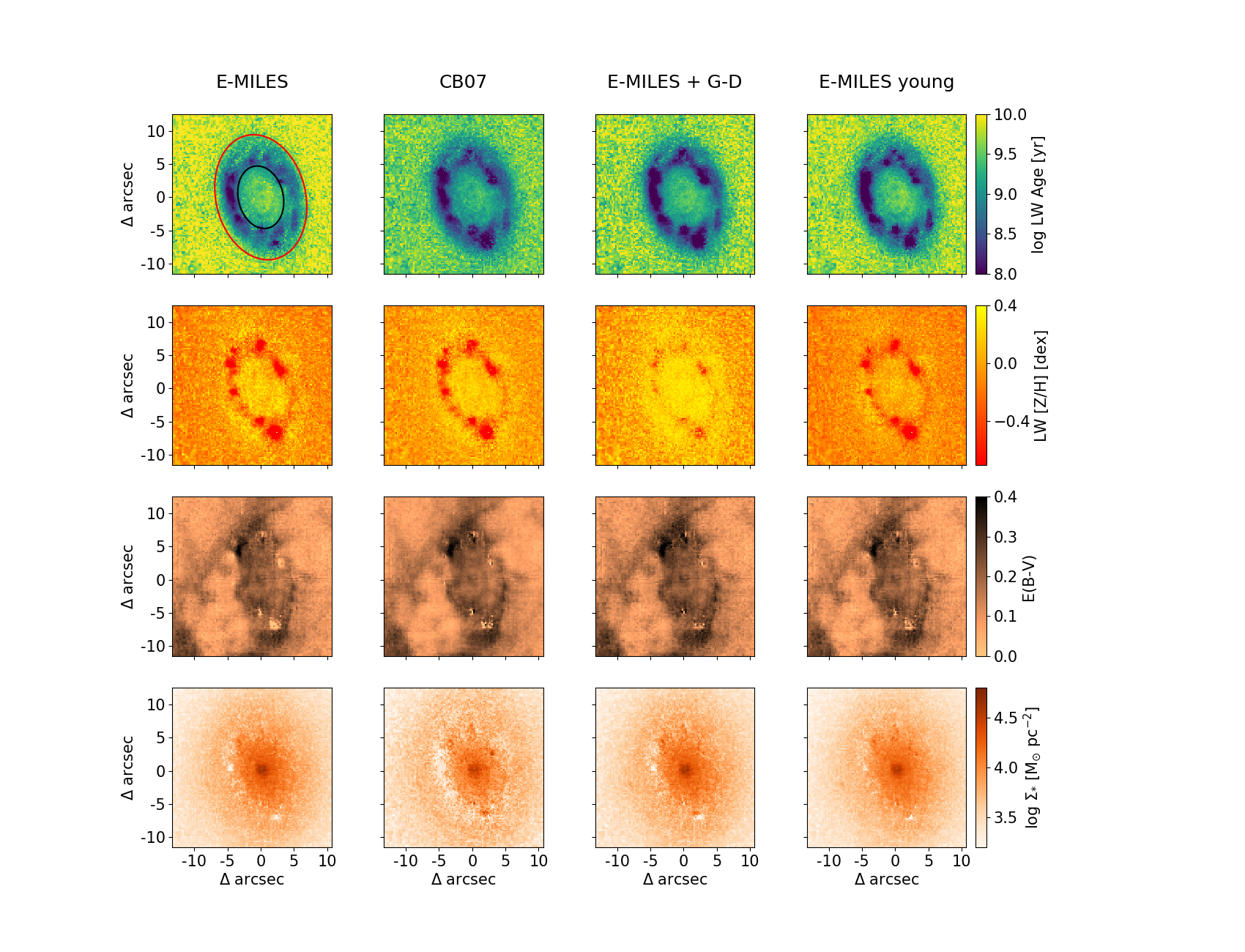}
     \caption[Age, metallicity, extinction and stellar mass surface density maps obtained for the central region of NGC\,3351, with the four different sets of templates tested]{Luminosity-weighted age (top row), luminosity-weighted [Z/H] (second row), stellar E(B-V) (third row), and stellar mass surface density (bottom row) maps derived for the central star-forming ring of NGC\,3351. Each column shows the maps obtained using the set of templates indicated at the top of each column. The maps obtained using our fiducial set of templates are shown in the left column. The area outside the red ellipse in the top-left panel is used to probe disk stellar population properties, and the area between the red and the black ellipses is used to probe stellar population properties of the young star-forming ring (see text).}
     \label{fig:base_case_tests}   
\end{figure*}

Figure~\ref{fig:base_case_tests} shows the maps of these four quantities, for each one of the set of templates tested here. It is clear that the addition of younger templates to the age grid is not sufficient to overcome the issue of extremely metal-poor young regions, as the low metallicity values in the inner ring of NGC3351 persist for all SSP templates.
To quantify variations, in terms of these four quantities, we consider their distributions within the ring (between the black and red ellipses in the top-right panel of Fig.~\ref{fig:base_case_tests}), and at radii larger than the ring (i.e., disk, outside the red ellipse). Specifically, we compare the 5$^\mathrm{th}$ percentile of the ring distribution, with the median of the disk distribution, for each one of these four quantities. The reason for this difference is that young clusters within the ring are young, often  ``erroneously'' characterized by low metallicity, low extinction, and low stellar masses. Therefore, the low-value tail of these ring distributions encodes information about these problematic regions. On the other hand, the distribution beyond the ring is relatively smooth, and its median is a robust metric for characterizing it.

\begin{figure*}
 	\includegraphics[width=0.9\textwidth, trim=0 0 0 0, clip]{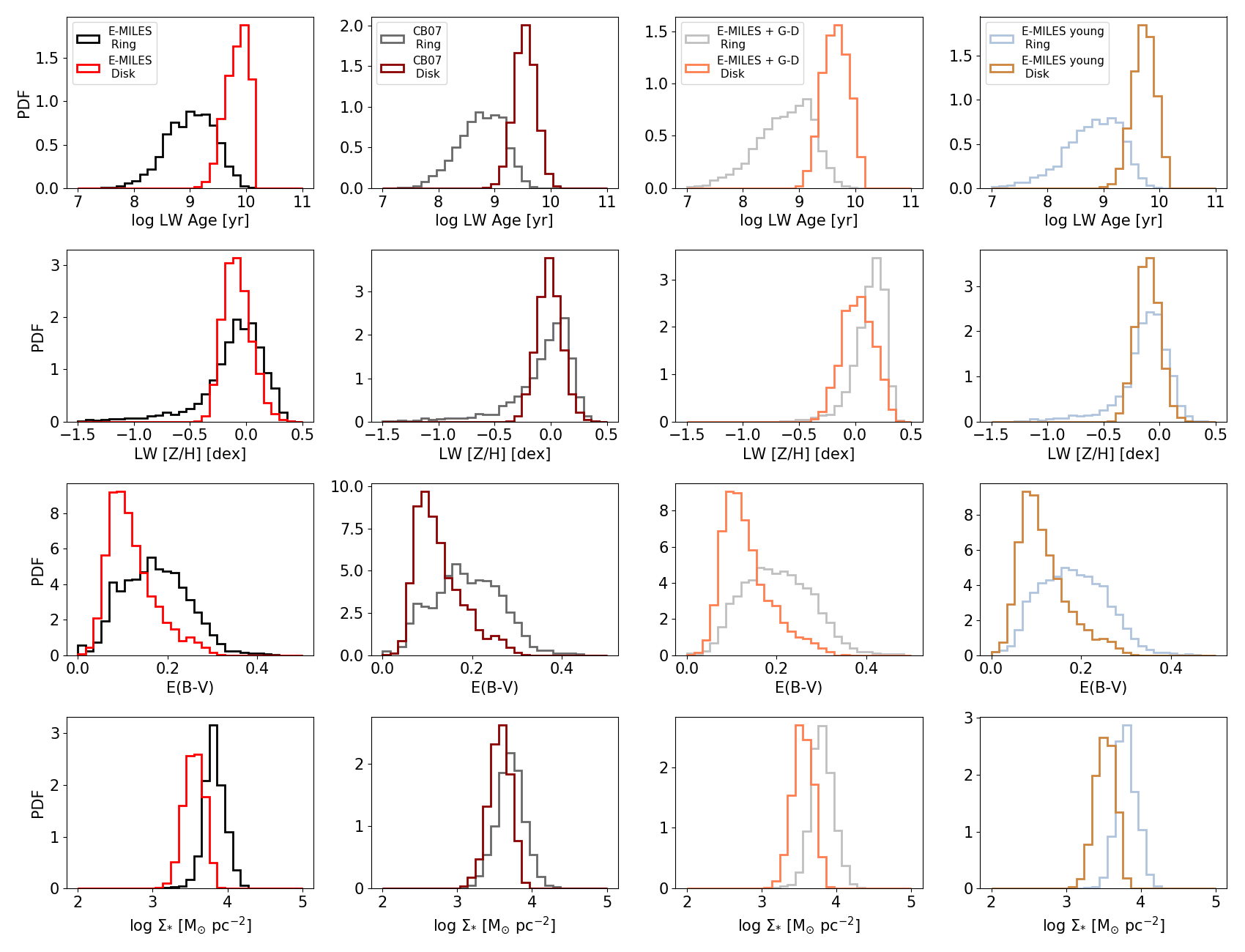}
     \caption[Age, metallicity, extinction and stellar mass surface density distributions of the central star-foming ring of NGC\,3351, and its surrounding, obtained with the four different sets of templates tested]{Luminosity-weighted age (top row), luminosity-weighted [Z/H] (second row), stellar E(B-V) (third row), and stellar mass surface density (bottom row) normalized distributions for the star-forming ring (black) and disk (red), delimitated in the top-left panel of Fig.~\ref{fig:base_case_tests}. Each column shows the distributions obtained using the set of templates indicated in the legends of the top-row panels. The distributions obtained using our fiducial set of templates are shown in the left column.}
     \label{fig:base_case_tests_hist}   
\end{figure*}

Figure~\ref{fig:base_case_tests_hist} shows the ring and disk distributions for each of these four quantities, for each set of templates used to produce Fig.~\ref{fig:base_case_tests}. 
We define:
\begin{equation}
    \label{eq:Delta_metric}
    \Delta \mathrm{X} = \mathrm{X}^{\mathrm{Ring}}_{p5} - \mathrm{X}^{\mathrm{Disk}}_{p50} 
\end{equation}
for X in [log age, [Z/H], E(B-V), log $\Sigma_{*}$], where $\mathrm{X}^{\mathrm{Ring}}_{p5}$ stands for the 5$^\mathrm{th}$ percentile of the ring distribution for the quantity X, and $\mathrm{X}^{\mathrm{Disk}}_{p50}$ stands for the 50$^\mathrm{th}$ percentile (median) of the disk distribution for the same quantity.

Hence, in terms of these four quantities, a good solution should yield to: (i) lower (negative) $\Delta \log$ age, that is, star-forming clusters are expected to be younger than the surrounding disk, (ii) higher (positive or $\sim$0) $\Delta$[Z/H], or, in other words, young clusters should not be associated with very metal-poor metallicities, compared to the disk, (iii) not strongly negative $\Delta$E(B-V) values, since holes in the E(B-V) map associated with these young regions are unphysical, and finally, (iv) positive (or $\sim 0$) $\Delta \log \Sigma_{*}$, due to the expected smooth underlying stellar mass surface density gradient.

In the following, we show the output maps generated with these four set of templates, under different considerations, and at the end of the section, we present the final comparison in terms of $\Delta \log$ age, $\Delta$[Z/H], $\Delta$E(B-V), and $\Delta \log \Sigma_{*}$.

\subsection{Is the stellar continuum shape driving the low-metallicity feature?}
\label{sec:stellarcont_test}
Low-metallicity SSPs have a continuum bluer than their high-metallicity counterparts, at a fixed stellar age. Thus, a natural feature to explore the origin of this problem is the stellar continuum shape. One possible explanation for the young and metal-poor feature is that the solution is being driven by the continuum shape, as the software might be struggling to find blue enough templates to reproduce the observed spectra of these young regions \citep[i.e., age-$Z$ degeneracy, see, e.g., ][]{Worthey1994}. To test if the shape of the stellar continuum is responsible for the metal-poor feature, we have repeated our measurements after subtracting the continuum of both, the templates and the observed spectra of each Voronoi bin. We have used the \texttt{specutils}\footnote{https://specutils.readthedocs.io/en/stable/} python package to calculate the continuum shape for a given spectrum, iteratively fitting a 5$^{\mathrm{th}}$ order polynomial to account for the stellar continuum, removing outlier absorption and emission features in each iteration. This step is repeated until no more pixels are identified as outliers and rejected from the fit.

\begin{figure*}
 	\includegraphics[width=0.9\textwidth, trim=0 0 0 0, clip]{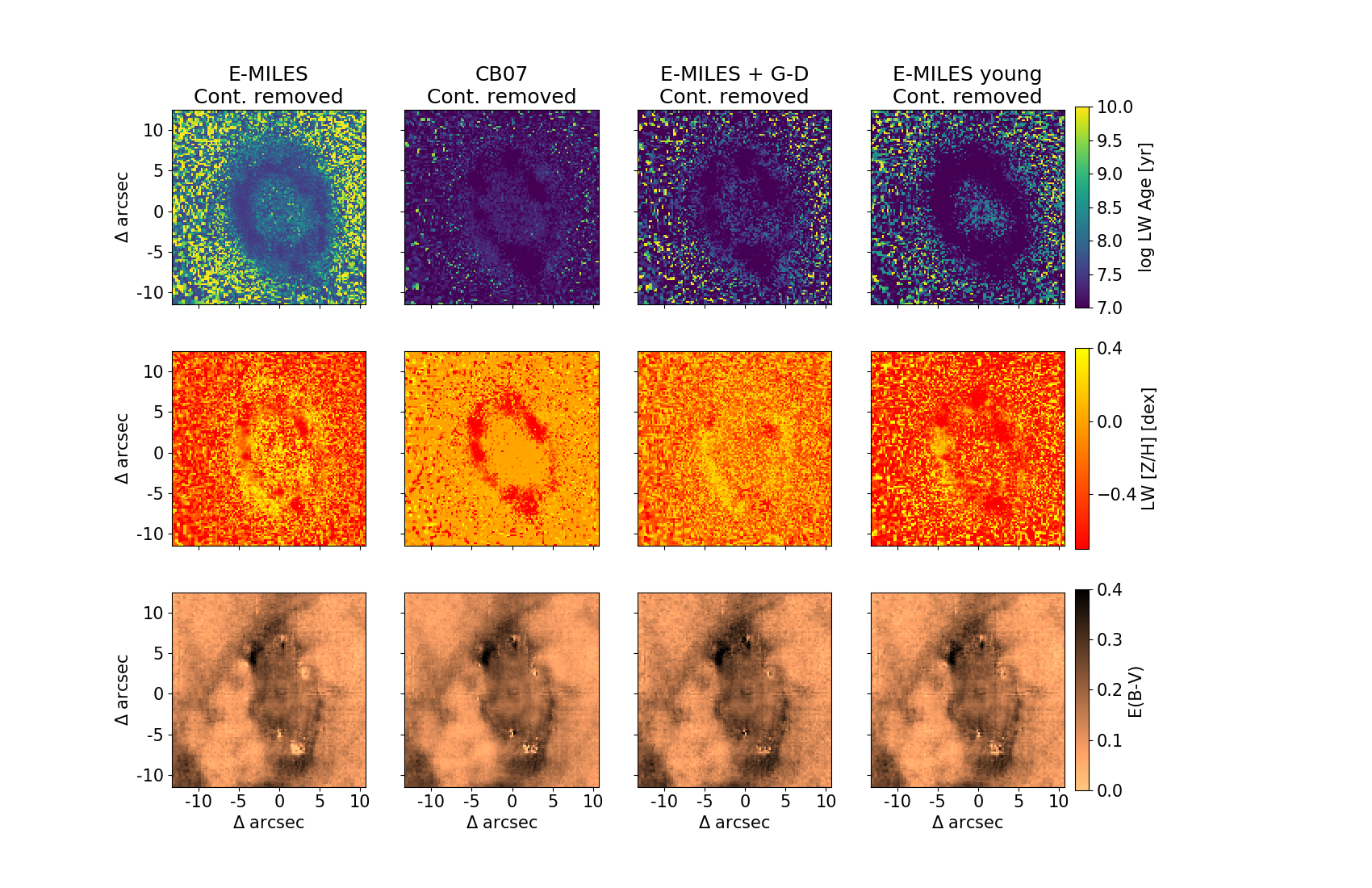}
     \caption[Age, metallicity, and extinction maps obtained for the central region of NGC\,3351, with the four different sets of templates tested, once the stellar continuum has been removed]{Luminosity-weighted age (top row), luminosity-weighted [Z/H] (middle row), and stellar E(B-V) (bottom row) maps derived for the central star-forming ring of NGC\,3351, once the stellar continuum has been removed from the observed spectra and the templates. Each column shows the maps obtained using the set of templates indicated at the top of each column. The maps obtained using our fiducial set of templates (after removing the stellar continuum) are shown in the left column. We note that the E(B-V) maps are identical to those in Fig.~\ref{fig:base_case_tests_hist}, as the stellar extinction is measured before subtracting the stellar continuum.}
     \label{fig:cont_removed_tests}   
\end{figure*}

Figure~\ref{fig:cont_removed_tests} shows the maps obtained with the four template sets under study when the stellar continuum is subtracted before performing the fit. The extinction is calculated before subtracting the continuum, therefore this presents no change with respect to the fiducial case. The stellar mass surface density maps can not be computed once the continuum is subtracted, and hence, the last row has been omitted from the figure. 
We highlight two aspects from Fig.~\ref{fig:cont_removed_tests}.  First, the solution in age and metallicity is much noisier, compared to the fiducial case. This is because the unsupervised continuum subtraction is not perfect, and creates artifacts in the spectra, often associated with absorption features or residual emission lines perturbing the definition of the continuum level.  Nevertheless, the ring is still recovered as a young and metal-poor feature for 3 out of the 4 libraries investigated, and we conclude that the shape of the continuum is not the primary driver of the low metallicity solutions for young regions.

\subsection{Could nebular emission be the responsible?}
\label{sec:nebcont_test}
High-energy ultraviolet photons emitted from nearby hot (and young) stars can ionize the surrounding interstellar medium, leading to nebular emission. Nebular emission is composed of nebular line emission and nebular continuum. While the former is primarily produced by radiative recombination processes and emission from specific line transitions, the latter  is a continuous emission spectrum that consists of free-free (Bremsstrahlung), free-bound (recombination continuum), and two-photon emission \citep{Byler2017}

The strength of emission from these two components depends on both the ionizing radiation field and the metallicity of the gas. The amount of nebular emission thus varies from galaxy to galaxy, and can evolve with cosmic time. As stated in Sec.~\ref{sec:stellarpops_maps}, for our analysis we have masked emission lines, as they make measurements of the underlying absorption features unreliable. However, nebular continuum emission might also have a pronounced impact on the SSP fitting. In this regard, \citet{Cardoso2022} very recently showed that ignoring the presence of nebular continuum emission in the spectra can lead to important biases in the stellar population properties derived through full spectral fitting.

\begin{figure}
\centering
 	\includegraphics[width=\columnwidth]{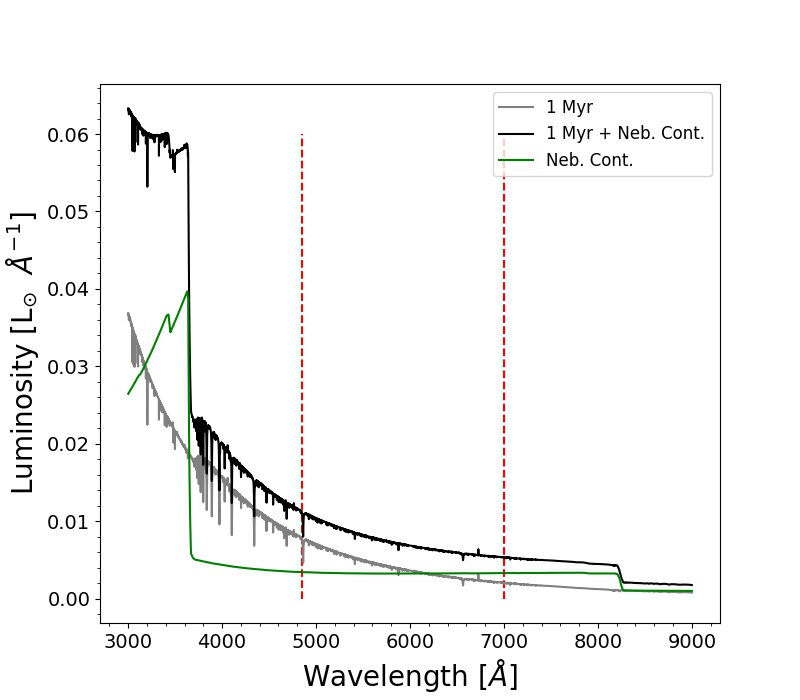}
     \caption[Example of the impact of nebular continuum emission on the SED of a 1 Myr age and solar metallicity SSP.]{Example of the impact of nebular continuum emission on the SED of a 1 Myr old solar metallicity SSP. The sudden decrease of nebular emission (green line) at $\sim3646 \AA$ and $\sim8207 \AA$ correspond to the Balmer and Paschen breaks, respectively. The vertical red dashed lines enclose the wavelength range considered for the full spectral fitting [4850,7000] $\AA$.}
     \label{fig:Nebular_cont_example}   
\end{figure}

Figure~\ref{fig:Nebular_cont_example} shows the spectrum  of a solar metallicity SSP at an age of 1 Myr, with and without the inclusion of nebular emission. The nebular component is computed with \texttt{FSPS} \citep[Flexible Stellar Population Synthesis; ][]{Conroy2009, Conroy2010}, a python package that generates spectra and photometric predictions for arbitrary stellar populations, using Cloudy \citep{Ferland2013, Byler2017, Byler2018} to calculate the emission produced by ionized gas. 

\begin{figure}
\centering
 	\includegraphics[width=\columnwidth]{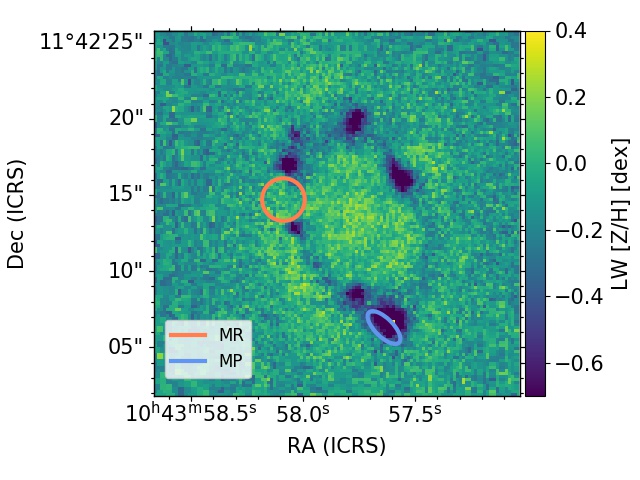}
     \caption[Definition of two young regions within the central star-formin ring of NGC\,3351 with different metallicity values]{Two very young regions within the central star-forming ring of NGC\,3351 with different inferred metallicity values. The orange circle encloses the metal-rich (MR) region, and the blue ellipse encloses the metal-poor (MP) region. The background map shows the luminosity-weighted stellar metallicity in the central region .}
     \label{fig:2regions_NGC3351}   
\end{figure}

In order to test the influence of  nebular continuum emission in our fitting, we first choose two representative regions within the young ring in the center of NGC\,3351, one of them is found to be extremely metal-poor, while the other is relatively metal-rich in our fiducial fitting. For each region, we extract a high-SN spectrum, and we use these spectra for further tests. We will refer to these two young regions as MR (metal-rich) and MP (metal-poor). Figure~\ref{fig:2regions_NGC3351} shows these regions, and the stellar metallicity map determined with E-MILES templates as background. The MR and MP regions are indicated with an orange circle and a blue ellipse, respectively.

We use \texttt{FADO} \citep[Fitting Analysis using Differential evolution Optimization; ][]{Gomes2017}, a SED-fitting software (similar to pPXF) that aims for recovering the best-fitting SFH,  while consistently accounting for the observed nebular emission characteristics of a star-forming (SF) galaxy, to fit the spectra of these two regions, and quantify the contribution expected from nebular emission.

\begin{figure*}
\centering
 	\includegraphics[width=0.8\textwidth]{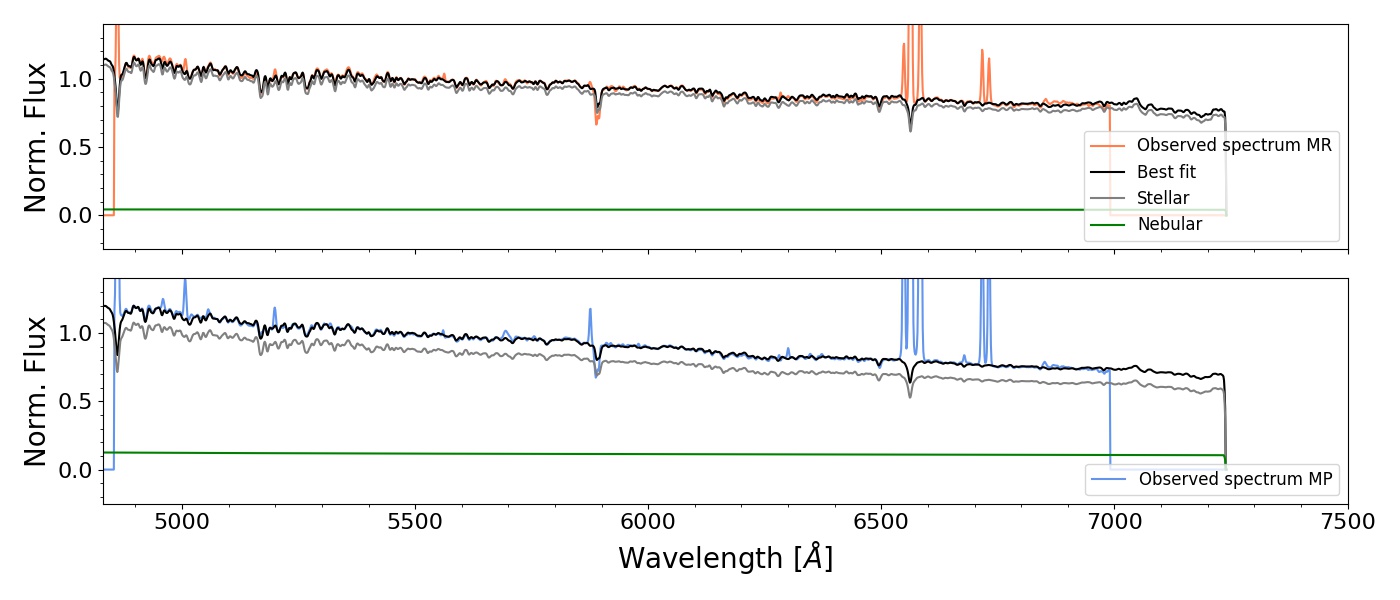}
     \caption[Best SED-fitting obtained with FADO for two young regions with different metallicity values]{Best SED-fitting obtained with FADO for the MR and MP regions, defined in Fig.~\ref{fig:2regions_NGC3351}. The orange and blue lines represent the MR and MP observed spectra, respectively. The black line show the best-fitting model, and the gray and green curves show separately the stellar and nebular continuum contribution, respectively. It is clear that the MP region (bottom) requires a more significant contribution of nebular continuum than the MR (top) region.}
     \label{fig:MR_MP_FADO}   
\end{figure*}

Figure~\ref{fig:MR_MP_FADO} shows that both spectra contain some degree of nebular continuum emission contribution. This is expected, as both of them correspond to young SF regions. However, region MP hosts a much larger nebular continuum contribution than MR, while also being much younger. Moreover, the metal-poor peaks in Fig.~\ref{fig:2regions_NGC3351} spatially correlate well with peaks in the H$\alpha$ flux surface density map. This suggests that not accounting properly for nebular continuum contribution in the spectra of the youngest regions could be responsible for the low metallicity obtained for the MP spectrum. Indeed, in the wavelength range of interest, the nebular continuum can be represented, to zero$^{\mathrm{th}}$ order, as a flat spectrum (see Fig.~\ref{fig:Nebular_cont_example}). This boost of the continuum yields weaker absorption features (i.e., shallower relative to the continuum level). This effect can be further quantified by measuring different line-strength indices for different contributions of nebular continuum emission.

\begin{figure*}
\centering
 	\includegraphics[width=0.9\textwidth]{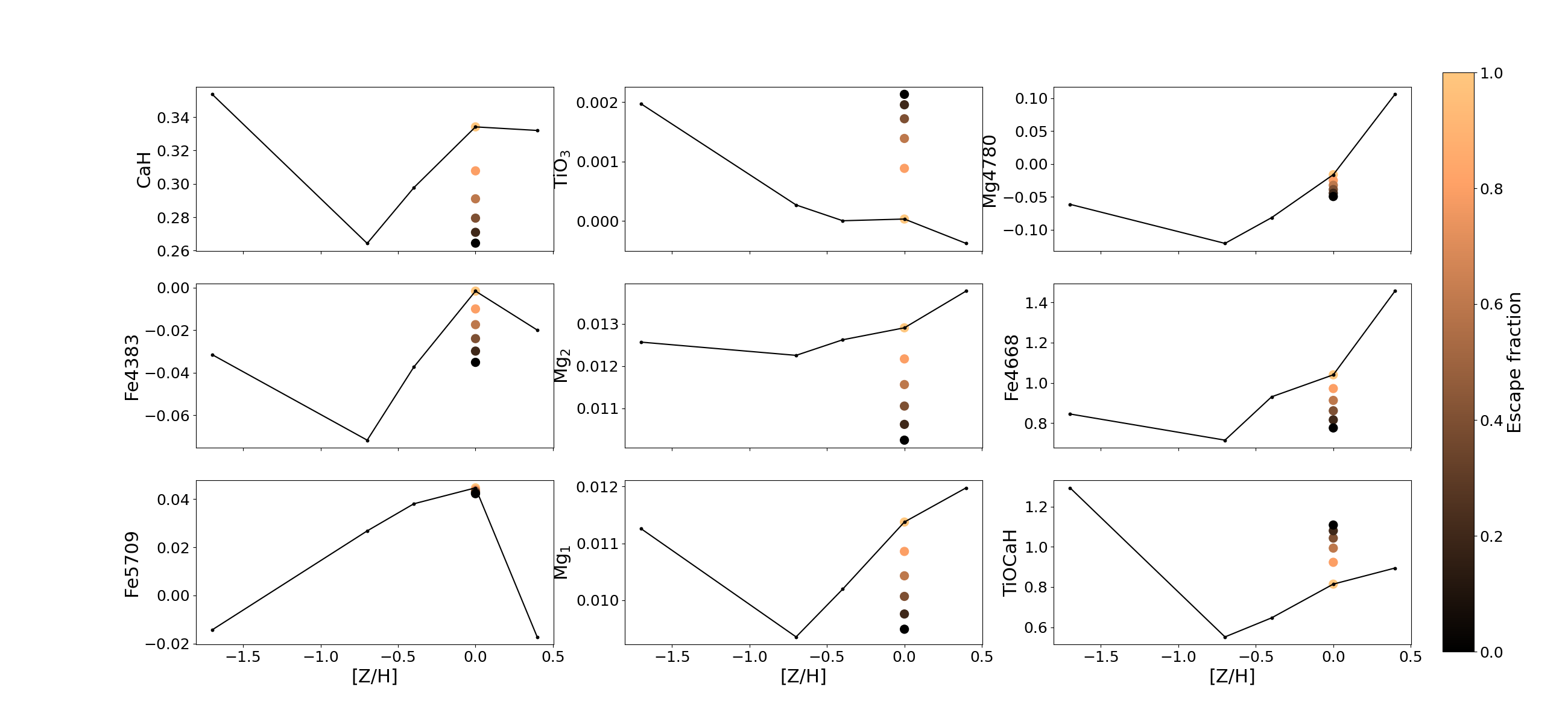}
     \caption[Set of line-strength indices commonly used as stellar metallicity tracers, measured for the SED of a 1 Myr SSP model for the full range of metallicity values available]{Set of line-strength indices commonly used as stellar metallicity tracers, measured for the SED of a 1 Myr SSP model from the CB07 library, for the full range of metallicity values available. For solar metallicity, we show the measurement for different levels of nebular continuum contribution. It is clear that a stronger nebular continuum contribution (lower escape fraction) yields a line-strength index measurement consistent with lower metallicities (with TiOCaH being the only exception, bottom right panel). Also, we note that the most metal-poor metallicity bin often appears as a strong outlier in the line-strength-[Z/H] relation.}
     \label{fig:Lick}   
\end{figure*}

Figure~\ref{fig:Lick} shows a set of line-strength indices \citep{Burstein1984, Faber1985} as a function of stellar metallicity measured for the CB07 templates, at a fixed age of $1$ Myr. It also shows the measurement at solar metallicity obtained under different assumptions for the nebular continuum contribution, from an escape fraction of 1 (i.e., no nebular continuum), to an escape fraction of 0 (i.e., the full contribution of nebular emission as computed using \texttt{FSPS}). 

For most line-strength indices that correlate with stellar metallicity (neglecting the most metal-poor bin), the contribution of nebular emission causes a shift in the index, consistent with a more metal-poor metallicity. Therefore, in the next tests, we explore the impact that including nebular continuum emission in the SSP fitting has in the output maps. 

It is worth noting that although \texttt{FADO} does account for nebular emission, it is currently not implemented in python, and thus, we opt for different approaches to account for this contribution using pPXF. To include nebular emission in the fitting, we investigate two different options:

\begin{itemize}
    \item Include nebular contribution in the templates, as computed by \texttt{FSPS}: this approach has the advantage of including an accurate description of the nebular emission in the fitting. However, it has the drawback that it is only useful for libraries that already include very young templates (i.e., $\lesssim 5$ Myr).  We will refer to this approach as ``Neb. A'' in the following text.
    
    \item Tie the nebular continuum contribution to the emission line measurement: as they are originating from the same source, it is expected that the amount of nebular continuum scales with the flux of emission lines \citep[e.g.,][]{Byler2017}. Specifically, we model the nebular continuum emission as a flat spectrum with its flux equal to $0.5\%$ of the H$\beta$ flux, and we subtract this contribution before proceeding with the SSP fit. As this is a very simplistic representation of the nebular contribution, we do not perform additional corrections (i.e., we do not correct by extinction), which could, in principle, lead to an underestimation of the nebular continuum emission. Despite the simplicity of this approach, it offers the advantage of being applicable to any of the libraries of templates used.  We will refer to this approach as ``Neb. B'' in the following text.
    
\end{itemize}

\subsubsection{Impact of neglecting nebular continuum contribution in synthetic spectra}
\label{sec:mock_data_test}
Here we use a set of mock spectra built from the CB07 library to assess how problematic it can be to not account for existing nebular emission during the full spectrum fitting. To evaluate this, we define our mock-spectra library to harbor a fixed contribution of 100 solar masses of an old ($\sim10$ Gyr) and metal-poor ([Z/H] = -0.7) component, a fixed contribution of 20 solar masses of a intermediate age ($\sim1$ Gyr) and medium metallicity ([Z/H] = -0.4) component, and a variable contribution (30 values covering the range from 0 to 1.5 solar masses) of a young (2 Myr) and solar metallicity component. 
We have chosen such a configuration aiming for a realistic representation of the combination of multiple stellar populations, in which most of the mass is contributed by old and metal-poor stars (formed from unprocessed gas), and subsequent stellar generations formed from a recycled (and thus, metal-enriched) interstellar medium. Finally, young stars are subdominant in terms of mass, but their light strongly influences the spectrum of the combined population. We build this mock library using young templates that include nebular continuum contribution (computed using FSPS), and then we fit them to recover their original SFH, removing the nebular contribution from the templates used in the fitting. Since we want to explore exclusively the impact of neglecting the contribution of existing nebular emission in the mock spectra, we assume an infinite S/N regime, in order to remove noise-driven differences.

\begin{figure*}
\centering
 	\includegraphics[width=0.8\textwidth]{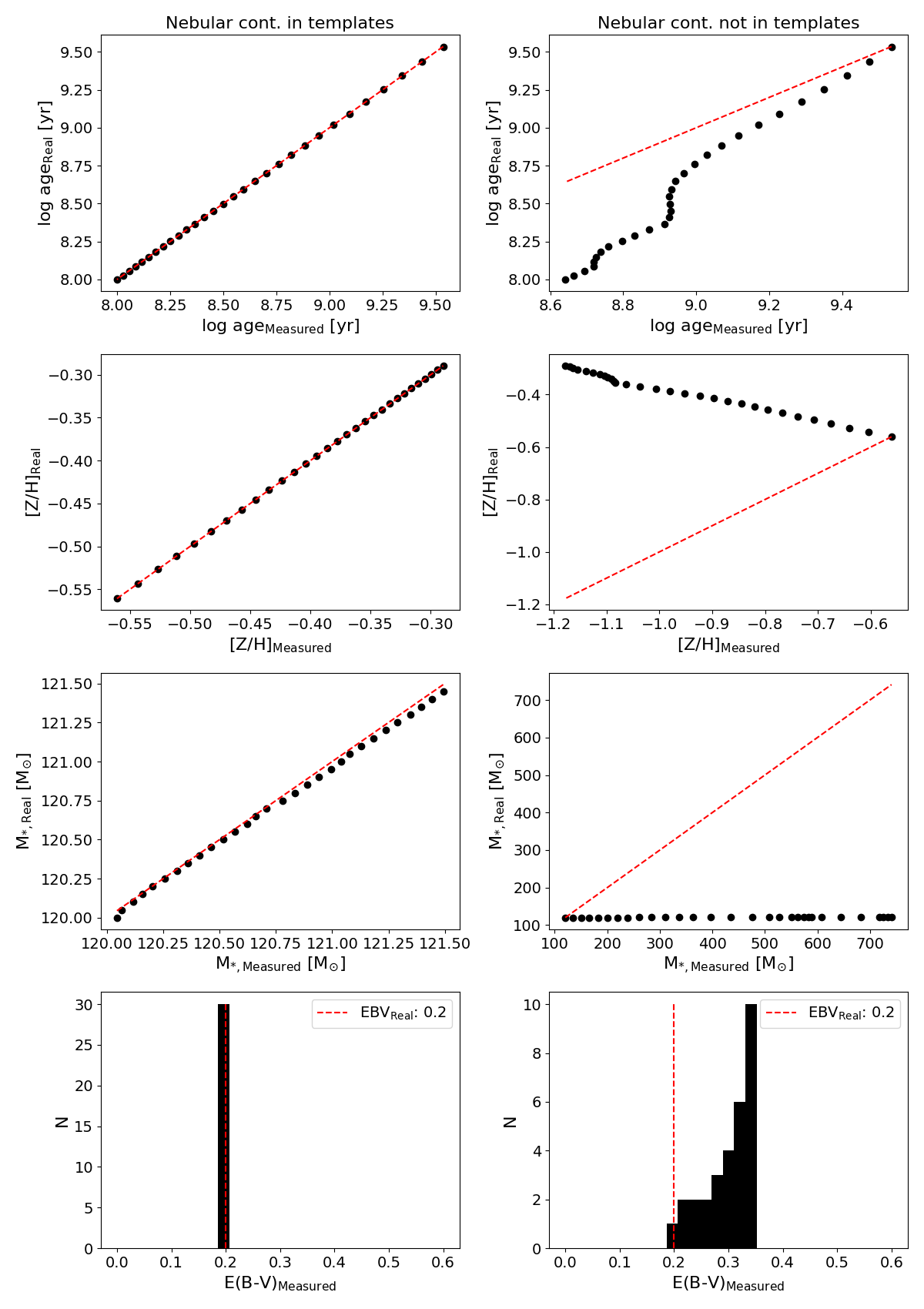}
     \caption[Mock-spectra test designed to investigate the impact on the outcome of the SSP fitting when nebular continuum emission is present in the data, but neglected in the analysis]{Mock-spectra test designed to investigate the impact on the outcome of the SSP fitting when nebular continuum emission is present in the data, but neglected in the analysis. We created a synthetic sample of 30 linear combinations of SSP, with a constant contribution of old (metal-poor), and intermediate age (intermediate metallicity) SSPs, and a varying contribution of a young (metal-rich) SSP that include nebular continuum contribution (computed using FSPS, i.e., Neb. A approach). Both columns show the real age (top row), metallicity (second row), and stellar mass (third row) of each mock spectra in the $y$-axis, and the recovered value in the $x$-axis. The bottom panels show an histogram of the recoved E(B-V) values, and the vertical dashed red line shows the real E(B-V) applied to the mock spectra (0.2). The left column shows the results obtained when nebular continuum emission is accounted for in the templates used to fit the data, and the right column shows the same results when nebular continuum emission is neglected from the fitting. It is remarkable that when existing nebular emission is neglected in the fitting, we obtain an anticorrelation between the real metallicity and its recovered values.}
     \label{fig:mock_spectra_test}   
\end{figure*}

Figure~\ref{fig:mock_spectra_test} shows the real luminosity-weighted mean age, luminosity-weighted mean metallicity, total mass, and extinction values as a function of their recovered values, accounting for nebular emission (left column) and neglecting nebular emission (right column) in the fitting. It is remarkable that when existing nebular emission is neglected in the fitting, we recover an anticorrelation between the real metallicity and its measured values. This effect is similar to the trend revealed by the line-strength indices; the nebular continuum contribution reduces the equivalent width of absorption features, mimicking lower metallicities. Another interesting effect is that the stronger continuum is interpreted as a higher total stellar mass, when the templates lack this contribution. This is a direct consequence of the higher level of the continuum due to the nebular continuum emission. However, in our data, we see that the stellar mass surface density is underestimated in the young regions (see, e.g., Fig.~\ref{fig:base_case_tests}), that is, the observed bias in stellar mass in regions expected to have a strong nebular continuum emission contribution is in the opposite direction than that found in this test. A higher extinction is not surprising, as the recovered low-metallicity of the young templates will lead to a bluer spectrum. Thus, a higher extinction value is needed to match the spectral shape of the observed spectrum.

\subsubsection{Effect of accounting for nebular continuum contribution in the output maps}

\begin{figure*}
\centering
 	\includegraphics[width=0.9\textwidth, trim=0 0 0 0, clip]{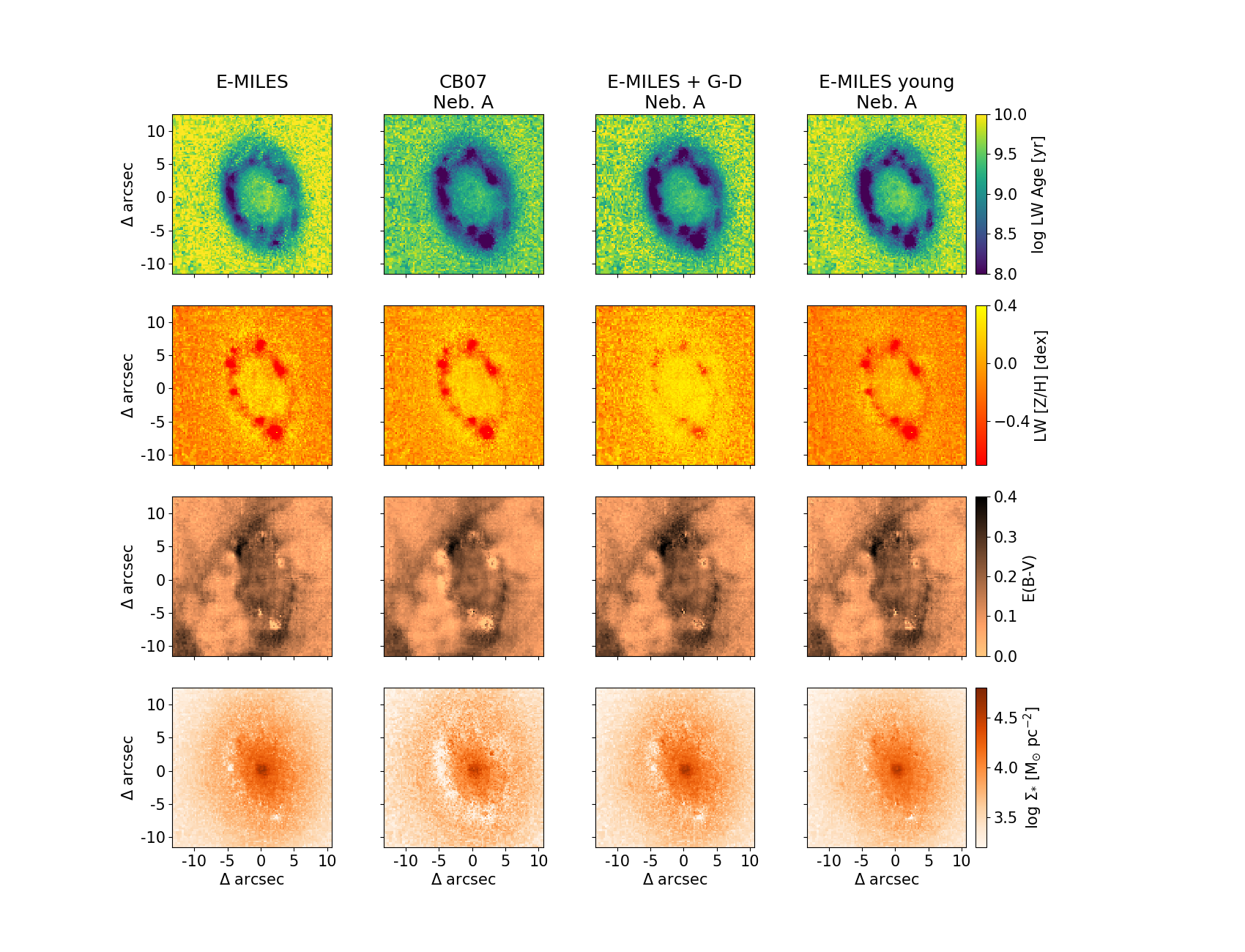}
     \caption[Age, metallicity, extinction and stellar mass surface density maps obtained for the central region of NGC\,3351, with the four different sets of templates tested, accounting for nebular continuum emission using the approach Neb. A]{Luminosity-weighted age (top row), luminosity-weighted [Z/H] (second row), stellar E(B-V) (third row), and stellar mass surface density (bottom row) maps derived for the central star-forming ring of NGC\,3351. Each column shows the maps obtained using the set of templates indicated at the top of each column, accounting for nebular continuum emission using the approach Neb. A (see main text). The maps obtained using our fiducial set of templates are shown in the left column.}
     \label{fig:NebA_test}   
\end{figure*}

\begin{figure*}
\centering
 	\includegraphics[width=0.9\textwidth, trim=0 0 0 0, clip]{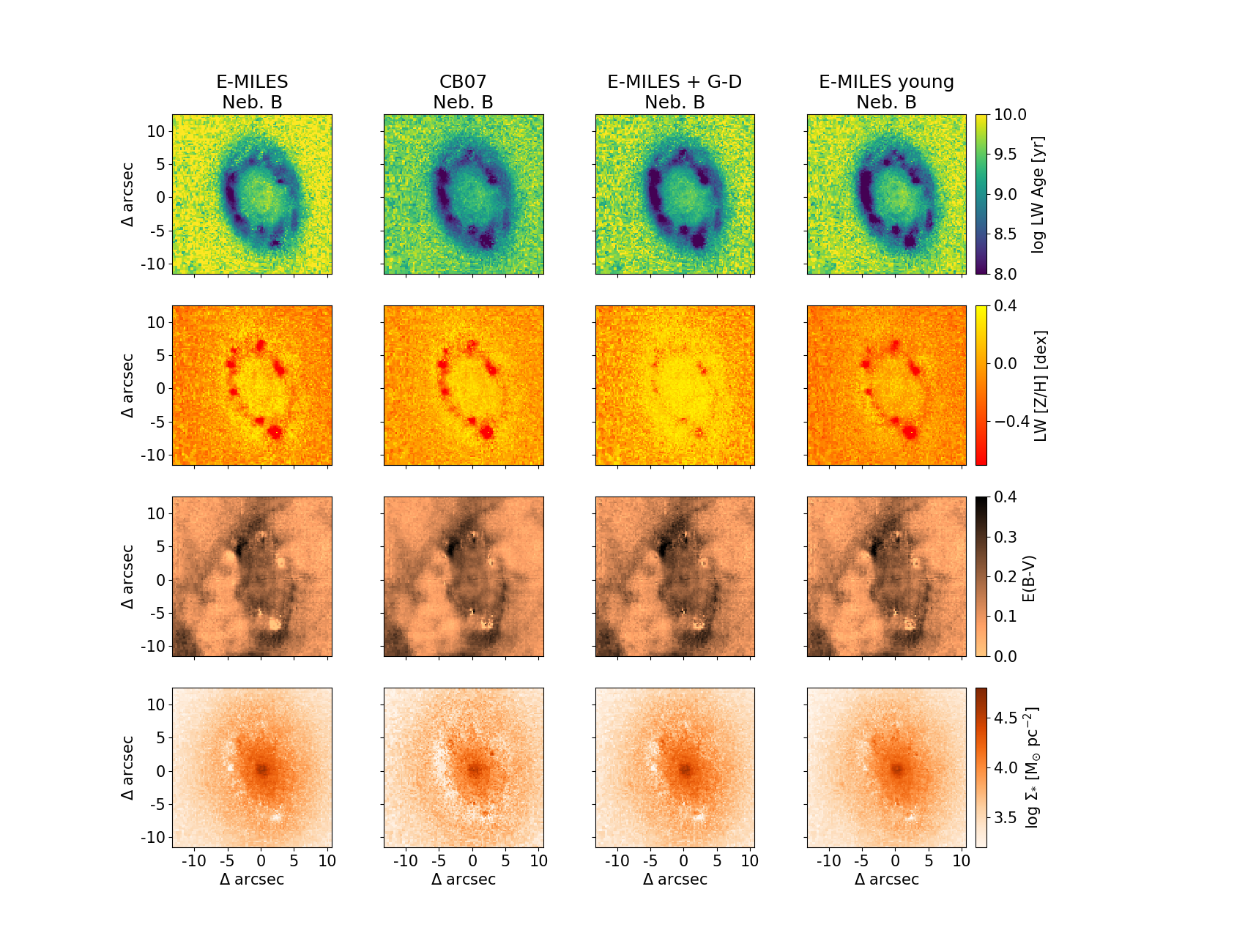}
     \caption[Age, metallicity, extinction and stellar mass surface density maps obtained for the central region of NGC\,3351, with the four different sets of templates tested, accounting for nebular continuum emission using the approach Neb. B]{Luminosity-weighted age (top row), luminosity-weighted [Z/H] (second row), stellar E(B-V) (third row), and stellar mass surface density (bottom row) maps derived for the central star-forming ring of NGC\,3351. Each column shows the maps obtained using the set of templates indicated at the top of each column, accounting for nebular continuum emission using the approach Neb. B (see main text). The maps obtained using our fiducial set of templates (with the modifications described) are shown in the left column.}
     \label{fig:NebB_test}   
\end{figure*}

Figures~\ref{fig:NebA_test} and \ref{fig:NebB_test} show the output maps obtained when nebular emission is included in the SSP fitting using the two approaches described earlier (Neb. A and Neb. B, respectively). Despite the impact that nebular continuum emission is expected to have on the spectra of SSPs, particularly in young populations, its addition to the fitting does not lead to a visually significant improvement of the low-metallicity feature.

\subsection{Do we trust our full metallicity grid?}
\label{sec:metal_test}
In the previous paragraphs, it has become clear that the issue of the young metal-poor feature issue persists, even when we add younger templates to the age grid, remove the stellar continuum of the spectra, or account for nebular continuum emission. 

In the next set of tests, we explore if this metal-poor feature could be due to the lack of many hot stars in the stellar libraries used to build the models, that makes the predictions uncertain \citep{Villaume2017, Conroy2018}. A hint for potential a mismatch between the metal-poor templates and the true galaxy spectra is provided by Fig.~\ref{fig:Lick}, which shows that the most metal-poor bin is often an outlier in the line-strength index versus [Z/H] correlation. Furthermore, metallicity values such as -1.49 (MILES), -1.7 (CB07) and -1.33 (MILES young extensions) are extremely low, and unlikely to be significantly present in the stellar disk of massive star-forming galaxies, where the interstellar medium is being constantly recycled and enriched by star formation and its subsequent feedback \cite[see, e.g.,][]{Ho2017}. Thus, we repeat our measurements, removing these extremely low metallicity bins from our grid. We note that the MILES+G-D library already lacks metallicities lower than $-1.0$ for ages younger than 30 Myr. We will refer to the libraries lacking their most metal-poor metallicity bin as NoMP.

\begin{figure*}
\centering
 	\includegraphics[width=0.9\textwidth, trim=0 0 0 0, clip]{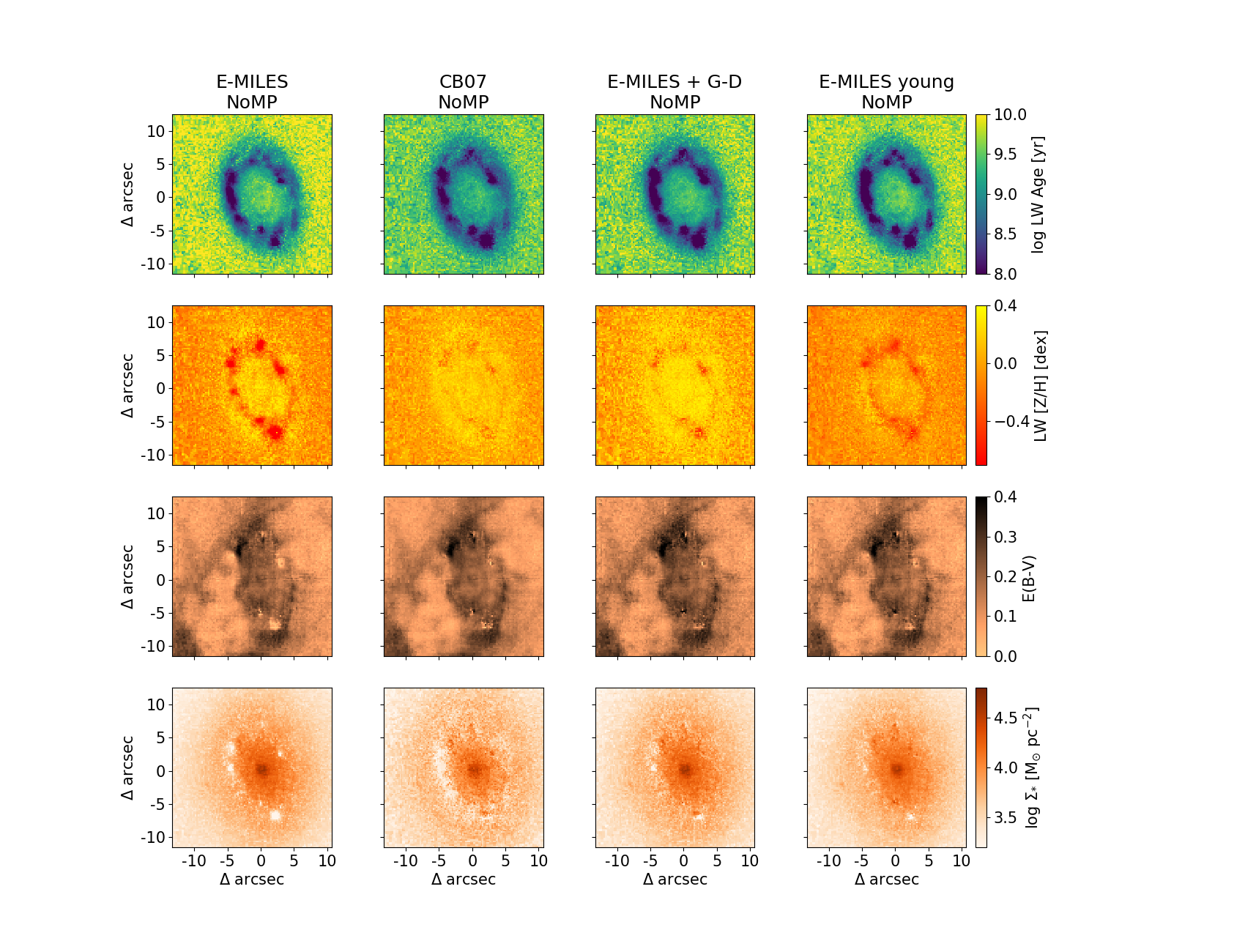}
     \caption[Age, metallicity, extinction and stellar mass surface density maps obtained for the central region of NGC\,3351, with the four different sets of templates tested, removing the lowest metallicity bins from the metallicity grid]{Luminosity-weighted age (top row), luminosity-weighted [Z/H] (second row), stellar E(B-V) (third row), and stellar mass surface density (bottom row) maps derived for the central star-forming ring of NGC\,3351. Each column shows the maps obtained using the set of templates indicated at the top of each column, removing the lowest metallicity bins from the metallicity grid (see main text). The maps obtained using our fiducial set of templates (with the modifications described) are shown in the left column.}
     \label{fig:NoMP_test}   
\end{figure*}

Figure~\ref{fig:NoMP_test} shows the output maps after removing the most metal-poor metallicity bin from each stellar library. The young metal-poor feature is drastically improved after implementing this change in the metallicity prior, meaning that the most metal-poor values get a significant weight in the fitting of the spectra of the young regions, when they are present.

With this in mind, and for completeness, we explore once again if the inclusion of nebular contribution has a positive impact on the fitting, after removing the most metal-poor bin of each library, by examining the output maps obtained under this combination of settings.

\begin{figure*}
\centering
 	\includegraphics[width=0.9\textwidth, trim=0 0 0 0, clip]{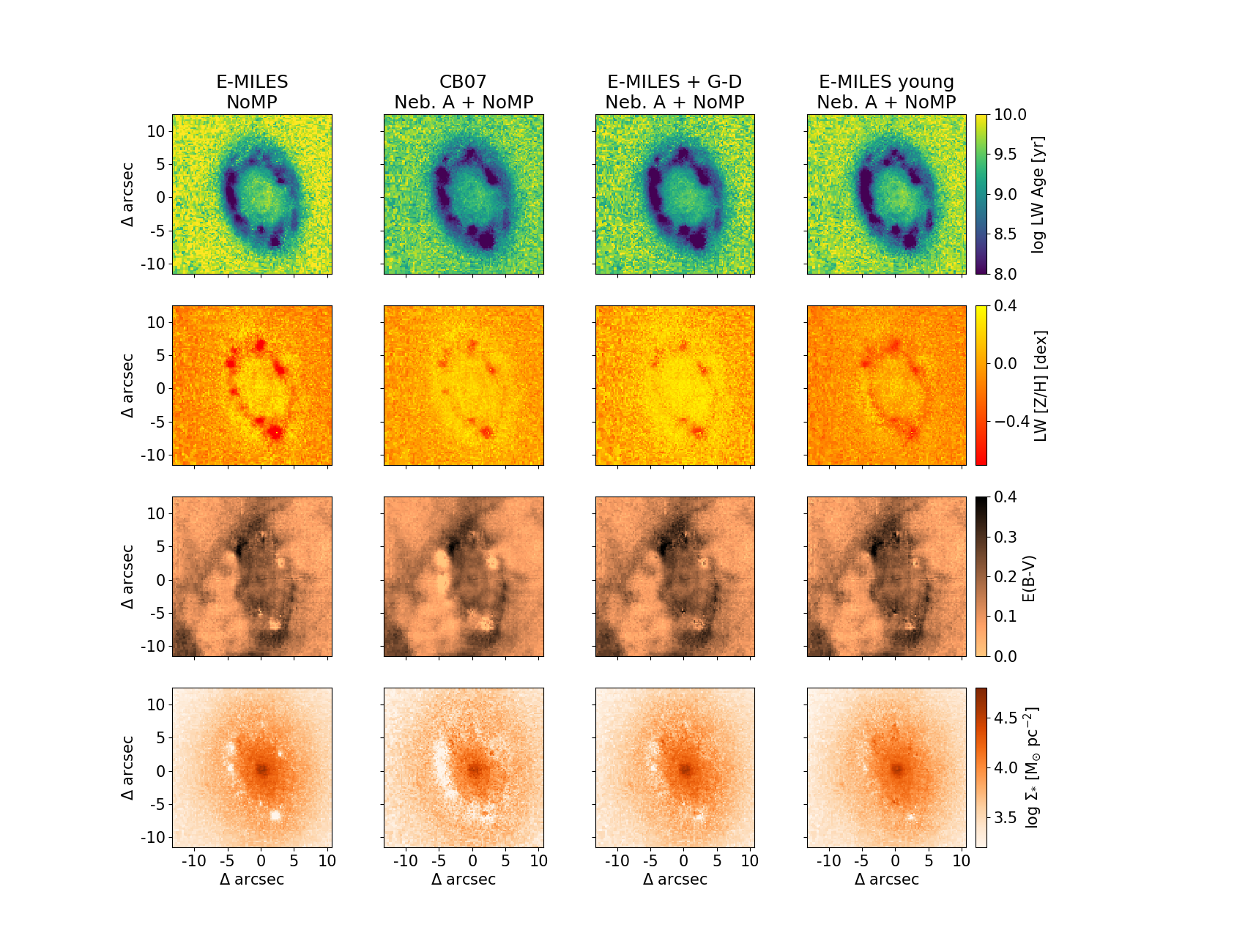}
     \caption[Age, metallicity, extinction and stellar mass surface density maps obtained for the central region of NGC\,3351, with the four different sets of templates tested, accounting for nebular continuum emission using the approach Neb. A, and removing the lowest metallicity bins from the metallicity grid]{Luminosity-weighted age (top row), luminosity-weighted [Z/H] (second row), stellar E(B-V) (third row), and stellar mass surface density (bottom row) maps derived for the central star-forming ring of NGC\,3351. Each column shows the maps obtained using the set of templates indicated at the top of each column, accounting for nebular continuum emission using the approach Neb. A, and removing the lowest metallicity bins from the metallicity grid (see main text). The maps obtained using our fiducial set of templates (with the modifications described) are shown in the left column.}
     \label{fig:NoMPNebA_test}   
\end{figure*}

\begin{figure*}
\centering
 	\includegraphics[width=0.9\textwidth, trim=0 0 0 0, clip]{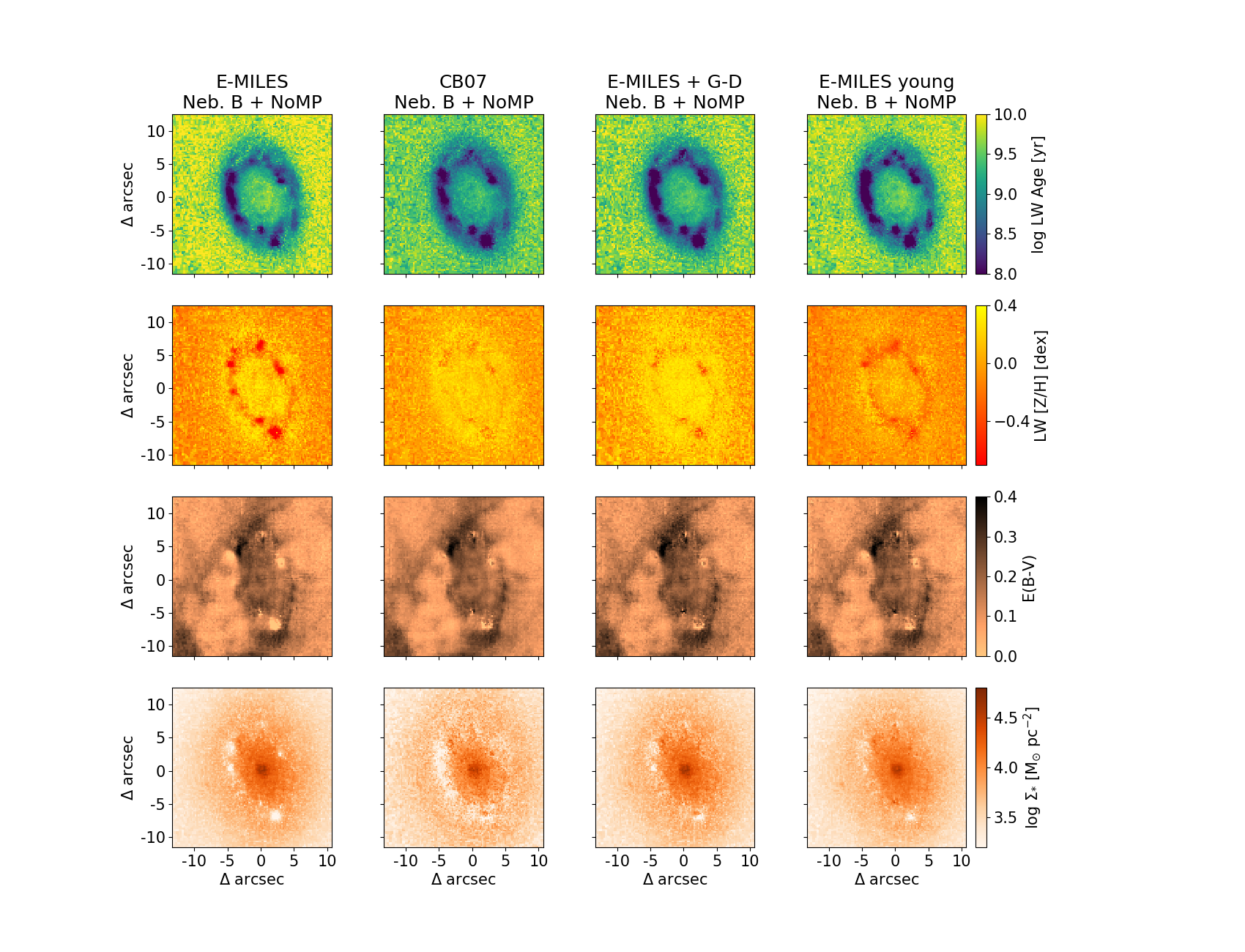}
     \caption[Age, metallicity, extinction and stellar mass surface density maps obtained for the central region of NGC\,3351, with the four different sets of templates tested, accounting for nebular continuum emission using the approach Neb. B, and removing the lowest metallicity bins from the metallicity grid]{Luminosity-weighted age (top row), luminosity-weighted [Z/H] (second row), stellar E(B-V) (third row), and stellar mass surface density (bottom row) maps derived for the central star-forming ring of NGC\,3351. Each column shows the maps obtained using the set of templates indicated at the top of each column, accounting for nebular continuum emission using the approach Neb. B, and removing the lowest metallicity bins from the metallicity grid (see text). The maps obtained using our fiducial set of templates (with the modifications described) are shown in the left column.}
     \label{fig:NoMPNebB_test}   
\end{figure*}

Figures~\ref{fig:NoMPNebA_test} and \ref{fig:NoMPNebB_test} show the maps obtained when nebular emission is included in the fit, together with the removal of low-metallicity templates, using the Neb. A and Neb. B descriptions, respectively. The metallicity maps show weaker low-metallicity features under the Neb. B + NoMP approach,  compared to the Neb. A + NoMP one. Nevertheless, none of these tests exhibit a visually significant improvement compared to the NoMP scenario. Thus, it seems that the inclusion of nebular continuum emission does not significantly improve the quality of the output maps, even when the most metal-poor templates have been removed.

\subsection{How the different physical conditions in which young stars are embedded bias the results of our SSP fitting?}

Young stellar populations are embedded in deeper layers of gas and dust, remnants of the cloud from which they formed, than old stellar populations. In this work, we have assumed a single extinction value for all the stellar populations present in a given region, parametrized by a \citet{Calzetti2000} extinction law. Although young stars disperse molecular clouds on relatively short time scales of a few Myr \citep{Chevance2020}, young stellar populations in star-forming regions are likely affected by greater extinction values than their older counterparts. The impact of this different extinction across coexisting stellar populations on the measured SFHs has not been determined yet. To evaluate the effect that this feature would have on the SSP fitting, we have repeated the test with mock data described in Appendix~\ref{sec:mock_data_test}, but this time, applying the extinction curve (E(B-V) = $0.2$) only to the SED of the youngest stellar population.

\begin{figure*}
\centering
 	\includegraphics[width=0.8\textwidth]{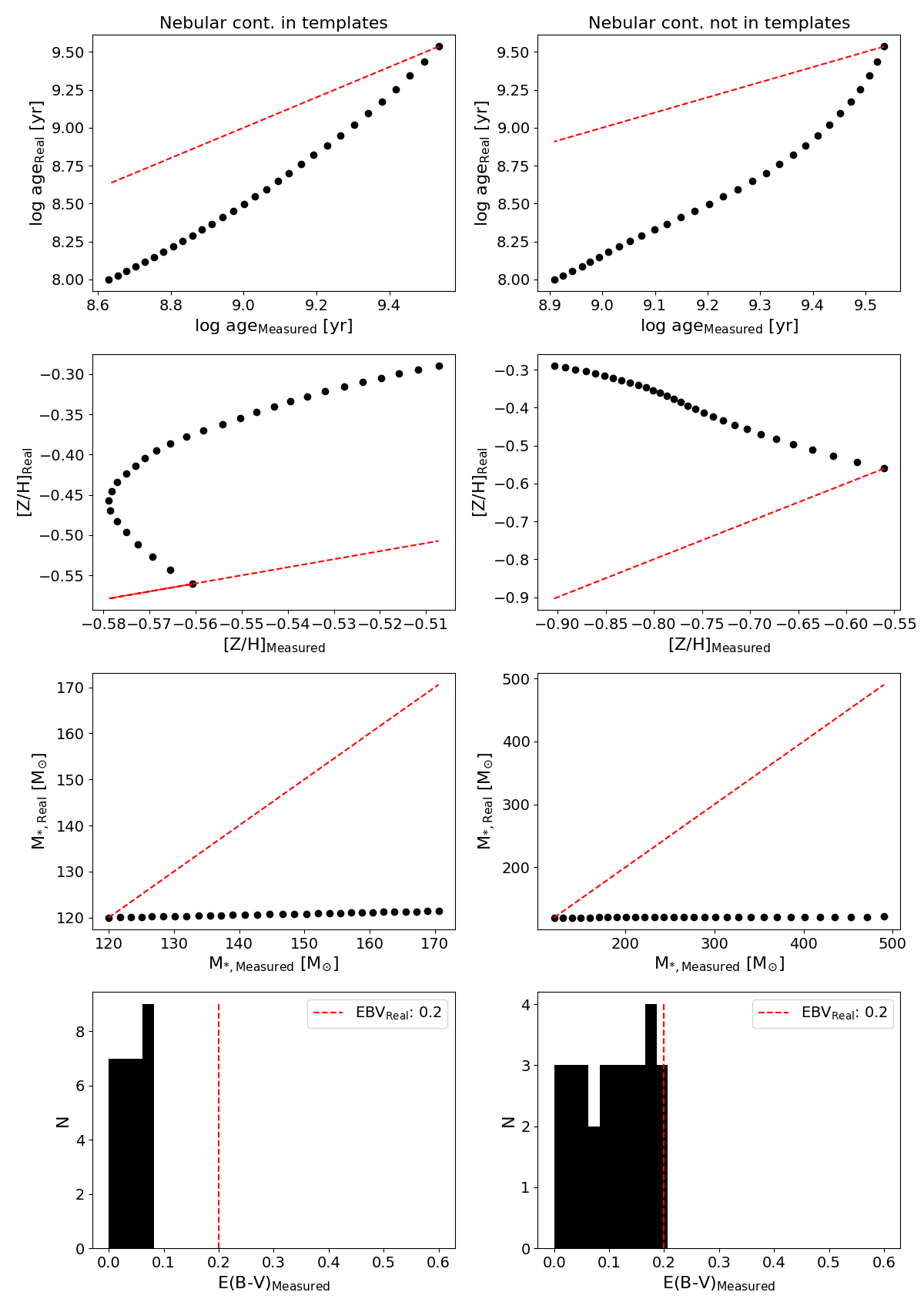}
     \caption{Analogous to Fig.~\ref{fig:mock_spectra_test}, but in this case, dust extinction (E(B-V) = $0.2$) has been applied only to the SED of the youngest stellar population. The left column shows the results obtained when nebular continuum emission is accounted for in the templates used to fit the data, and the right column shows the same results when nebular continuum emission is neglected from the fitting. The figure shows that even when the templates used to build the mock spectra are identical to the templates used to measure their stellar properties, the measured age, metallicity, mass, and extinction values are severely biased respect to the real values.}
     \label{fig:mock_spectra_test2}   
\end{figure*}

Figure~\ref{fig:mock_spectra_test2} is analogous to Fig.~\ref{fig:mock_spectra_test}, except that in this case, each input spectrum is composed by constant a linear combination of unextincted old ($\sim10$ Gyr) and intermediate age ($\sim1$ Gyr) stellar population, and a variable contribution of a dust-extincted young (2 Myr) stellar population (see Appendix~\ref{sec:mock_data_test} for details). The figure shows that even when the templates used to build the mock spectra are identical to the templates used to measure their stellar properties (left column), the measured age, metallicity, mass, and extinction are severely biased respect to the real values. Moreover, the bias in the metallicity is particularly complex, being always toward more metal-poor values, but showing two clearly different regimes. Lower contributions of young stars lead to an anticorrelation between the measured and the true metallicity, whereas for higher contribution of young stars, we see a nearly linear correlation between the measured metallicity and the true metallicity. Nevertheless, the dynamical range of the measured metallicity is small, implying that overall, an extinction curve applied exclusively to young (and metal-rich) stellar populations lead to apparently lower metallicities. 

We stress here that although the specific magnitude of the biases found for the stellar population properties driven by this effect might depend on the assumptions made in this test (e.g., SSP models used to build our mock data), the general trend is preserved under different assumptions. Thus, this effect could indeed contribute to the unphysically low stellar metallicities measured for very young regions.

Finally, implementing a fitting routine that accounts for this feature (e.g., using a different extinction value for each kinematic component, in the case of pPXF) is beyond the scope of this paper. However, this test shows that addressing this issue would lead to more reliable SFHs, and thus improve the utility of these spectral fitting tools.

\subsection{Deciding on best fitting results}
\label{sec:best_fitting_decision}
In the previous sections, we have shown the output log LumW age, LumW [Z/H], E(B-V), and log $\Sigma_{*}$ maps obtained using different spectral libraries and approaches to fit our data in order to obtain a visualization of the resulting maps that these different approaches yield. We parametrize quantitatively the quality of these results, in terms of $\Delta \log$ age, $\Delta$[Z/H], $\Delta$E(B-V), and $\Delta \log \Sigma_{*}$, introduced earlier.

\begin{figure*}
\centering
 	\includegraphics[width=0.9\textwidth, trim=0 0 0 0, clip]{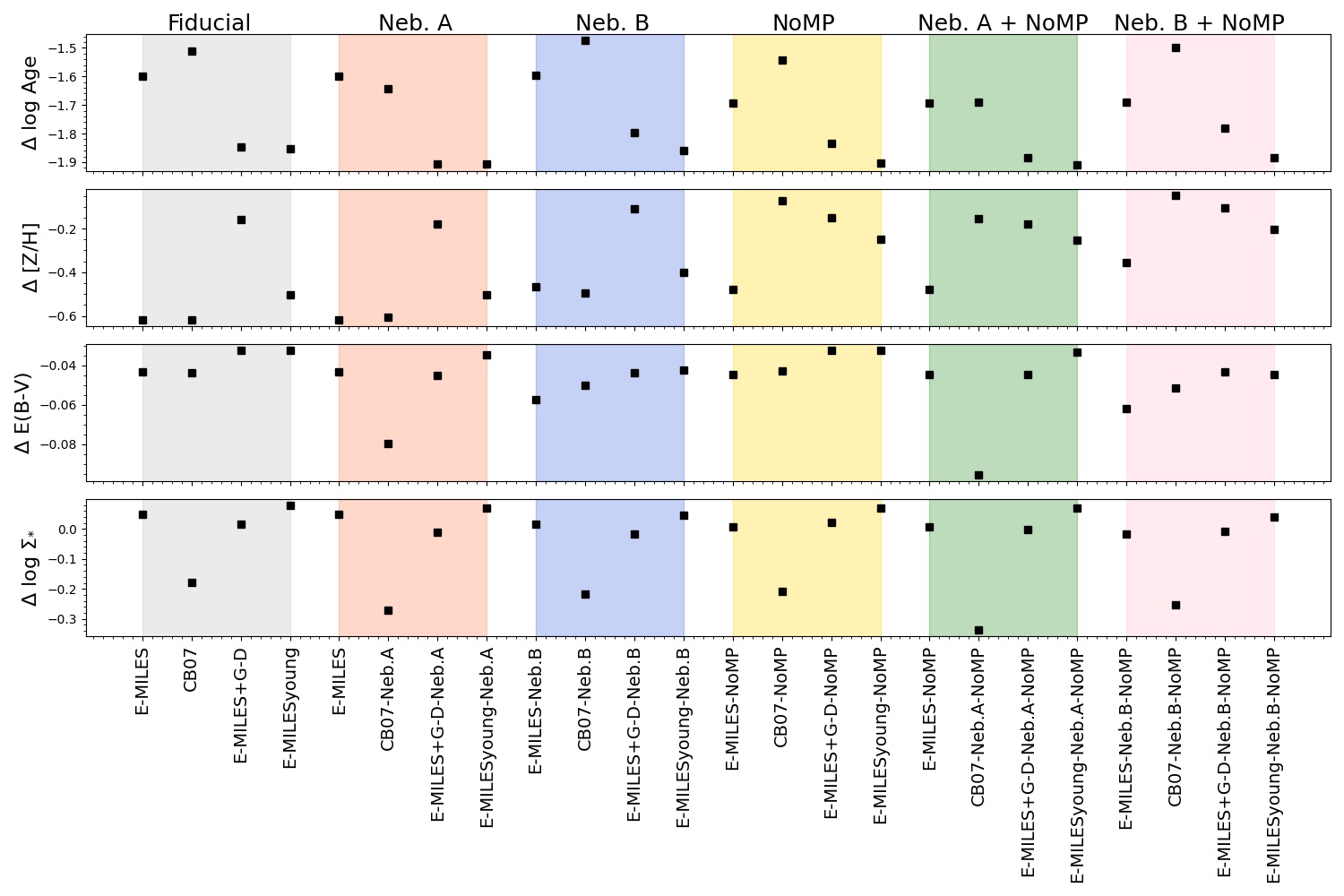}
     \caption[Values measured for $\Delta \log$ age, $\Delta$Z/H, $\Delta$E(B-V), and $\Delta \log \Sigma_{*}$, from the distribution of each one of these four quantities within the star-forming ring of NGC\,3351, and in its surrounding disk]{Values measured for $\Delta \log$ age, $\Delta$[Z/H], $\Delta$E(B-V), and $\Delta \log \Sigma_{*}$, following Eq.~\ref{eq:Delta_metric}, from the distribution of each one of these four quantities within the star-forming ring of NGC\,3351, and in its surrounding disk, using each one of the sets of templates described in Sec.~\ref{sec:improving_young}. We find that the MILES young templates, with the most metal-poor metallicity bin removed from the grid offers a good compromise between improving the young and metal-poor feature, reaching younger ages than our fiducial set of templates, and not leading to noticeable additional artifacts neither in the stellar mass surface density, nor in the stellar extinction map. See discussion in Sec.~\ref{sec:best_fitting_decision}.}
     \label{fig:quantify_test}   
\end{figure*}

Figure~\ref{fig:quantify_test} shows the values calculated for these quantities, for each one of the sets of templates used, under each different approach. The fiducial case for the four libraries is indicated in gray. The two implementations of nebular emission are marked in red and blue. The tests removing the most metal-poor templates are marked in yellow, and finally, the tests that include a combination of including nebular emission, and removing the most metal-poor templates are indicated with green and pink backgrounds.

The figure clearly shows that the highest (closest to zero) $\Delta$[Z/H] values (2nd row) are achieved when the most metal-poor templates are removed from the metallicity grid (yellow, green, and pink background). In particular, the CB07 templates yield the highest $\Delta$[Z/H] value among the set of templates tested. This is visually consistent with what can be seen in Fig.~\ref{fig:NoMP_test}. However, the last row shows that the CB07 templates also yield low stellar mass surface density values in the young regions (i.e., low $\Delta \log \Sigma_{*}$). This low-mass feature is also clearly visible in Fig.~\ref{fig:NoMP_test}. Furthermore, in some cases, the CB07 templates also yield particularly low E(B-V) values in the young regions. Out of the remaining set of templates (MILES + G-D and MILES young, with metal-poor metallicity bins removed), both show similarly low $\Delta \log$ age, and high $\Delta$[Z/H], without any noticeable artifacts neither in E(B-V) nor in stellar mass surface density. Since the MILES + G-D templates suffer from a low spectral resolution in the youngest ages, and this different spectral resolution is not properly treated by the DAP in its current state, we finally choose the MILES young templates (NoMP) for future analysis of the stellar populations. 

\begin{figure*}
\centering
 	\includegraphics[width=0.8\textwidth, trim=0 0 0 0, clip]{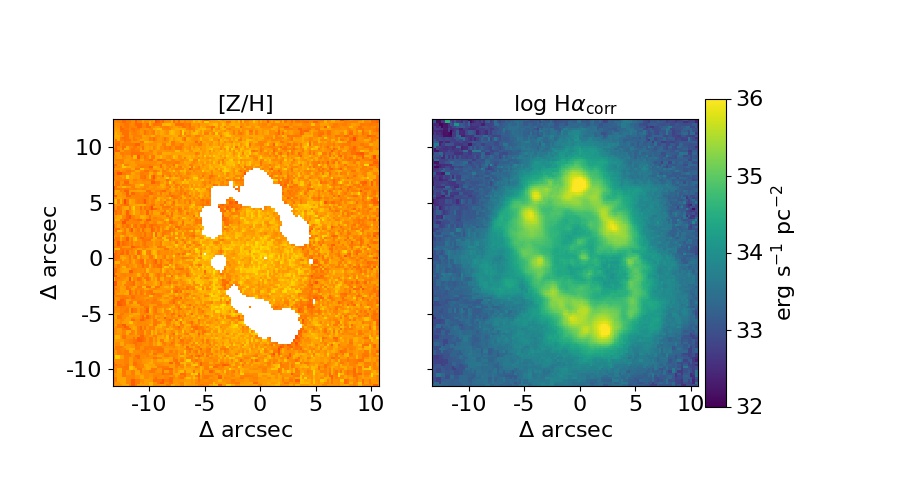}
     \caption[Extinction-corrected H$\alpha$ and metallicity map of the central region of NGC\,3351, after masking according to a extinction-corrected H$\alpha$ surface density emission threshold]{Stellar metallicity (left) map of the center of NGC\,3351, after masking pixels with an extinction-corrected H$\alpha$ surface density (right) emission higher than $10^{-35}$ erg s$^{-1}$ pc$^{-2}$. This value successfully characterizes the youngest and most problematic regions.}
     \label{fig:masking_example}   
\end{figure*}

Despite choosing the best set of templates according to our criteria, among those libraries tested, we acknowledge that the fitting of the youngest regions is still suboptimal, as the metal-poor metallicities persist for these regions. We consider these metallicities unreliably, and therefore mask them in further analysis. We have tested and adopted an extinction-corrected H$\alpha$ \footnote{We de-reddend the H$\alpha$ fluxes, assuming that under the absence of dust, H$\alpha$/H$\beta$ = 2.86, as appropriate for a case B recombination, temperature T = $10^4$ K, and density $n_e$ = 100 cm$^{-2}$, adopting a \cite{ODonnell1994} extinction law. See also \citet{Emsellem2021} for a detailed description on how the emission line measurements have been made.} surface density emission threshold for our masking scheme, as it traces ongoing star-formation and thus, the presence of young stellar populations. Specifically, we masked out from the metallicity maps those pixels in which we measure $\log \mathrm{H}\alpha_\mathrm{corr} >  35 $ erg s$^{-1}$ pc$^{-2}$\footnote{or $\log$ SFR $\gtrsim-0.63$ M$_{\odot}$ yr$^{-1}$ pc$^{-2}$, following \citet{Calzetti-book}}. We find that this value successfully identifies the youngest and most problematic regions. Figure~\ref{fig:masking_example} shows the extinction-corrected H$\alpha$ and LumW [Z/H] maps of the central region of NGC\,3351 after applying our masking scheme. The low metallicity pixels with high H$\alpha$ emission values are thus removed from our sample. Although this masking scheme can bias the measured mean metallicity in environments with significant presence of young regions (e.g., sp. arms and rings), it is necessary in order to keep our analysis free from this unphysical values.

There are ongoing efforts to further improve the SSP fitting in the regions dominated by young stellar populations. It is known that stellar populations of different ages are also characterized by different kinematic conditions, with older stars being dynamically hotter than younger stars \citep[see, e.g., ][]{Tarricq2021}. In this regard, Zhang et al. (in prep.) will present a detailed exploration on how these differences in the kinematic conditions across different stellar populations can bias the recovered SFHs. This will potentially lead to improvements in the methodology currently employed in the PHANGS-MUSE pipeline, by including multiple stellar kinematic components to reproduce the observed spectra. An additional possibility to significantly improve our SSP fitting is using available photometric data from the PHANGS-HST survey to constrain bluer wavelength ranges, valuable to shed light on the properties of young stellar populations \citep[see, e.g., ][]{GonzalezDelgado2005}. However, the inclusion of additional photometric data to the spectral fitting technique is not trivial. It requires further development of the currently available tools, and thus, it is beyond the scope of the analysis presented here.

\end{appendix}

\end{document}